\documentclass[]{aa}
\usepackage[utf8]{inputenc}
\usepackage[english]{babel}
\usepackage{graphicx}
\usepackage{graphbox}
\usepackage{xcolor}
\usepackage{soul}
\usepackage{booktabs} 
\usepackage{amssymb,amsmath}
\usepackage{array}
\usepackage{siunitx}
\usepackage{textgreek}
\usepackage{natbib}
\usepackage[varg]{txfonts}
\usepackage[normalem]{ulem}
\usepackage{url}
\usepackage[unicode=true]{hyperref}
\bibpunct{(}{)}{;}{a}{}{,} 
\hypersetup{
    colorlinks=true,
    linkcolor={blue!50!black},
    citecolor={blue!50!black}, 
    urlcolor={blue!80!black}   
}
\usepackage{etoolbox}
\begin{document}

\title{Simulating observable structures due to a perturbed interstellar medium in front of astrospheric bow shocks in 3D MHD}
\titlerunning{Simulating observable structures due to a perturbed ISM in front of astrospheric bow shocks in 3D MHD}
\author{L. R. Baalmann\inst{\ref{tp4}} \and K. Scherer\inst{\ref{tp4},\ref{pci}} \and J. Kleimann\inst{\ref{tp4}} \and H. Fichtner\inst{\ref{tp4},\ref{pci},\ref{rapp}} \and D. J. Bomans\inst{\ref{pci},\ref{rapp},\ref{astro}} \and K. Weis\inst{\ref{astro}}}
\institute{Ruhr-Universität Bochum, Fakultät für Physik und Astronomie, Institut für Theoretische Physik IV, 44780 Bochum, Germany\label{tp4}\\
e-mail: \texttt{lb@tp4.rub.de} \and
Ruhr-Universität Bochum, Research Department, Plasmas with Complex Interactions, 44780 Bochum, Germany\label{pci} \and
Ruhr Astroparticle and Plasma Physics (RAPP) Center, 44780 Bochum, Germany\label{rapp} \and 
Ruhr-Universität Bochum, Fakultät für Physik und Astronomie, Astronomisches Institut, 44780 Bochum, Germany\label{astro}}
\date{Received 3 November 2020 / Accepted 26 March 2020}
\abstract{While the shapes of many observed bow shocks can be reproduced by simple astrosphere models, more elaborate approaches have recently been used to explain differing observable structures.}{By placing perturbations of an otherwise homogeneous interstellar medium in front of the astrospheric bow shock of the runaway blue supergiant \textlambda~Cephei, the observable structure of the model astrosphere is significantly altered, providing insight into the origin of perturbed bow shock images.}{Three-dimensional single-fluid magnetohydrodynamic (MHD) models of stationary astrospheres were subjected to various types of perturbations and simulated until stationarity was reached again. As examples, simple perturbations of the available MHD parameters (number density, bulk velocity, temperature, and magnetic field) as well as a more complex perturbation were chosen. Synthetic observations were generated by line-of-sight integration of the model data, producing H\textalpha, $70\,\si{\micro m}$ dust emission, and bremsstrahlung maps of the perturbed astrosphere's evolution.}{The resulting shock structures and observational images differ strongly depending on the type of the injected perturbation and the viewing angles, forming arc-like protrusions or bifurcations of the bow shock structure, as well as rings, arcs, and irregular structures detached from the bow shock.}{} 
\maketitle

\section{Introduction}\label{sec:intro}

Three-dimensional (3D) magnetohydrodynamics (MHD) have recently come into focus as a viable tool for further examining and understanding the structures of astrospheres \citep[e.g.][]{katushkina18,gvaramadze18,scherer2020}. By changing these models' orientations with respect to an observer's line of sight (LOS), the generation of synthetic observations produces images similar to genuine data of actual bow shocks \citep[BSs; e.g.][]{meyer17,katushkina18,baalmann2020}. However, the shapes of many BSs \citep[see e.g.][]{kobulnicky16} look decidedly dissimilar to those created by the above approach. By introducing inhomogeneities and perturbations to the interstellar medium \citep[ISM; cf. e.g.][]{borrmann2005}, further observations of BSs can be reproduced on a case-by-case basis \citep[e.g.][]{gvaramadze18}. 

However, because of the variety of possible perturbations (see Sects.~\ref{sec:pshape} and \ref{sec:ptype}) and the lack of observational data that could constrain them, this approach requires a great deal of effort and insight for each observational image that is to be reproduced. To facilitate this process, the influence of different perturbations on a model astrosphere's structure must be understood. As a first step, in this paper simple inhomogeneities of all available MHD quantities are placed in the ISM in front of the model's BS; a more complex perturbation is examined as well. While the simple inhomogeneities (see Sect.~\ref{sec:simpleblobs}) are motivated by computational reasons, resulting in increases in the number density, temperature, bulk speed, or magnetic field by a factor of ten, the characteristics of the complex perturbation (see Sect.~\ref{sec:moreblobs}) correspond to those common for tiny scale atomic structures (TSASs) and tiny scale ionised structures (TSISs; cf. \citealt{stanimirovic2018} and references therein).

Because a significant number of ISM perturbations are the cause of (M)HD shocks, investigations into colliding shocks \citep{courant1948} are also of interest to this examination (see e.g. \citealt{fogerty2017}). While these studies are generalisable and applicable to various MHD scenarios, their origin lies outside the scenario of astrospheres, instead focusing on, for example, coronal mass ejections \citep[e.g.][]{lugaz2005}, interplanetary magnetic clouds \citep{xiong2009}, or colliding winds of binary systems \citep{kissmann2016}.

Within the scope of this work, \textlambda~Cephei is used as a prototype star to generate a large-scale astrosphere. As a runaway blue supergiant of spectral type O6.5, \textlambda~Cephei features a high velocity relative to a dense ISM as well as a strong stellar wind (SW), resulting in a BS distance on the scale of several parsecs. The simulated structures are examined via contour plots of the MHD values and by generating synthetic sky maps in H\textalpha, in $70\,\si{\micro m}$ dust emission, and, for selected cases, in bremsstrahlung.

An overview of the methodology used for simulating and examining the model astrospheres is given in Sect.~\ref{sec:methods}. The results are presented in Sect.~\ref{sec:res} and compared to the classification of wind-ISM interactions by \citet{cox12} in Sect.~\ref{sec:obs}. Conclusions are drawn in Sect.~\ref{sec:conc}.

\section{Methodology}\label{sec:methods}

A brief overview of the computational model is given in Sect.~\ref{sec:comp}, followed by details of the perturbations' computation (Sect.~\ref{sec:perturb}), shapes (Sect.~\ref{sec:pshape}), and types (Sect.~\ref{sec:ptype}). The methodology for analysing the simulations, including synthetic observations, is presented in Sect.~\ref{sec:analysis} and applied to the unperturbed models in Sect.~\ref{sec:basis}. Numerical diffusion as the expected main source of errors is examined in Sect.~\ref{sec:numdif}.

\subsection{Computational model}\label{sec:comp}

To model the astrosphere in 3D single-fluid MHD, the semi-discrete finite-volume \textsc{Cronos} code \citep{kissmann18} was used, solving the equations of ideal MHD, 
\begin{align}
 \frac{\partial n}{\partial t} + \nabla \cdot \left( n \vec{u}\right) &= 0 &,\label{eq:mhdcont}\\
 \frac{\partial}{\partial t} \left( mn \vec{u} \right) + \nabla \cdot \left( m n \vec{u} \otimes \vec{u}\right) + \nabla p + \frac{1}{\mu_0} \vec{B} \times \left( \nabla \times \vec{B} \right) &= \vec{0} &,\label{eq:mhdimp}\\
 \frac{\partial e}{\partial t} + \nabla \cdot \left[ \left( e + p + \frac{1}{2\mu_0} \left| \vec{B} \right|^2 \right) \vec{u} - \frac{1}{\mu_0}\left( \vec{u} \cdot \vec{B} \right) \vec{B}\right] &= 0 &,\label{eq:mhdenergy}\\
 \frac{\partial \vec{B}}{\partial t} - \nabla \times \left( \vec{u} \times \vec{B} \right) &= \vec{0} &,\label{eq:mhdindu} 
\end{align}
with an HLL Riemann solver and a second-order Runge-Kutta scheme, where: $n, \vec{u}, p, \vec{B}$, and $e$ are the number density, fluid velocity, thermal pressure, magnetic induction, and total energy density, respectively; $m$ and $\mu_0$ are the proton mass and the vacuum permeability, and $\otimes$ represents the dyadic product. The system was closed by the additional equations
\begin{align}
 e &= \frac{p}{\gamma -1} + \frac{1}{2}mn \left| \vec{u} \right|^2 + \frac{1}{2\mu_0}\left| \vec{B} \right|^2 &,\label{eq:mhdgas}\\
 \nabla \cdot \vec{B} &= 0 &,
\end{align}
where $\gamma=5/3$ is the polytropic index of a mono-atomic ideal gas. For examining the model, the temperature $T= p / (2n k_{\mathrm{B}})$ was evaluated instead of the thermal pressure, assuming an ideal gas with an additional factor $1/2$ from the assumption of thermodynamic equilibrium between electrons and protons ($p=p_{\mathrm{e}}+p_{\mathrm{p}}=2p_{\mathrm{p}}$), where $k_{\mathrm{B}}$ is the Boltzmann constant. Internally, however, \textsc{Cronos} uses the thermal energy density $e_{\mathrm{th}}=p/(\gamma-1)$ instead of the temperature $T$, which is reflected by the physical parameters available for the later perturbations. 

To allow for heating processes, the term
\begin{equation}\label{eq:heat}
 \Gamma=n^2 G_0 + n G_1
\end{equation}
was added to the right-hand side of Eq.~(\ref{eq:mhdenergy}) as described in \citet{kosinski2006}, following \citet{reynolds1999}. Here, $G_0=1\times10^{-24}\,\si{erg\,cm^3\,s^{-1}}$ accounts for photoionisation heating, and $G_1=1\times10^{-25}\,\si{erg\,s^{-1}}$ accounts for photoelectric heating by dust, Coulomb collisions with cosmic rays, and the dissipation of interstellar turbulence. Because $n\gg0.1\,\si{cm^{-3}}$ outside the astropause (AP; cf. Sect.~\ref{sec:basis}), the term $nG_1$ is negligible there. As with the heating, a cooling term following \citet{schure09} was used to incorporate radiative cooling for temperatures $T\in[10^{3.8}, 10^{8.16}]\,\si{K}$. For $T<10^4\,\si{K}$, no cooling was implemented. For $T\geq 10^{8.16}\,\si{K}$, the cooling value for $T=10^{8.16}\,\si{K}$ was used; however, no such temperatures occurred in any of the models.

The computational grid was arranged in spherical coordinates $(r, \vartheta, \varphi)$, producing a spherical polyhedron with equidistant intervals in the radius $r$ and equiangular intervals in the polar and azimuthal angles, $\vartheta$ and $\varphi$. While the entirety of the angular ranges $\vartheta\in[-90\si{\degree}, 90\si{\degree}]$ and $\varphi\in[-180\si{\degree}, 180\si{\degree}]$ were covered by the grid, the radial range $r\in[r_{\mathrm{SW}}, r_{\mathrm{ISM}}]$, $r_{\mathrm{SW}}> 0$ excluded the central volume of the star and its surroundings. The number of grid cells, $N_{r,\vartheta,\varphi}$, in each coordinate direction and the associated cell sizes,  $\Delta(r,\vartheta,\varphi)$, of the model were changed for different purposes, as displayed in Table~\ref{tab:boxvals}, corresponding to a low-resolution but high-performance grid (\texttt{lores}) and a high-resolution but low-performance one (\texttt{hires}) for more detailed examinations. A third grid with a more similar resolution in $r$ and $\vartheta,\varphi$ (\texttt{eqres}) was used as a reference for the examination of numerical diffusion (cf. Sect.~\ref{sec:numdif}).

The astrospheres of different stars can be simulated by choosing different boundary conditions of the modelling equations for the inner and outer edges of the grid (at $r=r_{\mathrm{SW, ISM}}$), corresponding to the spherical SW and the ISM boundary. In this paper, \textlambda~Cephei is used as an example of a particularly strong ISM inflow and SW \citep[cf.][]{baalmann2020}; the corresponding boundary conditions are summarised in Table~\ref{tab:modvals}. The ISM was assumed to be homogeneous with an oblique magnetic field orientated with respective angles $\vartheta_{B}$ and $\varphi_{B}$. By setting the angles of the ISM fluid inflow to $\vartheta_u=90\si{\degree}$ and $\varphi_u=180\si{\degree}$, the simulation's $x$-axis was aligned with the stellar motion (i.e. the upwind direction). Accordingly, the negative $x$-axis points in the downwind or tail direction. The SW was set up as a spherically symmetric outflow with only a radial component, its magnetic field following Parker's spiral \citep{parker1958}. For both the SW and the ISM, $n$ and $T$ were fixed at the respective boundaries, $r_{\mathrm{SW}}$ and $r_{\mathrm{ISM}}$. More details about the simulation setup are presented in \citet{scherer16}.

\begin{table}
\caption{\label{tab:boxvals}Simulation sizes and resolutions of the various models.}
\centering
\begin{tabular}{rllll}
\toprule\toprule
\multicolumn{2}{l}{Parameter} & \texttt{lores} & \texttt{hires} & \texttt{eqres} \\ \midrule
$N_r$ & [cells] & 1140 & 2048 & 512\\
$N_{\vartheta}$ & [cells] & \phantom{00}30 & \phantom{00}64 & 128\\
$N_{\varphi}$ & [cells] & \phantom{00}60 & \phantom{0}128 & 256\\ \midrule
$\Delta r$ & $[\si{mpc}]$ & $4.3421$ & $2.4170$ & $9.66797$\\
$\Delta(\vartheta,\varphi)$ & $[\si{\degree}]$ & $6$ & $2.8125$ & $1.40625$ \\
\bottomrule
\end{tabular}
\tablefoot{Number of cells $N_{r,\vartheta,\varphi}$ in $r$, $\vartheta$, $\varphi$, as well as radial ($\Delta r$) and angular ($\Delta(\vartheta,\varphi)$) resolutions of the different grids.}
\end{table}

\begin{table}
\caption{\label{tab:modvals}Boundary values chosen for the model of \textlambda~Cephei.}
\centering
\begin{tabular}{lllr}
\toprule\toprule
 \multicolumn{2}{l}{Variable} & SW & ISM  \\  \midrule
$r$&$\left[\si{pc}\right]$ & $0.05$ & $5$ \\
$n$&$\left[\si{cm^{-3}}\right]$ & 3.4 & 11 \\
$u$&$\left[\si{km/s}\right]$ & 2500 & 80 \\
$T$&$\left[10^3\,\si{K}\right]$ & 10 & 9 \\
$B$&$\left[\si{nT}\right]$ & $3\times 10^{-3}$ & 1 \\ 
$\vartheta_u$&$\left[\si{\degree}\right]$ & & 90  \\
$\varphi_u$&$\left[\si{\degree}\right]$ & & 180  \\
$\vartheta_B$&$\left[\si{\degree}\right]$ & & 30  \\
$\varphi_B$&$\left[\si{\degree}\right]$ & & 150 \\
\bottomrule
\end{tabular}
\tablefoot{$n,T,u$, and $B$ are the number density, temperature, bulk speed, and magnetic flux density, respectively. $r$ is the radius at which the respective boundary conditions are met. $\vartheta_{u,B}$ and $\varphi_{u,B}$ are the angles of the homogeneous ISM.}
\end{table}

\subsection{Computation of perturbations}\label{sec:perturb}

Once the \textsc{Cronos} run reached stationarity, perturbations were injected into the ISM by externally changing the parameter values of pre-selected cells. The modified simulation file was then used as the initial set of values for a new \textsc{Cronos} run, either computed until the perturbed model reached stationarity again or until another custom condition was satisfied. 

Because the model of \textlambda~Cephei is rotationally symmetric about its inflow axis to a reasonable degree \citep{baalmann2020}, the axis parallel to the homogeneous ISM inflow that passes through the star, examining its ecliptic plane (cf. Sect.~\ref{sec:basis}), is a useful approach for understanding the entire 3D structure. Therefore, all injected perturbations were centred in the ecliptic plane ($\vartheta=0\si{\degree}$). 

\subsection{Shapes of perturbations}\label{sec:pshape}

Two basic shapes were chosen as the domain in which the ISM was to be perturbed: boxes and spheres. The box represents the small-scale limiting case of any shape on a finite, rectangular grid. Because the grid is spherical and not Cartesian, a box with coordinate lengths $(N_{r,1},N_{\vartheta,1},N_{\varphi,1})$ is not strictly cuboid but slightly trapezoid, its volume and shape depending on the coordinates $(r_1, \vartheta_1, \varphi_1)$ at which it is placed. As a small-scale limiting case, $N_{\vartheta,1}=N_{\varphi,1}=1$ was used for all box-shaped perturbations, corresponding to angular extents of $\Delta(\vartheta,\varphi)_1 =\Delta(\vartheta,\varphi)=6\si{\degree}$ for the low-resolution models (cf.~Table~\ref{tab:boxvals}), and $N_{r,1}$ was varied to gain different shapes (cf. Table~\ref{tab:blobs}). For spherical perturbations, all model cells within a radius $R_1$ around the perturbation's centre $(r_1, \vartheta_1, \varphi_1)$ were set to the perturbed values. 

\subsection{Types of perturbations}\label{sec:ptype}

Inside the perturbation area, the available physical parameters of the number density $n$, the thermal energy density $e_{\mathrm{th}}$, the bulk velocity $\vec{u}$, and the magnetic flux density $\vec{B}$ were set to their perturbed values. However, not all perturbations are useful for physical modelling. From a physical standpoint, injecting a perturbation at an arbitrary position inside the homogeneous ISM necessitates it either being stable enough to have moved to that position or having been generated there. For example, a perturbation of the thermal energy density $e_{\mathrm{th}}$ without any further changes to $n,\vec{u},$ or $\vec{B}$ would thus correspond to a perturbation of the temperature $T$ that would dissolve after a short amount of time. A perturbation of the magnetic field $\vec{B}$ must be divergence-free yet would generally still dissolve without any accompanying changes to other parameters due to magnetic pressure. 

A change to the bulk velocity $\vec{u}$ might naively be assumed to correspond to a clump of material moving at a different velocity through the ISM. However, because of the single-fluid approach, the entire content of a model cell would move with this new velocity, causing an immediate accumulation of material in front of as well as a cavity behind the perturbed area, in regards to the moving direction. Also, velocities must firmly lie in the non-relativistic regime because relativistic processes are not yet supported for the simulation.

Mathematically, a perturbation would remain stable if it did not change the MHD Eqs. (\ref{eq:mhdcont})-(\ref{eq:mhdindu}) from the unperturbed case. In HD, this is true for any perturbation of the number density $n$ because it only functions as a scaling parameter. However, in MHD, the magnetic field in Eqs.~(\ref{eq:mhdimp}) and (\ref{eq:mhdenergy}) generally rules this out. For a weak magnetic field (i.e. if the magnetic pressure is much lower than the thermal pressure), a perturbation in $n$ is still reasonably stable. 

The MHD equations would likely allow for other stable configurations as well. However, finding these configurations was not part of this investigation. Instead, if a perturbation was found to be unstable, it was assumed to be generated in situ, for example by turbulence. 

\subsection{Model analysis}\label{sec:analysis}

The models were examined via contour plots, commonly in the ecliptic plane, via synthetic observations as described in \citet{baalmann2020}, and via viewing their 3D structure. While the last approach gives an overview of the entire model, it is insufficient for properly quantifying the results, making the other methods necessary. 

For comparisons with observational images, synthetic observations of the various models are indispensable. Synthetic sky maps were created by shifting and rotating the model to the desired position and orientation and then dividing the covered area of the virtual sky into pixels on a two-dimensional (2D) grid. By calculating the emission of each model cell and their respective flux densities through the pixels, the models were projected. However, because taking the full geometry of every model cell into account would require an unreasonable amount of computing power, only the positions of each model cell's centre were regarded when dividing the projection into the 2D grid. Thus, model cells that would be large enough to appear in two or more pixels were only counted for the pixel in which their centres were located. Because of the large number of model cells, it was assumed that the effects of multiple model cells would roughly cancel out; the resulting errors are therefore insignificant \citep[cf.][]{baalmann2020}. The resolution of the 2D pixel grid is dependent on the largest distance of next-neighbour cells in the 3D model and is thus limited by the angular resolution of the model. To improve the 2D pixel resolution, the 3D model was linearly interpolated with a factor of four in the radial directions before generating synthetic sky maps. This does, however, introduce visible artefacts with the \texttt{lores} grid that do not correspond to physical structures, for example a ring of higher flux density with a radius of $0.2\si{\degree}$ around the model centre and sharp jags perpendicular to the radial direction at the BS (cf. Sect.~\ref{sec:basis}). Depending on the geometry, Moir\'e patterns typical for two overlaying grids are visible (cf. Sect.~\ref{sec:obs}).

The primary observable calculated with this approach was the H\textalpha\ flux,
\begin{equation}\label{eq:rec}
    j_{\text{H\textalpha}}=\alpha_{\mathrm{eff},3\rightarrow 2}(T) \cdot h\nu_{\text{H\textalpha}} n^2 \cdot \frac{V}{4\pi d^2}\ ,
\end{equation}
where $V$ and $d$ are the volume and distance to the observer of the respective model cell, $h$ is the Planck constant, $\nu_{\text{H\textalpha}}=456.81\,\si{THz}$ is the H\textalpha\ frequency, and $\alpha_{\mathrm{eff},3\rightarrow 2}$ is the temperature-dependent effective recombination rate coefficient, calculated via
\begin{equation}
    \alpha_{\mathrm{eff},3\rightarrow 2}(T) \approx \alpha_3(T) + \sum\limits_{n=4}^{16} \alpha_n(T)\, P_{n,3} \ ,
\end{equation}
with $\alpha_i(T)$ the temperature-dependent recombination rate coefficient of state $i$, and $P_{i,k}$ the branching ratio from state $i$ through state $k$. The $\alpha_i(T)$ were taken from \citet{mao16} and the $P_{i,k}$ computed as in \citet{baalmann2020} with data from \citet{wiese2009}. The resulting curve for the effective recombination rate coefficient is valid for temperatures $T\in[10^1, 10^8]\,\si{K}$ and decreases strictly monotonously from $\alpha_{\mathrm{eff},3\rightarrow 2}(10^1\,\si{K})\approx 3.6\times 10^{-12}\,\si{cm^3/s}$ to $\alpha_{\mathrm{eff},3\rightarrow 2}(10^8\,\si{K})\approx 1.1\times 10^{-18}\,\si{cm^3/s}$.

\begin{figure}
    \centering
    \includegraphics[width=\linewidth]{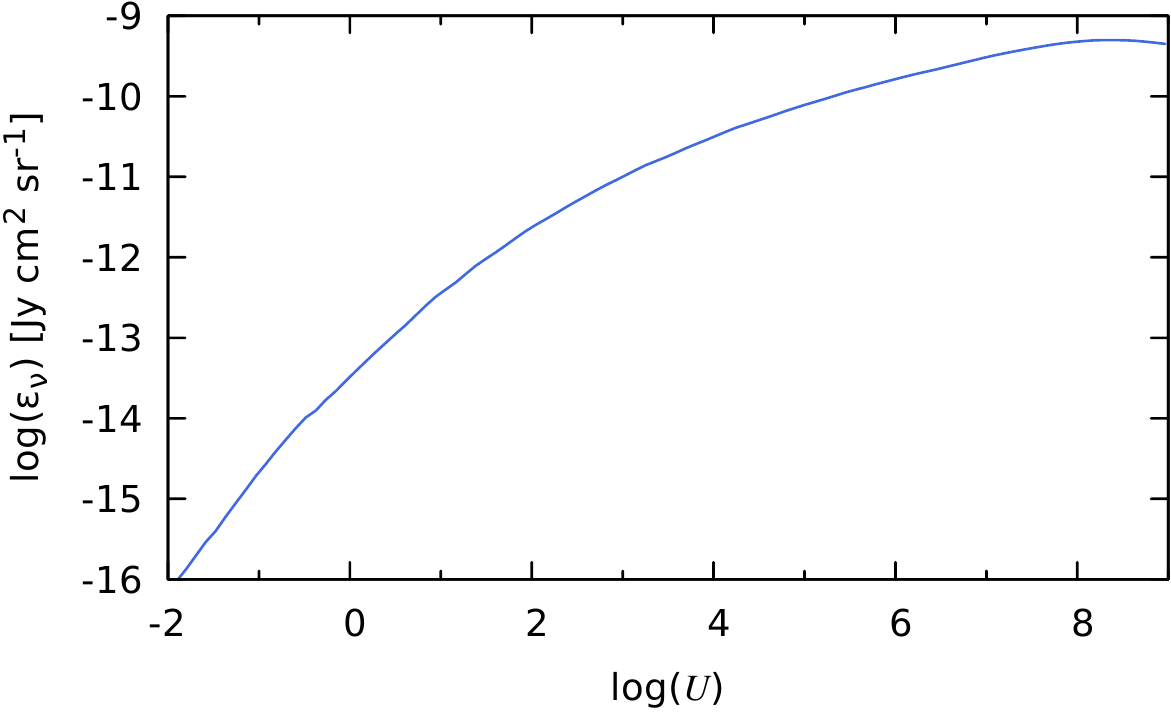}
    \caption{Spectral emissivity of dust at $70\,\si{\micro m}$ as a function of the dimensionless radiation field, $U=L_{\text{\textlambda\,Cep}}/(4\pi r^2 k)$, reproduced following \citet{henney2019}.}
    \label{fig:henneydust}
\end{figure}

All models were also analysed in the spectral emission due to dust at a wavelength of $70\,\si{\micro m}$. The requisite spectral emissivity $\varepsilon_{\nu}$ is plotted in Fig.~\ref{fig:henneydust} as a function of the dimensionless radiation field $U$, reproduced following \citet{henney2019}, where a variety of models were computed for the spectral energy distribution of an OB star using the photoionisation code Cloudy \citep[e.g.][]{ferland+2017}. At low $U$, the emissivity was extrapolated with data from \citet{draine2007}, adapted by \citet{henney2019}. The dimensionless radiation field $U=L_{\text{\textlambda\,Cep}}/(4\pi r^2 c u_{\mathrm{MMP83}})$ depends only on the distance $r$ from the star; $L_{\text{\textlambda\,Cep}} = 6.3\times 10^5 L_{\odot} = 2.424\times 10^{39}\,\si{erg/s}$ is the stellar luminosity of \textlambda~Cephei, $c$ the speed of light, and $u_{\mathrm{MMP83}}=0.0217\,\si{erg\,s^{-1}\,cm^{-2}}/c$ the energy density of the interstellar radiation field for wavelengths $\lambda<8\,\si{\micro m}$ in the solar neighbourhood (\citealp{MMP1983}, \citealp{henney2019}). The thus calculated spectral emissivity $\varepsilon_{\nu}$ is the emitted spectral power per hydrogen atom and unit solid angle. The spectral flux for dust emission was calculated by
\begin{equation}
 j_{\nu}=\varepsilon_{\nu} n V \frac{4\pi}{4\pi d^2} = \varepsilon_{\nu} n \frac{V}{d^2} \ .
\end{equation}
This approach is only valid under the assumption that the distribution of gas, or more specifically of ionised hydrogen, can be used as a tracer for the distribution of dust. In general, one must allow for the partial decoupling of gas and dust \citep{katushkina2019}. However, for BSs around luminous blue supergiants with an extent of multiple parsecs, decoupling does not occur \citep{henney2019ii}.

\begin{figure}
    \centering
    \includegraphics[width=\linewidth]{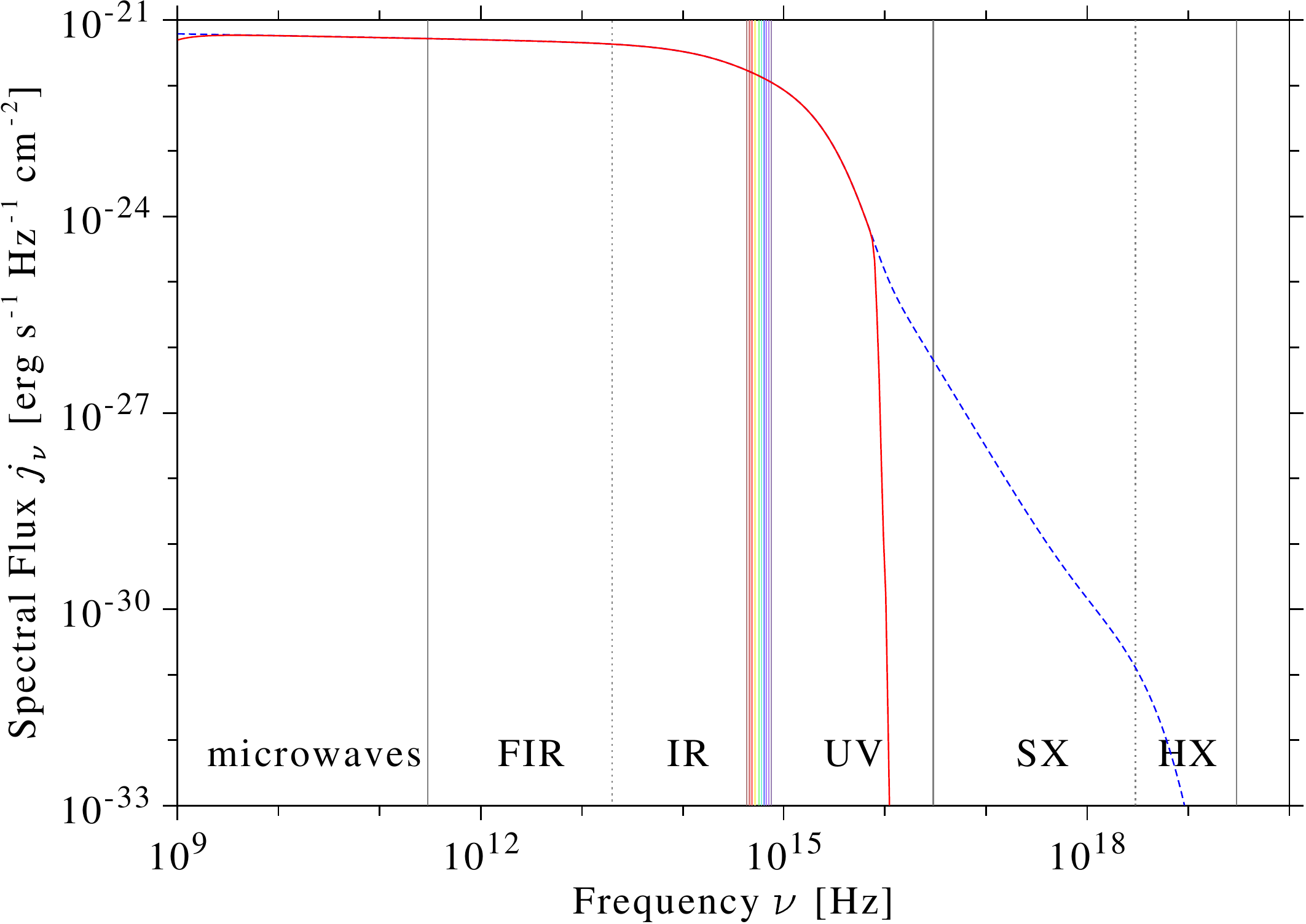}
    \caption{Bremsstrahlung spectrum of the \texttt{lores} grid's final time step, unmodulated (dashed blue curve) and modulated by a homogeneous ISM between the model and the observer with $T=T_{\mathrm{ISM}}$ (solid red curve). Spectral domains of microwaves, far infrared (FIR), infrared (IR), ultraviolet (UV), soft X-ray (SX), and hard X-ray (HX) are delimited by vertical lines; the shaded area corresponds to the optical.}
    \label{fig:brsspec}
\end{figure}

For select models, the thermal bremsstrahlung flux,
\begin{equation}\label{eq:brs}
    j_{\mathrm{brs}}=1.435\times 10^{-27}\,\si{\frac{erg\,cm^3}{s\,\sqrt{K}}}\cdot \sqrt{T} n^2 g_{\mathrm{ff}}(T) \cdot \frac{V}{4\pi d^2}\ ,
\end{equation}
was calculated as well, with $g_{\mathrm{ff}}(T)$ the temperature-dependent Gaunt factor for free-free transitions \citep{rybicki1979}. The Gaunt factor was calculated following \citet{vanhoof14} and resulted in values $g_{\mathrm{ff}}(T)\in[1.0, 1.5[$, so the usual approximation $\bar{g}=1.2$ holds \citep[cf.][]{baalmann2020}. The $j_{\mathrm{brs}}$ is the flux over the entire electromagnetic spectrum without any modulation by, for example, (self-)absorption. As a simple approximation, the bremsstrahlung spectrum can be assumed to be flat up to a critical frequency,
\begin{equation}
    \nu_{\mathrm{crit}}=\frac{4\pi m_{\mathrm{e}} v^2}{h} \ ,
\end{equation}
where $v$ is the thermal electron speed of the respective model cell and $m_{\mathrm{e}}$ is the electron mass \citep{longair92}. The exact spectral flux follows as
\begin{equation}
    j_{\mathrm{brs}}^{\nu}=\kappa_{\nu} \cdot \frac{V}{4\pi d^2} \ ,
\end{equation}
with the spectral emissivity \citep{rybicki1979}:
\begin{align}
    \kappa_{\nu}=&\frac{64}{\sqrt{54}}\frac{\sqrt{\pi^3}q_{\mathrm{e}}^6}{c^3 m_{\mathrm{e}}^2} \sqrt{\frac{m_{\mathrm{e}}}{k_{\mathrm{B}}T}} g_{\mathrm{ff}}(\nu, T)\, n^2 \exp\left(-\frac{h\nu}{k_{\mathrm{B}}T}\right) \\
    \approx\,& 6.842\times 10^{-38}\,\si{erg\,cm^3\,\sqrt{K}}\cdot\nonumber\\
    &\cdot \sqrt{T^{-1}}g_{\mathrm{ff}}(\nu, T)\, n^2 \exp\left(-\frac{h\nu}{k_{\mathrm{B}}T}\right) \ , \nonumber
\end{align}
where $q_{\mathrm{e}}, \varepsilon_0, c$, and $k_{\mathrm{B}}$ are the elementary charge, electric constant, speed of light, and Boltzmann constant, $\nu$ is the frequency, and $g_{\mathrm{ff}}(\nu,T)$ is the frequency- and temperature-dependent Gaunt factor taken from \citet{vanhoof14}. Absorption was incorporated by
\begin{equation}
    j_{\mathrm{brs}}^{\nu,\mathrm{m}}=\frac{\kappa_{\nu}V}{4\pi d^2}\left(1-d \chi_{\nu}\right) \ ,
\end{equation}
where the term $d\chi_{\nu}$ describes the fraction of radiation absorbed by a homogeneous medium between source and observer at distance $d$; negative values of $j_{\mathrm{brs}}^{\nu,\mathrm{m}}$ were set to $0$. The absorption coefficient $\chi_{\nu}$, calculated following \citet{longair92} via
\begin{equation}
    \chi_{\nu}=\frac{\kappa_{\nu} c^2}{8\pi h\nu^3}\cdot\left[\exp\left(\frac{h\nu}{k_{\mathrm{B}}T_{\mathrm{ISM}}}\right) -1\right] \ ,
\end{equation}
was assumed to be constant from the respective model cell to the observer within a homogeneous ISM with temperature $T_{\mathrm{ISM}}$.
The resulting spectra for the \texttt{lores} grid's final time step (cf. Sect.~\ref{sec:basis}), both with and without modulation by absorption, are presented in Fig.~\ref{fig:brsspec}. While the spectrum is not flat, it does remain in the same order of magnitude up to a frequency of roughly $\log(\nu_{\mathrm{crit}}\,\si{[Hz]})\approx 14.5$.

Equations~(\ref{eq:rec}) and (\ref{eq:brs}) give the fluxes of the respective model cells; summation over all model cells that appear inside a pixel gives the total flux of that pixel. However, because the different resolutions of the \texttt{lores} and \texttt{hires} models lead to different 2D grid resolutions, the total fluxes at different grid resolutions of otherwise identical models differ by the ratio of the respective pixels' solid angles. To be able to compare models of different 2D grids, the respective flux densities were obtained by dividing the fluxes as calculated by Eqs.~(\ref{eq:rec}) and (\ref{eq:brs}) by the respective pixels' solid angles: $5.07\times10^{-8}\,\si{sr}$ for the \texttt{lores} grid and $1.00\times10^{-8}\,\si{sr}$ for the \texttt{hires} grid.

\subsection{Unperturbed models}\label{sec:basis}

\begin{figure}
    \centering
    \includegraphics[scale=0.15]{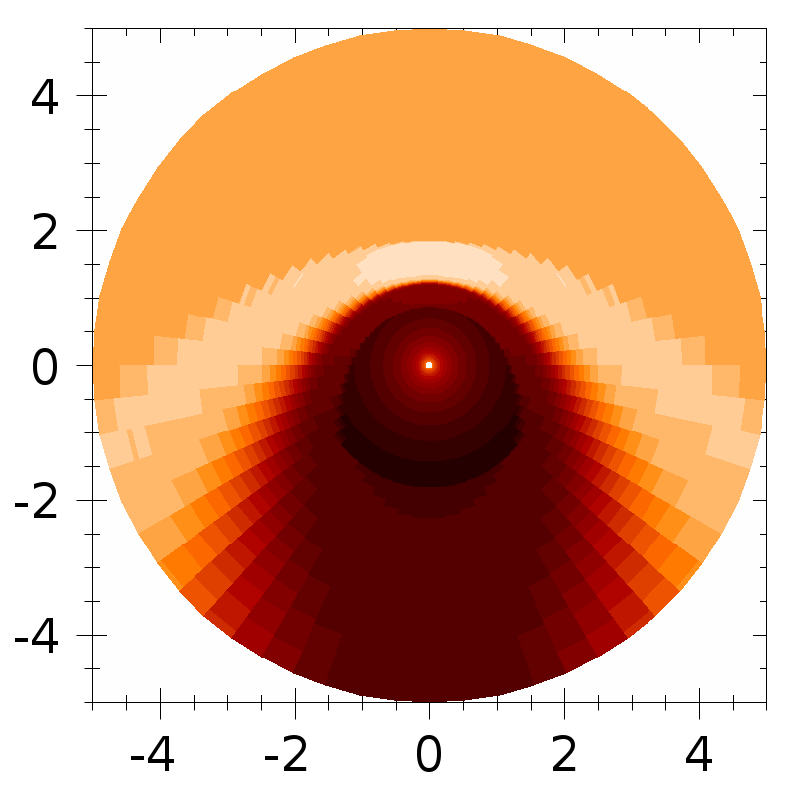}\hfill
    \includegraphics[scale=0.15,trim={26mm 0 0 0},clip]{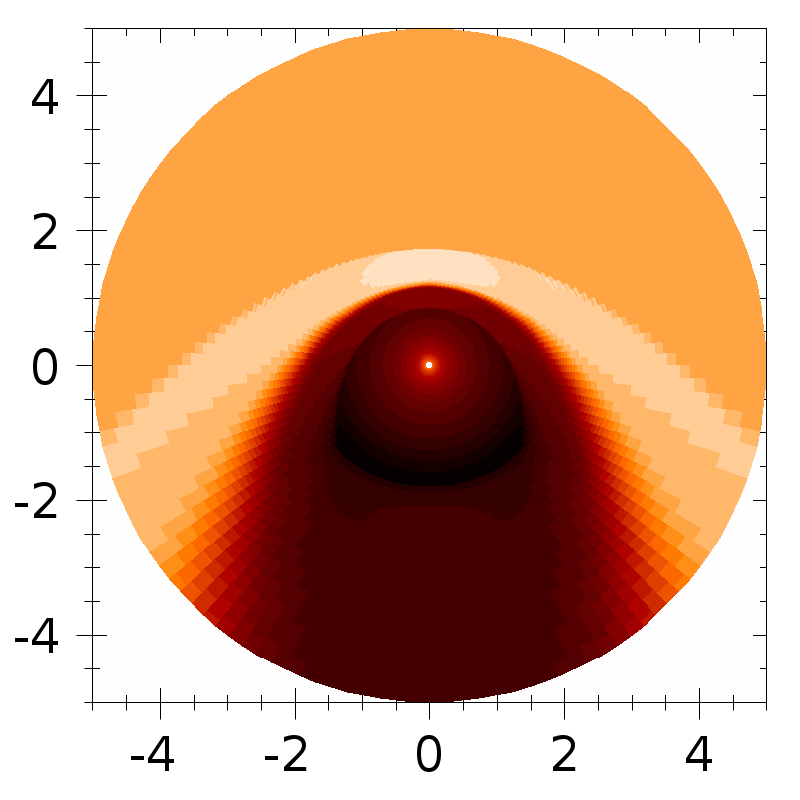}\hfill
    \includegraphics[scale=0.15]{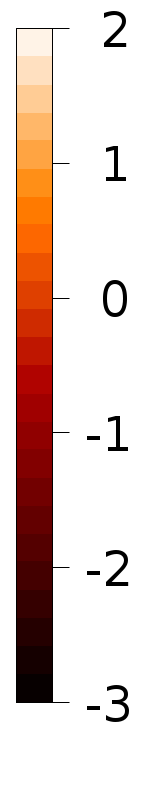}
    
    \raggedright
    \includegraphics[scale=0.275,align=c]{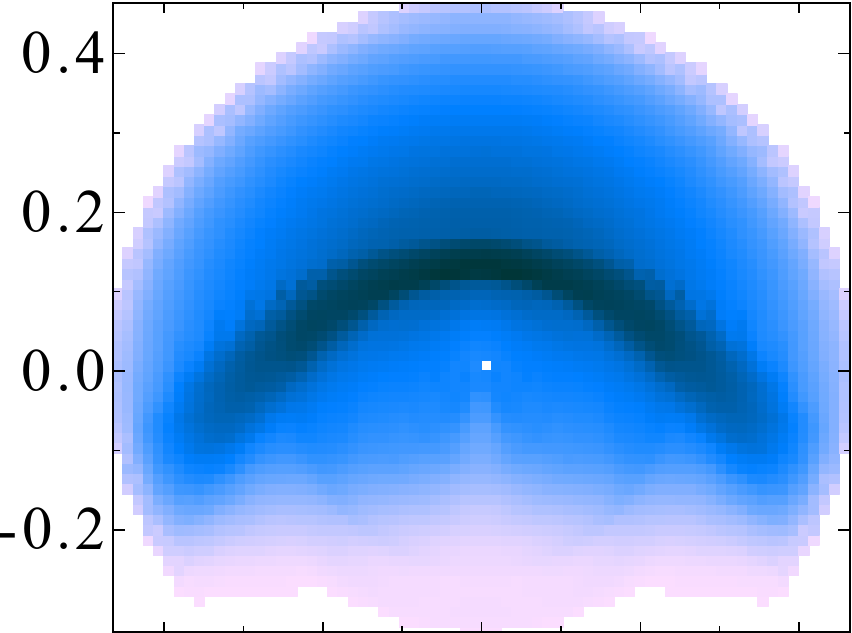}
    \includegraphics[scale=0.275,align=c]{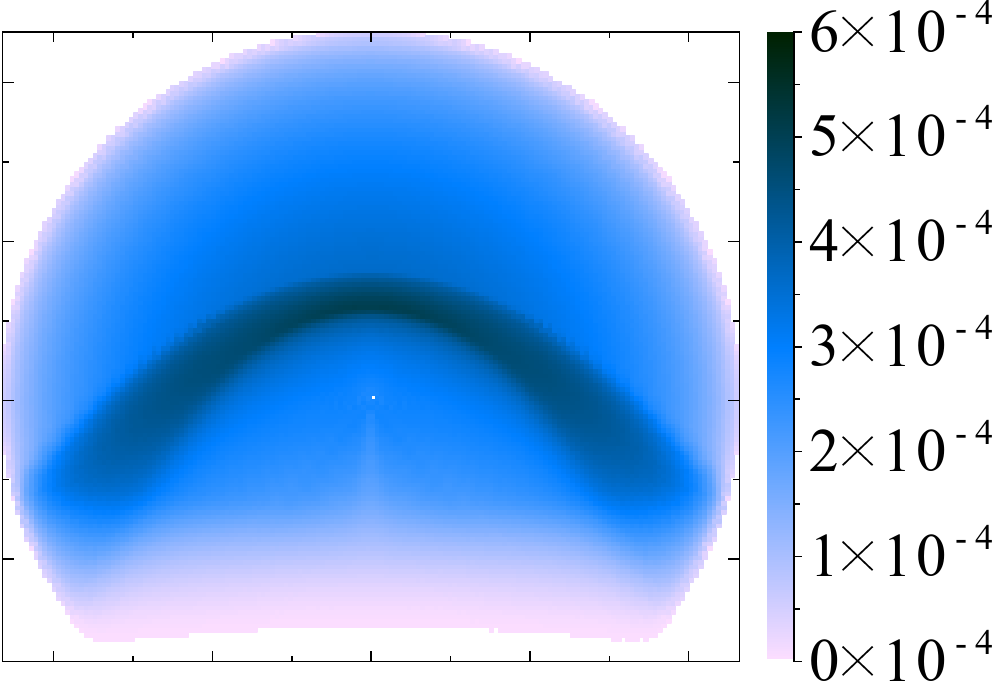}
    \includegraphics[scale=0.275]{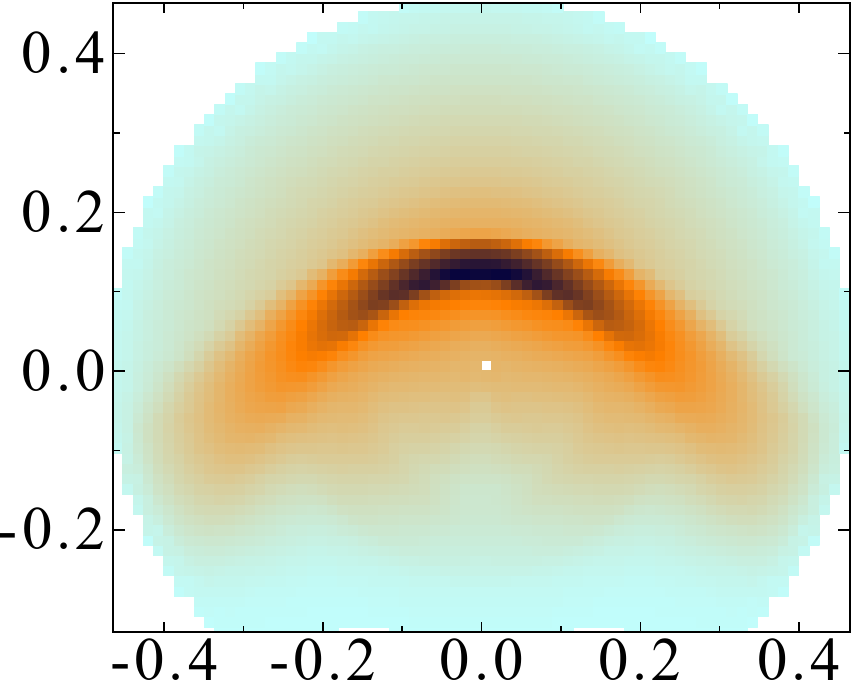}
    \includegraphics[scale=0.275]{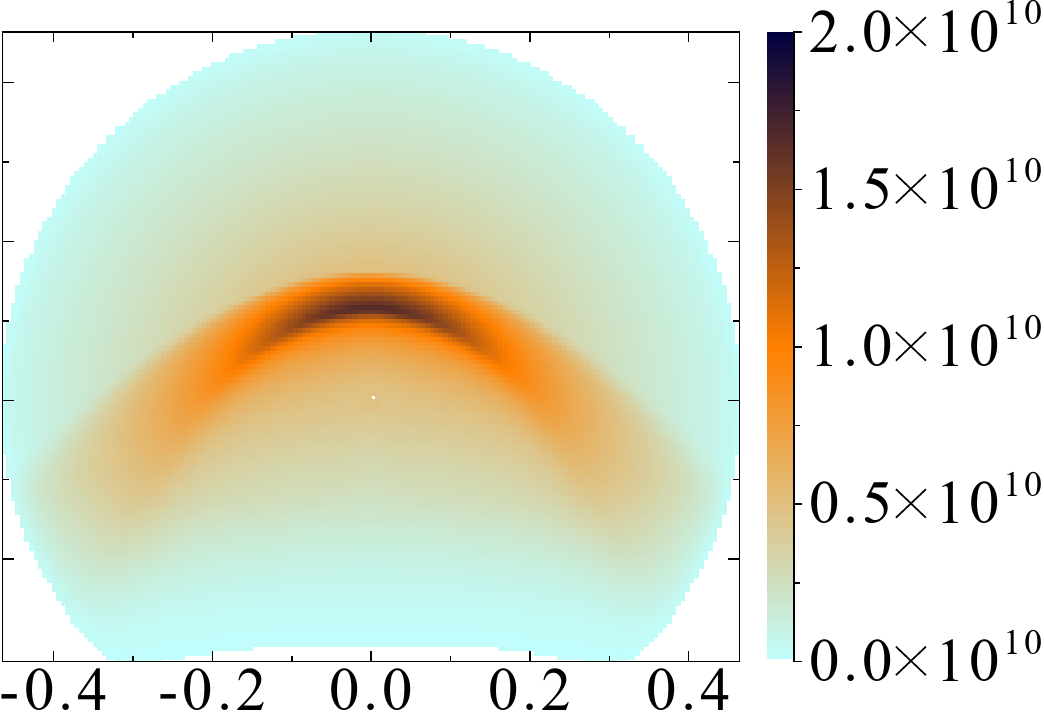}
    \caption{Stationary state of the unperturbed astrosphere. Top row: Contour plots of $\log(n [\si{cm^{-3}}])$ for the final time steps of the \texttt{lores} (left) and \texttt{hires} (right) grids in the ecliptic plane (distances in $\si{pc}$); the cornered structure reflects the model cells. Centre row: Synthetic face-on observations of the respective planes in H\textalpha\ $[\si{erg\,cm^{-2}\,s^{-1}\,sr^{-1}}]$. Bottom row: Synthetic observations in infrared dust emission at $70\,\si{\micro m}$, $[\si{Jy\,sr^{-1}}]$ (angular extent in degrees). At a distance of $617\,\si{pc}$, $1\si{\degree}$ corresponds to $10.8\,\si{pc}$. The solid angles per pixel are $5.07\times10^{-8}\,\si{sr}$ (\texttt{lores}) and $1.00\times10^{-8}\,\si{sr}$ (\texttt{hires}).}
    \label{fig:basis}
\end{figure}

Contour plots of the \texttt{lores} and the \texttt{hires} grids' final time steps in the respective ecliptic plane are presented in Fig.~\ref{fig:basis}. The difference in the tail direction (downwards from the centre) is due to the different times of the models. The \texttt{lores} model (left) has reached complete stationarity at $408\,\si{kyr}$, whereas the \texttt{hires} model (right) is not yet fully stationary in the tail direction at $262\,\si{kyr}$. 

In the contour plots of the models' ecliptic plane, displayed in the upper row of Fig.~\ref{fig:basis}, the star is located at the centre of each image, though it and its surrounding circular area with radius $r_{\mathrm{SW}}=0.05\,\si{pc}$ are not included in the model. The homogeneous ISM inflow comes from the topside of the image; this is the topmost domain in both models of $n=11\,\si{cm^{-3}}$, bordered at the bottom by the BS. Moving away from the star, the area with steadily decreasing density is the unperturbed SW, bordered at the top and sides by the termination shock (TS) and at the bottom by the Mach disk (MD). The area of higher densities compared to the homogeneous ISM, situated below the BS, is the outer astrosheath, while the area of higher densities compared to the SW's outer edge outside the TS and MD is the inner astrosheath. The inner and outer astrosheaths touch at the arc with $n\approx 1\,\si{cm^{-3}}$, which is the AP, separating stellar from interstellar material. Along the inflow axis, the TS is located at $r_{\mathrm{TS}}=0.86\,\si{pc}$, the AP at $r_{\mathrm{AP}}=1.23\,\si{pc}$, and the BS at $r_{\mathrm{BS}}=1.73\,\si{pc}$, measured for the \texttt{hires} model. For the \texttt{lores} model, these distances are slightly shifted outwards, to $0.87$, $1.25$, and $1.84\,\si{pc}$, respectively. The MDs are located at $r_{\mathrm{MD}}=1.80\,\si{pc}$ (\texttt{hires}) and $1.81\,\si{pc}$ (\texttt{lores}).
The centre and the bottom row of Fig.~\ref{fig:basis} display synthetic face-on observations of the same planes for the models at a distance of $617\,\si{pc}$ in H\textalpha\ (centre row) and in dust emission (bottom row). As described in \citet{baalmann2020}, the largest part of the H\textalpha~ emission comes from the outer astrosheath, resulting in the arc shape typically identified as the BS. Because the dust emission depends on the distance from the star in addition to the number density, the region of the outer astrosheath close to the AP near the inflow axis is brighter in dust emission than the regions farther from the star. This is in agreement with the results of \citet{katushkina18}, where dust was kinetically modelled. The ring of high flux densities in both radiation types with a radius of $0.2\si{\degree}$ around the model centre and the jags perpendicular to the radial direction at the BS are artefacts introduced by the interpolation (cf. Sect.~\ref{sec:analysis}).

\subsection{Numerical diffusion}\label{sec:numdif}

Because of the comparatively coarse resolution in $\vartheta$ and $\varphi$, numerical diffusion is expected to play a significant role and must be taken into account while examining the calculations. To this effect, a homogeneous and non-moving ISM ($\vec{u}=\vec{0}$, $\vec{B}=\vec{0}$) on the different grids was used with a boxy perturbation in the number density ($n_1=10n_0$) of $N_{r,1}\times N_{\vartheta,1}\times N_{\varphi,1}=50\times 1\times 1$ cells (\texttt{lores}) and $N_{r,1}\times N_{\vartheta,1}\times N_{\varphi,1}=90\times 2\times 2$ cells (\texttt{hires}) in size. As a reference, a third grid (\texttt{eqres}) with a higher angular but lower radial resolution was examined as well; here the perturbation was $N_{r,1}\times N_{\vartheta,1}\times N_{\varphi,1}=30\times 4\times 4$ cells in size. The perturbations were injected in the ecliptic plane close to the outer boundary of the model, centred at $4.65\,\si{pc}$ (\texttt{lores}), $4.61\,\si{pc}$ (\texttt{hires}), and $4.74\,\si{pc}$ (\texttt{eqres}), respectively.

\begin{figure}
    \centering
    \includegraphics[width=\linewidth]{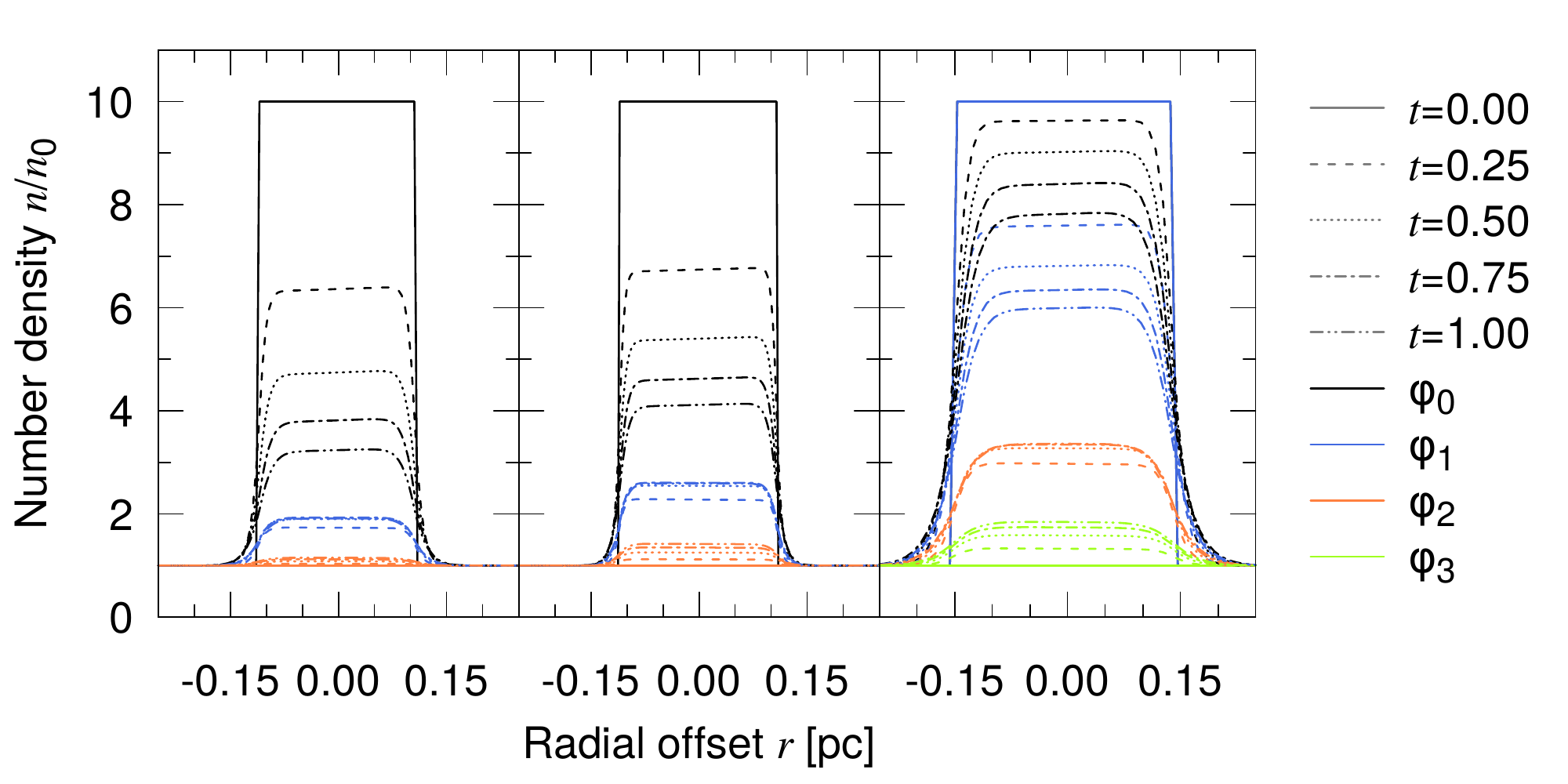}
    \caption{Number density profiles of the diffusion test for the \texttt{lores} (left), \texttt{hires} (centre), and \texttt{eqres} (right) grids; the number density is scaled by the unperturbed ISM value. Dash types denote the time since injection in $44828\,\si{kyr}$, and line colours denote the offset in the $\varphi$-direction, with $\varphi_i = \varphi_0 + i\; \Delta \varphi$.}
    \label{fig:numdif}
\end{figure}

Number density profiles of the progressing simulations are plotted in Fig.~\ref{fig:numdif}. As can be seen, numerical diffusion smoothed the initially sharp profile of the perturbation  even after short times. While this effect is negligible in the radial direction for the \texttt{lores} grid due to the high resolution, its magnitude in the angular directions is much graver. After $t=0.25\ \hat{=}\ 11.071\,\si{kyr}$, the number density at the perturbation's original position had decreased from the initial $n_1=10.0n_0$ to $n(\varphi_0)=6.4n_0$, while the neighbouring angular cells' number density had increased to $n(\varphi_1)=1.7n_0$. As time progressed, the perturbation's number density further decreased at its original position, while its angular extent increased. After $t=1.00\ \hat{=}\ 44.828\ \si{kyr}$, the number density at the perturbation's original position had decreased to $n(\varphi_0)=3.2n_0$ and increased in its flanks to $n(\varphi_1)=1.9n_0$ and $n(\varphi_2)=1.1n_0$. Farther out, the number density had slightly increased: $n(\varphi_3)=1.02n_0$ and $n(\varphi_4)=1.002n_0$. Taking only the broadening to $\varphi_2=12\si{\degree}$ into account, for the \texttt{lores} grid this corresponds at $r=4.65\,\si{pc}$ to a growth from a diameter of $0.5\,\si{pc}$ to $2.3\,\si{pc}$ perpendicular to the radial direction. 

The primary reason for the intensity of the angular numerical diffusion is the confinement of the perturbation to a single angular cell ($N_{\vartheta,1}=N_{\varphi,1}=1\,\text{cell}$). Whereas the radial numerical diffusion could smooth the perturbation's vertical flank to Gaussian-like wings while leaving the horizontal plateau untouched, the singular extent in the radial directions offered no such possibility. For a perturbation extending over a larger number of cells in the angular directions (realised by increasing either the physical size or the resolution), the angular numerical diffusion does not decrease the value of the perturbation's horizontal plateau as much. However, increasing the number of angular cells covered by the perturbation by only a small factor does not significantly lower angular numerical diffusion, as can be seen with the \texttt{hires} model (centre panel in Fig.~\ref{fig:numdif}). Here, the perturbation covers roughly the same physical region, which means double the angular cells due to the finer angular resolution. Because a plateau spanning two cells still offers no possibility of keeping the plateau intact while smoothing its wings, the number density profiles of the \texttt{hires} grid barely differ from those of the \texttt{lores} grid. Improving the angular resolution by another factor of two, as is the case on the \texttt{eqres} grid, finally yields significantly different results. Here, the initial plateau covered four angular cells, marked as offsets $\varphi_0$ and $\varphi_1$ in Fig.~\ref{fig:numdif}. While the number density in the initially unperturbed cells increases more strongly compared to the previously examined grids (compare \texttt{eqres} $\varphi_2$ to \texttt{hires} $\varphi_1$), the outer edge of the plateau (\texttt{eqres} $\varphi_1$, \texttt{hires} $\varphi_0$) decreases more slowly, and the inner plateau (\texttt{eqres} $\varphi_0$) even more slowly. However, the radial numerical diffusion on the \texttt{eqres} grid must be taken note of. 

Because the impact of numerical diffusion decreases not only with finer angular resolutions but also with larger perturbations, the significance of numerical diffusion diminishes for spread-out perturbations such as \texttt{v5blb} (cf. Sect.~\ref{sec:simpleblobs}), especially if computed on the \texttt{hires} grid like \texttt{b4blb} (cf. Sect.~\ref{sec:simpleblobs}) and \texttt{n2blb} (cf. Sect.~\ref{sec:moreblobs}). 
While a finer angular resolution would be preferable, the simulation's runtime unfortunately increases with approximately the fourth power of the angular number of cells: Keeping the two angular resolutions equal ($\Delta\varphi=\Delta\vartheta$) already amounts to a square increase in the number of computational cells; furthermore, the number of time steps per physical time is, to a reasonable approximation, proportional to the number of angular cells as well. This is because the length of a time step $\delta t$ is limited by the cell with the smallest value of $\delta t_\textrm{cell} = C_{\mathrm{CFL}} \Delta L/c_{\max}$, where $\Delta L$ is the spatial cell length, $c_{\max}$ is the largest characteristic speed, and $C_{\mathrm{CFL}}$ comes from the Courant-Friedrichs-Lewy condition \citep[see][]{kissmann18}; the last two values can be assumed as constant for the purpose of setting the grid's resolution. Because $\Delta L = \min\{ \Delta r, r_{\rm SW} \sin^2(\Delta \vartheta)/2\}$ and $r_{\mathrm{SW}}\sin^2(\Delta\vartheta)/2=0.025\Delta r$ for the \texttt{hires} grid, improving the angular resolution sharply decreases $\delta t$ as well. Thus, doubling the angular resolution necessitates an increase in computational resources by nearly a factor of 16. Even when the total number of cells is kept constant by impairing the radial resolution, as was done from \texttt{hires} to \texttt{eqres}, the computational cost still increases with the square of the factor of angular refinement.

\section{Results}\label{sec:res}

An overview over the computed perturbations is given in Table~\ref{tab:blobs}. Simple perturbations of a single physical variable (number density, temperature, velocity, or magnetic field) are examined in Sect.~\ref{sec:simpleblobs}, and larger and more complex perturbations are examined in Sect.~\ref{sec:moreblobs}.

\begin{table*}
\caption{\label{tab:blobs}Perturbation values.}
\centering
\begin{tabular}{lllll}
\toprule\toprule
 Ident. & Change in physical variables & Perturbation size [cells] & Injection position [cells] & Comment\\  \midrule
 \texttt{nblob} & $n\cdot 10$ & $30\times 1\times 1$ & outer edge ($r\in[1062,1091]$)& \\

 \texttt{t1blb} & $T\cdot 10$ & $50\times 1\times 1$ & 300 from BS &  \\
 \texttt{v5blb} & $\vec{u}\cdot 10$ & $50\times 1\times 1$ & 300 from BS &  \\ \midrule
 \texttt{b4blb} & $A_{\vartheta}=3.5$ & $90\times 2\times 2$ & 540 from BS & shaped like t1blb/v5blb \\ 
 \texttt{n2blb} & $n\cdot 10$ & $R=225$ & centre 300 from BS at $\varphi=3$& spherical shape\\ 
 \bottomrule
\end{tabular}
\tablefoot{A change in physical variables denoted as $n\cdot 10$, for example, signifies an increase by a factor of ten, $n_1=10n_0$.}
\end{table*}

\subsection{Simple perturbations}\label{sec:simpleblobs}

The \texttt{nblob} model describes a spatially small perturbation of the ISM, increasing the number density by a factor of ten to $n_1= 110\,\si{cm^{-3}}$ inside a volume of $N_{r,1}\times N_{\vartheta,1}\times N_{\varphi,1}=30\times 1\times 1\,\text{cells}$, injected close to the outer edge of the model ($r_1\in[1062, 1091]\,\si{cells}\ \hat{\approx}\ [4.61, 4.79]\,\si{pc}$) along the inflow axis. In physical terms, the volume of this pyramid frustum is $V_1=\frac{30}{3}(1062^2+1062\cdot1091+1091^2) \sin^2(\Delta\varphi) (\Delta r)^3 \approx 0.0311\,\si{pc^3}$, an excess of $V_1 (n_1-n_0)\approx 9.05\times10^{55}$ protons and electrons, and an excess mass of $0.0761M_{\odot}$. All other simulation values ($e_{\mathrm{th}}, \vec{u}, \vec{B}$) remained constant; from $T\propto e_{\mathrm{th}}/n$, it follows that $T_1=T_0/10=900\,\si{K}$. At this temperature, cooling holds little significance, but heating increases the energy density, according to Eq.~(\ref{eq:heat}), by $\Gamma\approx n_1 G_0\approx1.21\times10^{-20}\,\si{erg\,cm^{-3}\,s^{-1}}$. Typically, the timescale of heating $\tau_{\Gamma}$ is approximated by dividing the increase in thermal energy by the heating power, in this case $\tau_{\Gamma}=e_{\mathrm{th}}/\Gamma\approx 3k_{\mathrm{B}}(T_0-T_1)/(n_1G_0)=966.5\,\si{yrs}$. However, heating is incorporated into the model as a source term in Eq.~(\ref{eq:mhdenergy}). Thus, heating does not directly increase the thermal energy density but rather the total energy density, which is dominated by the kinetic energy density (i.e. the ram pressure): Within the perturbation right after its injection, the kinetic energy density amounts to $\frac{1}{2} m n_1 |\vec{u}_{\mathrm{ISM}}|^2\approx 5.89\times10^{-9}\,\si{erg/cm^3}$ ($99.2\%$ of the total energy density), the magnetic energy density to $|\vec{B}_{\mathrm{ISM}}|^2/(2\mu_0)\approx3.98\times10^{-12}\,\si{erg/cm^3}$ ($0.1\%$), and the thermal energy density to $3n_1k_{\mathrm{B}}T_1\approx4.10\times10^{-11}\,\si{erg/cm^3}$ ($0.7\%$). Over a time span of $\Delta t=4.4828\,\si{kyr}$, the heating term results in an increase in the energy density by $1.71\times 10^{-9}\,\si{erg/cm^3}$, which is $29.0\%$ of the total energy density at injection.
This additional energy does not remain inside the model cells or the perturbation region, and the heating term changes all MHD quantities to values different from their non-heated counterparts. While the perturbed region does heat up to $1347.512\,\si{K}$ after $\Delta t=4.4828\,\si{kyr}$, only $0.053\,\si{K}$ of this increase is due to the additional heating term, the rest stemming from pure MHD and numerical effects. Thus, the energy source term typically identified as `heating' actually causes only a negligible amount of heating but is nevertheless significant for the evolution of total energy within the perturbation. Furthermore, this implies in particular that the often used approximation for the heating timescale does not hold since it neglects the self-consistent conversion of thermal into other forms of energy.

\begin{figure*}
    \centering
    \includegraphics[scale=0.16]{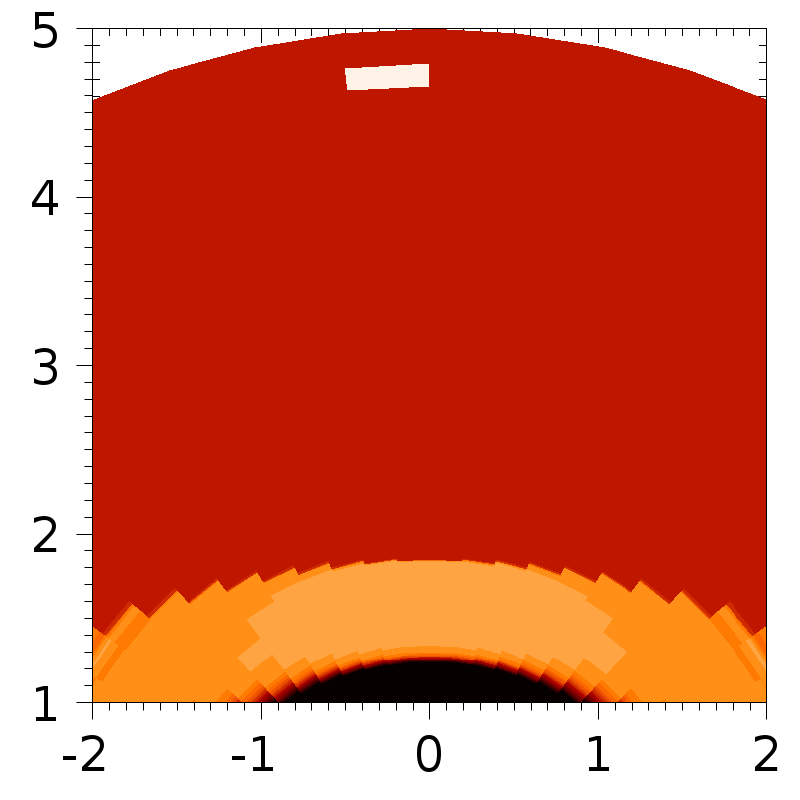}\hfill
    \includegraphics[scale=0.16,trim={2cm 0 0 0},clip]{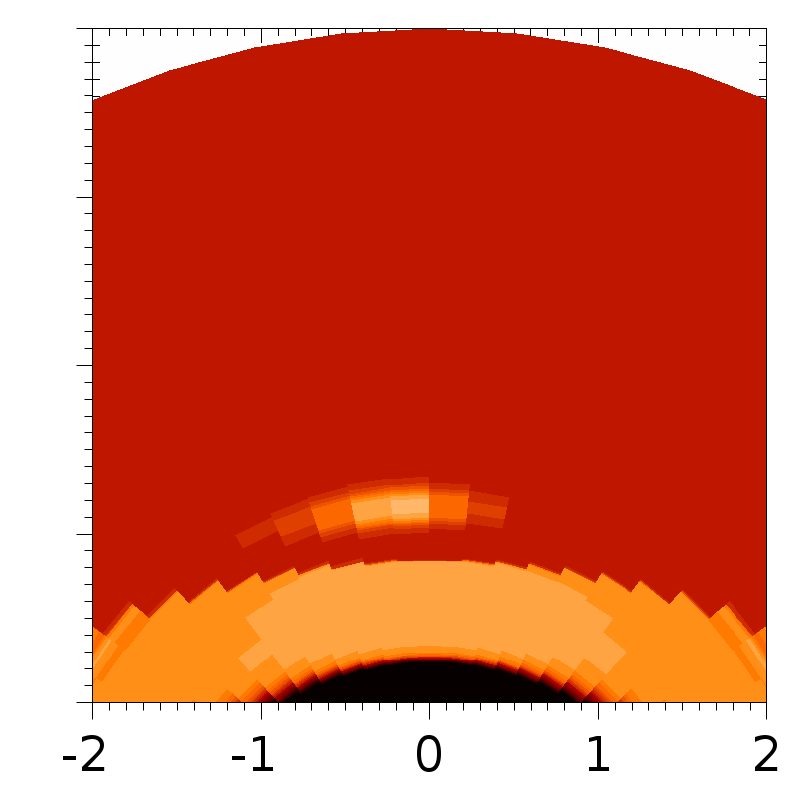}\hfill
    \includegraphics[scale=0.16,trim={2cm 0 0 0},clip]{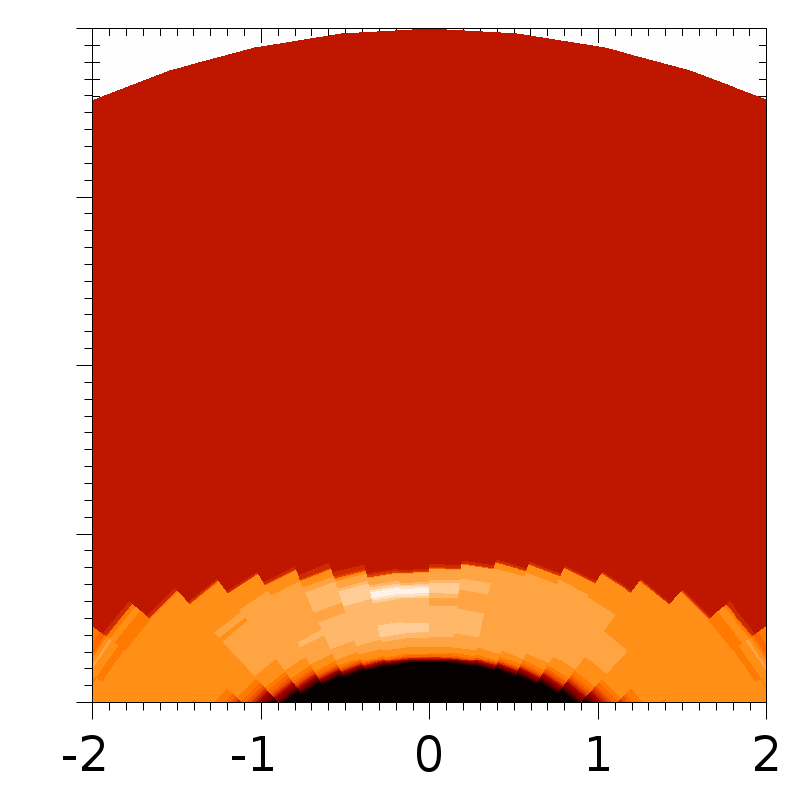}\hfill
    \includegraphics[scale=0.16,trim={2cm 0 0 0},clip]{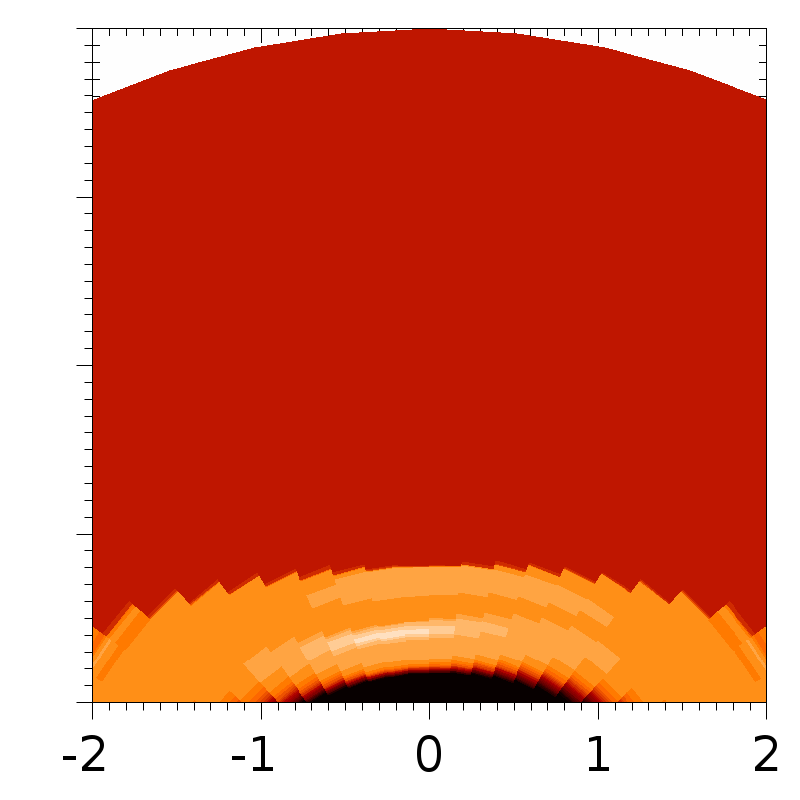}\hfill
    \includegraphics[scale=0.16]{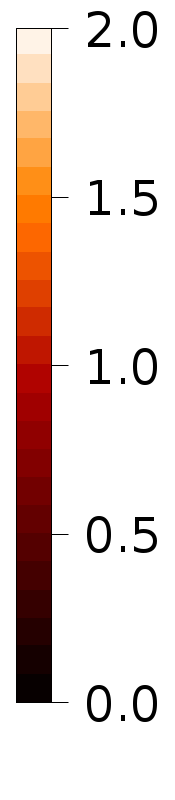}
    
    \raggedright
    \includegraphics[scale=0.27,align=c]{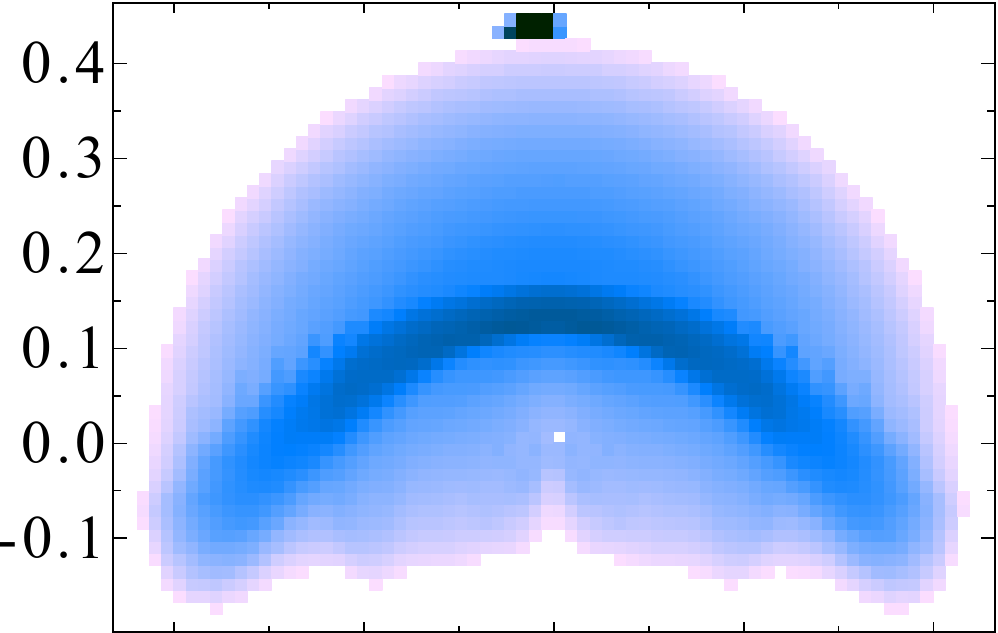}
    \includegraphics[scale=0.27,align=c]{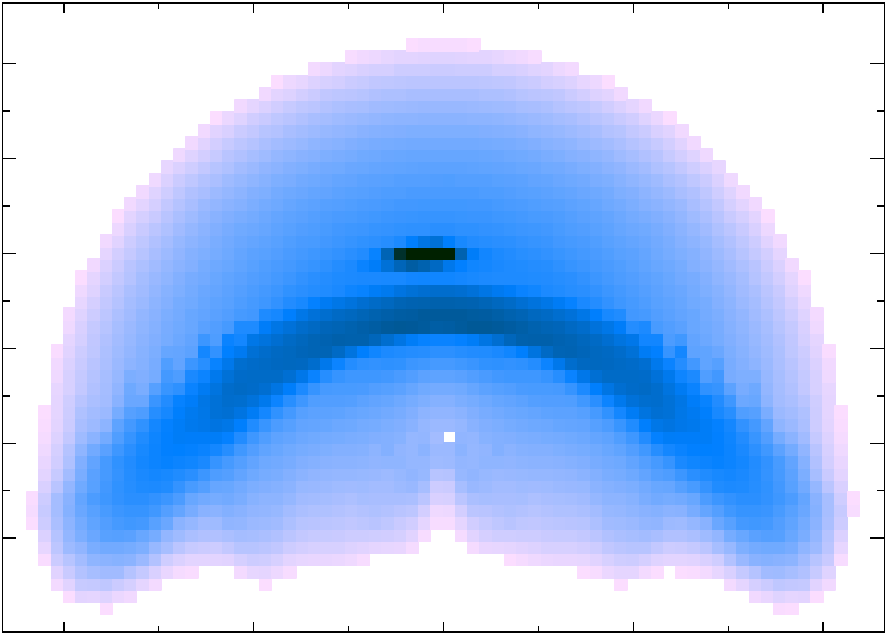}
    \includegraphics[scale=0.27,align=c]{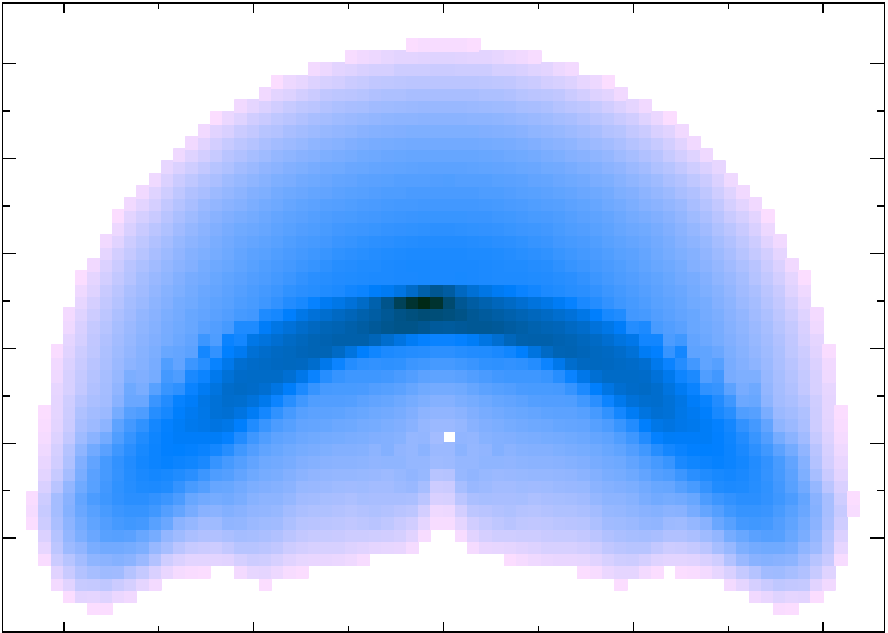}
    \includegraphics[scale=0.27,align=c]{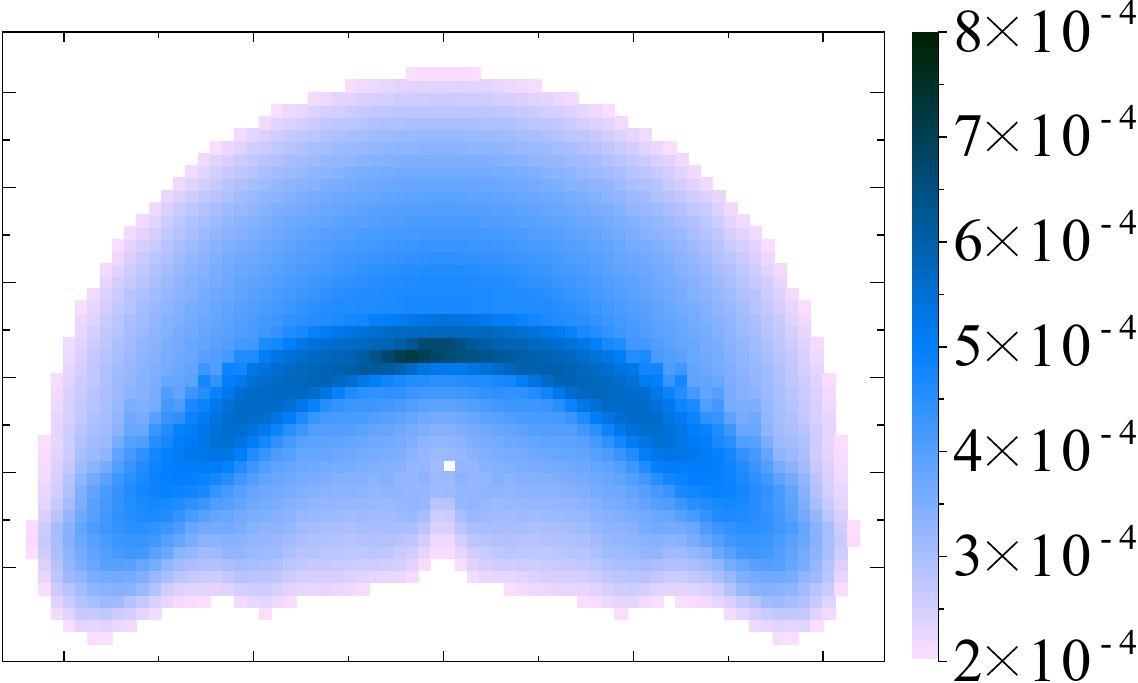}
    \includegraphics[scale=0.27]{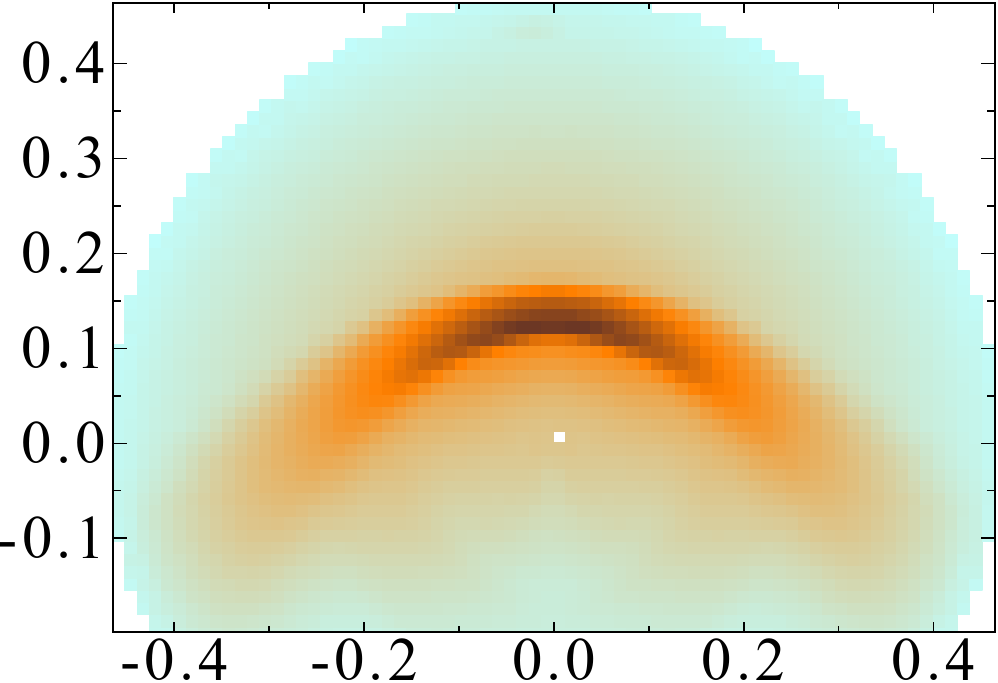}
    \includegraphics[scale=0.27]{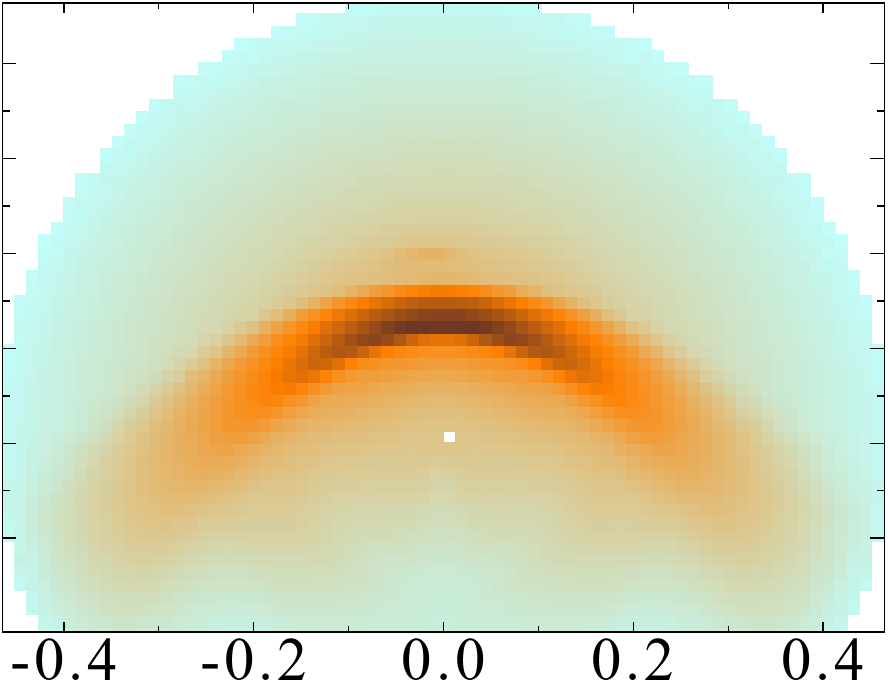}
    \includegraphics[scale=0.27]{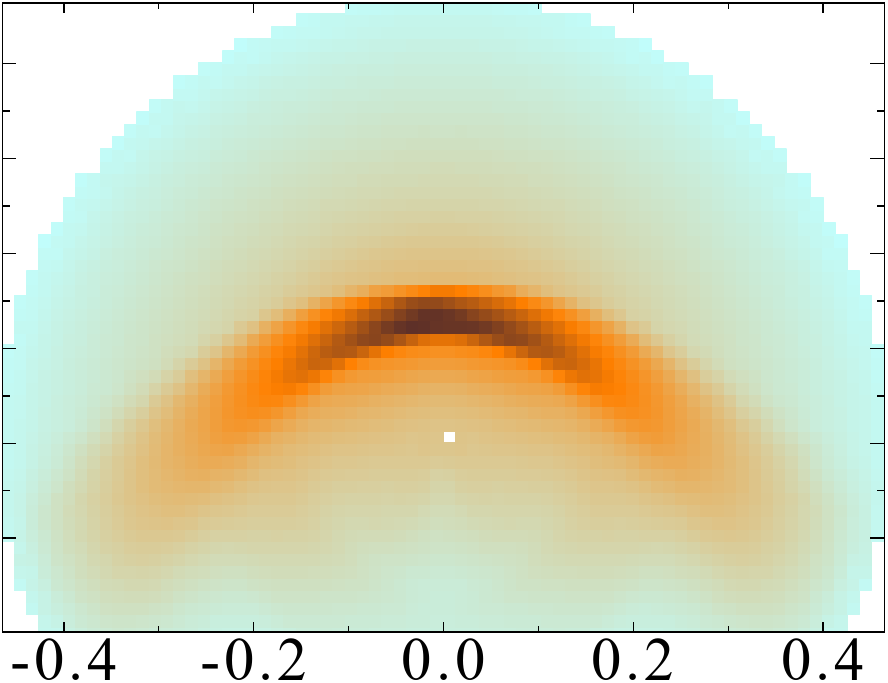}
    \includegraphics[scale=0.27]{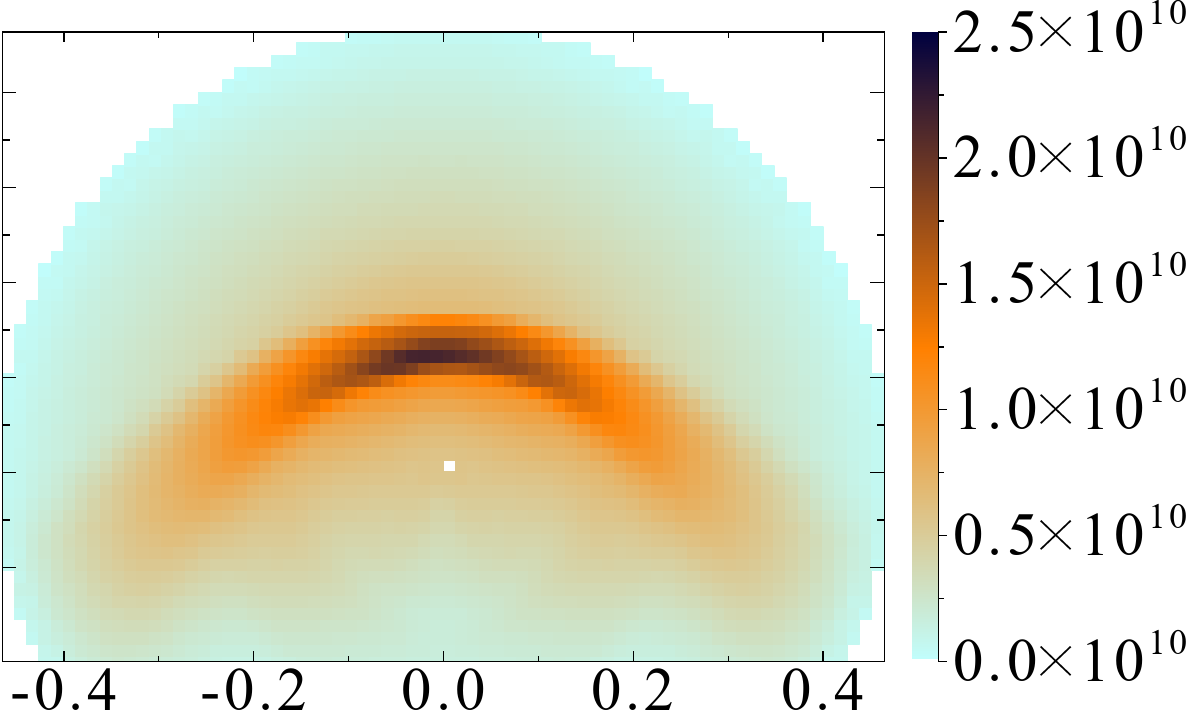}
    \caption{Evolution of the \texttt{nblob} perturbation. Top row: Cross-sections of the log number density, $\log(n \; [\si{cm^{-3}}]),$ in the \texttt{nblob} model's ecliptic plane at $\Delta t\in\{0,7,9,12\}\times 4.4828\,\si{kyr}$ after injecting the perturbation (distances in $\si{pc)}$. The displayed images show the $(r,\varphi)$-plane for $\vartheta=0\si{\degree}$, i.e. the ecliptic of the model. Centre row: Synthetic observations of the above models in H\textalpha\ $[\si{erg\,cm^{-2}\,s^{-1}\,sr^{-1}}]$ at a distance of $617\,\si{pc}$, face-on to the ecliptic (angular extent in degrees). The colour scale has been truncated on both ends; values below the minimum threshold are not displayed. Bottom row: Synthetic observations of the same configuration in $70\,\si{\micro m}$ dust emission $[\si{Jy\,sr^{-1}}].$ The solid angle per pixel is $5.07\times10^{-8}\,\si{sr}$.}
    \label{fig:nblob}
\end{figure*}

Cross-sections of the model's ecliptic at selected times are displayed in Fig.~\ref{fig:nblob} (top row), with corresponding synthetic observations in H\textalpha\ shown in the bottom row of the same figure. Directly after injection (leftmost column of panels in Fig.~\ref{fig:nblob}, $\Delta t=0\,\si{kyr}$), the perturbation distends to a Gaussian shape due to numerical diffusion (cf. Sect.~\ref{sec:numdif}) and slowly widens further (centre-left column, $\Delta t=7\times4.4828\,\si{kyr}$) until it impacts the BS. It enhances the number density within a small region inside the outer astrosheath by roughly a factor of two (centre-right column, $\Delta t=9\times4.4828\,\si{kyr}$) before dispersing inside the outer astrosheath (rightmost column, $\Delta t=12\times4.4828\,\si{kyr}$). The collision of perturbation and outer astrosheath causes a small indentation in the BS around the centre of impact (cf. centre-right column) that takes roughly $10\,\si{kyr}$ to fill out. 

While the perturbation is well visible in H\textalpha\ when still inside the homogeneous ISM, contrasting by roughly a factor of two even when partially dissolved (centre-left column), neither the enhanced density of the outer astrosheath after the impact nor the dent in the BS are readily apparent (centre-right column). This is in part due to the low resolution of the model but mostly stems from the poor contrast between the perturbation and the outer astrosheath shortly before the impact. A perturbation of higher density or larger size would cause a more notable indentation, both in size and density enhancement (cf. the \texttt{n2blb} model in Sect.~\ref{sec:moreblobs}).

Because dust emission is dependent on the distance to the star, the perturbation is nearly invisible at the time of injection (left column) and only faintly visible before impacting the BS (centre-left column). Because the post-impact density enhancement occurs at the BS surface of the outer astrosheath, where dust emission is weaker due to the larger distance from the star compared to the AP surface, the small density perturbation is of little significance for the observable dust.

The \texttt{t1blb} model similarly describes a small perturbation that increased the temperature from $T_{\mathrm{ISM}}=9\,\si{kK}$ to $T_1=90\,\si{kK}$ inside a volume of $N_{r,1}\times N_{\vartheta,1}\times N_{\varphi,1}=50\times 1\times 1$ cells, injected $300\,\text{cells}$ in front of the BS ($r_1\in [712, 761]\,\si{cells}$) along the model's inflow axis. In physical units, the perturbation's centre is located $1.41\,\si{pc}$ in front of the BS, with a radial thickness of $0.22\,\si{pc}$ and an average width of $0.33\,\si{pc}$.
This perturbation decreases its number density immediately after injection to regain thermal pressure equilibrium and disperses shortly afterwards, not causing any notable change in the astrosphere's structure or synthetic observations.

\begin{figure*}
    \centering
    \includegraphics[scale=0.125]{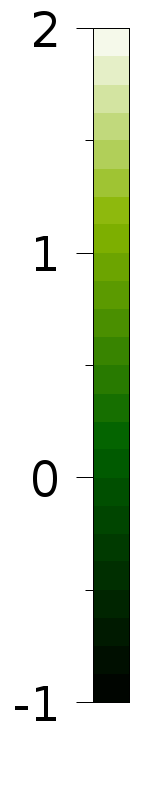}\hfill
    \includegraphics[scale=0.125]{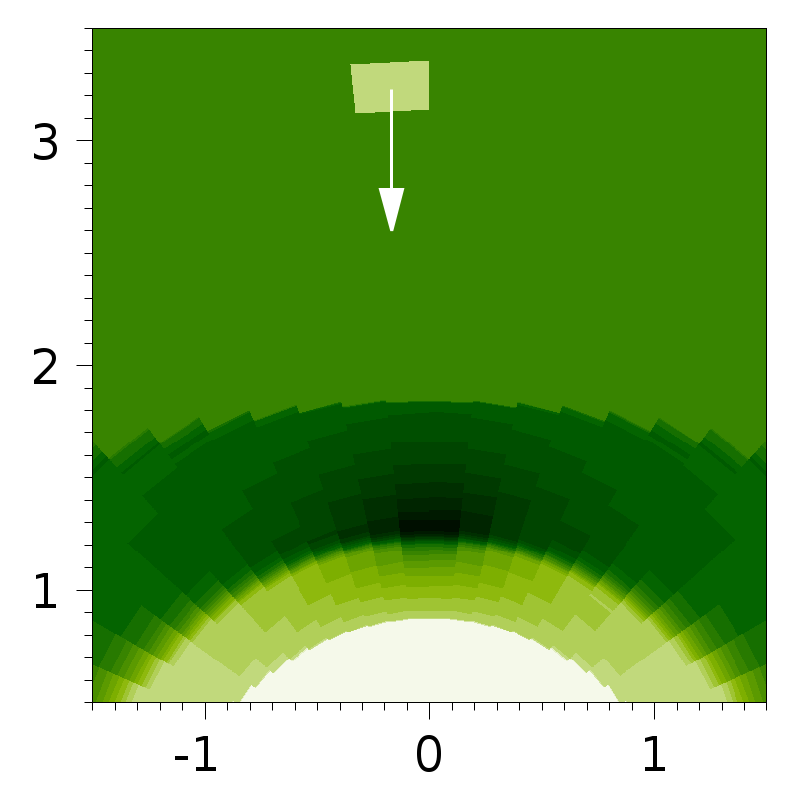}\hfill\vrule\hfill
    \includegraphics[scale=0.125,trim={18mm 0 0 0},clip]{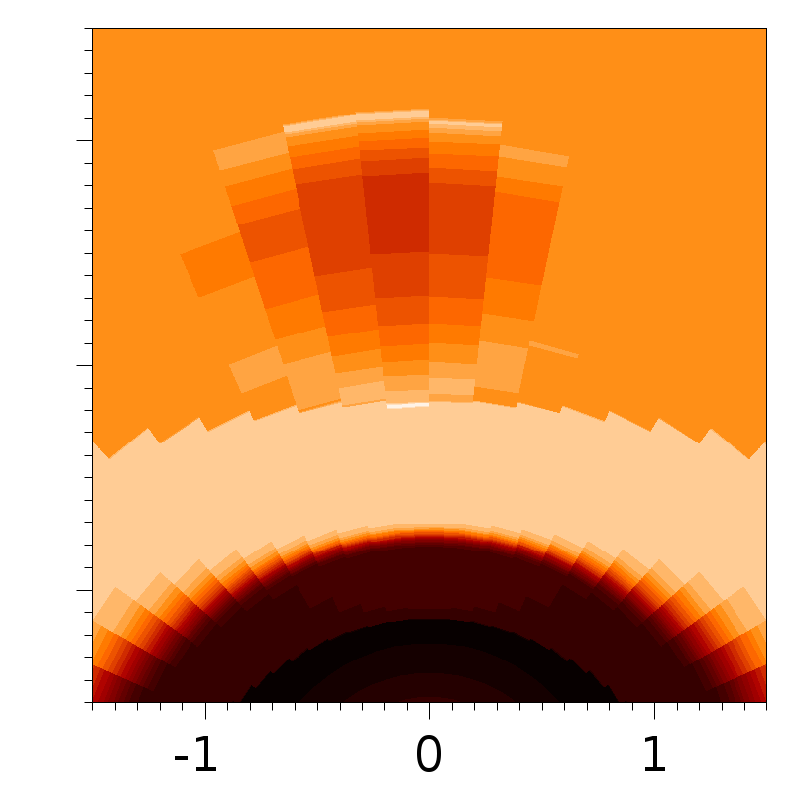}\hfill
    \includegraphics[scale=0.125,trim={18mm 0 0 0},clip]{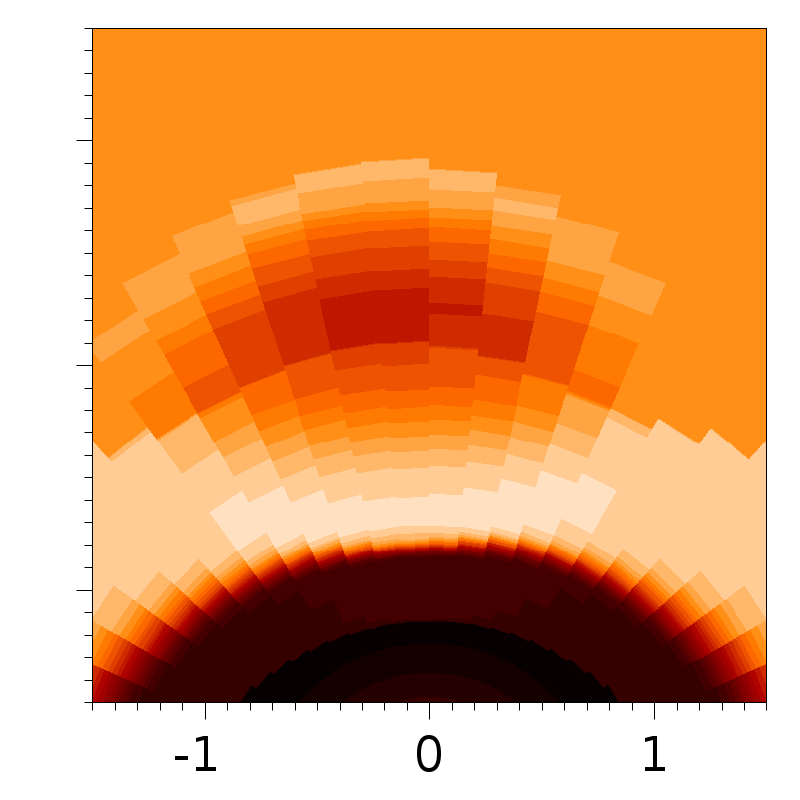}\hfill
    \includegraphics[scale=0.125,trim={18mm 0 0 0},clip]{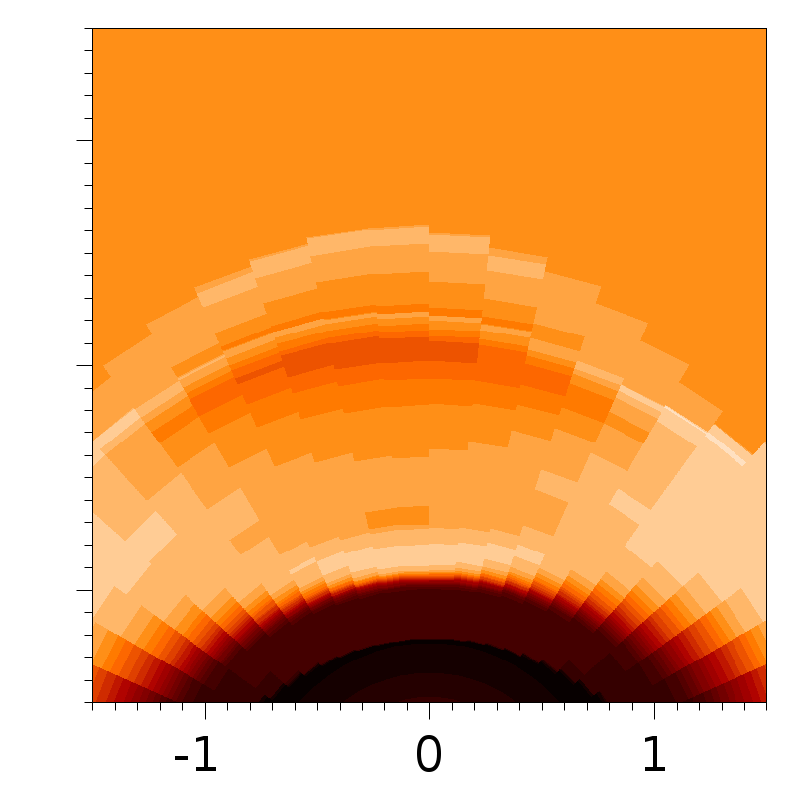}\hfill
    \includegraphics[scale=0.125,trim={18mm 0 0 0},clip]{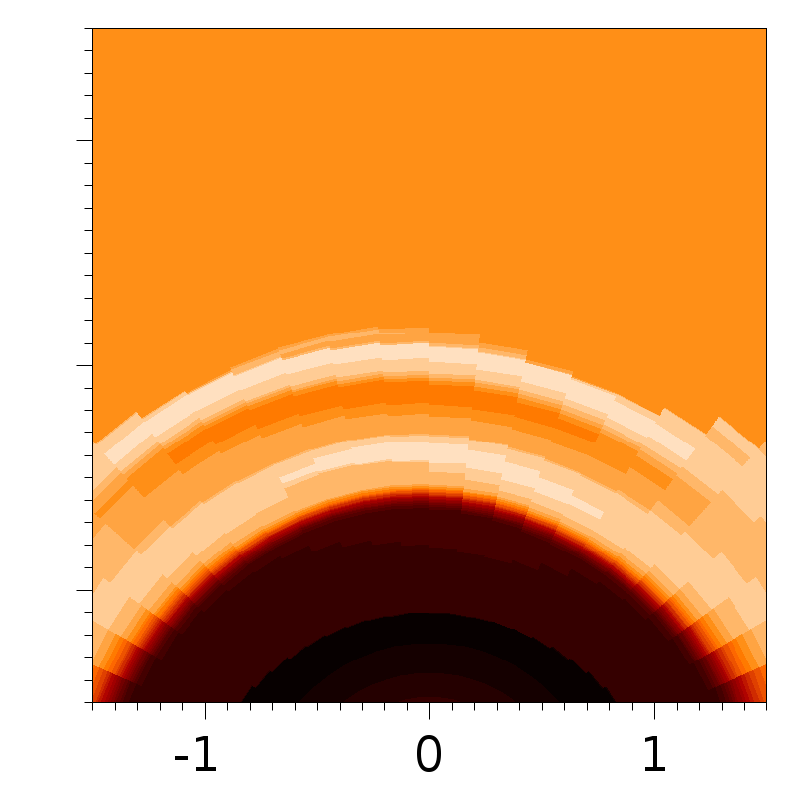}\hfill
    \includegraphics[scale=0.125]{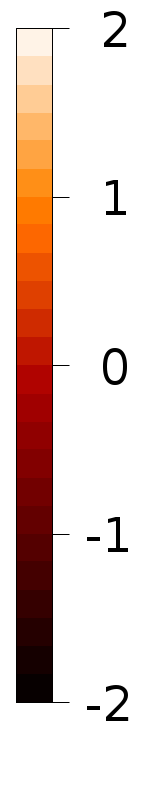}
    
    \raggedright
    \includegraphics[scale=0.217,align=c]{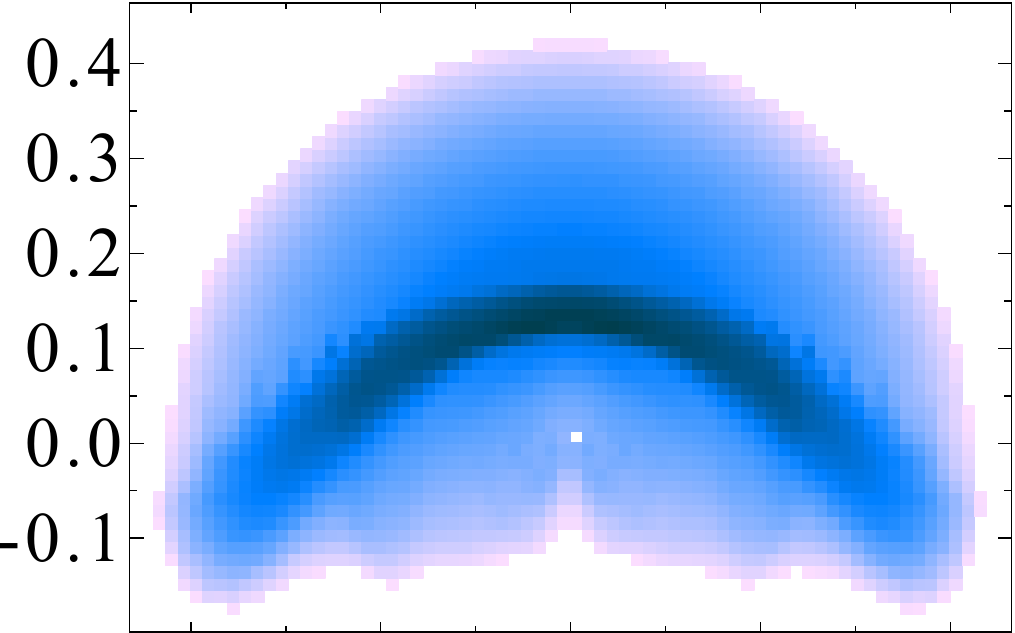}
    \includegraphics[scale=0.217,align=c]{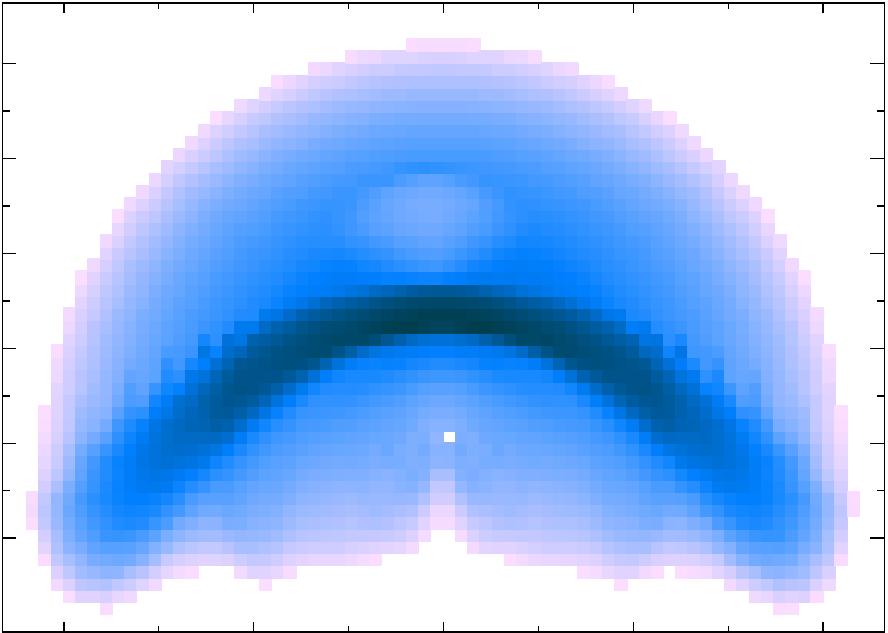}
    \includegraphics[scale=0.217,align=c]{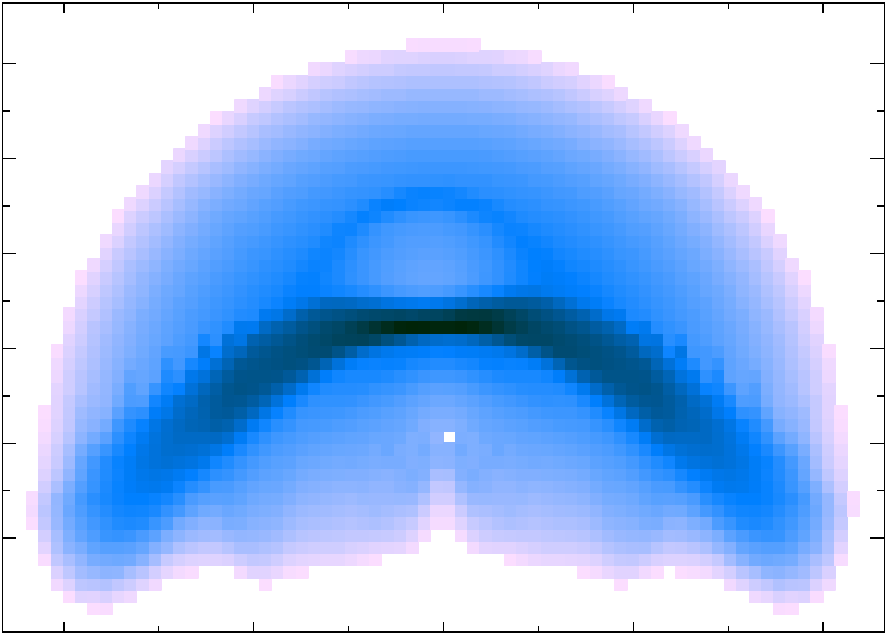}
    \includegraphics[scale=0.217,align=c]{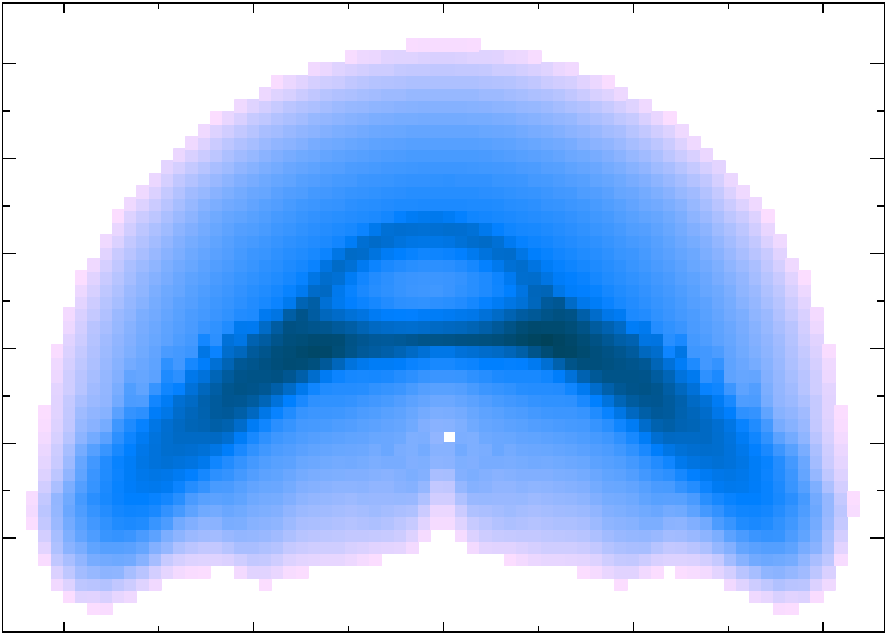}
    \includegraphics[scale=0.217,align=c]{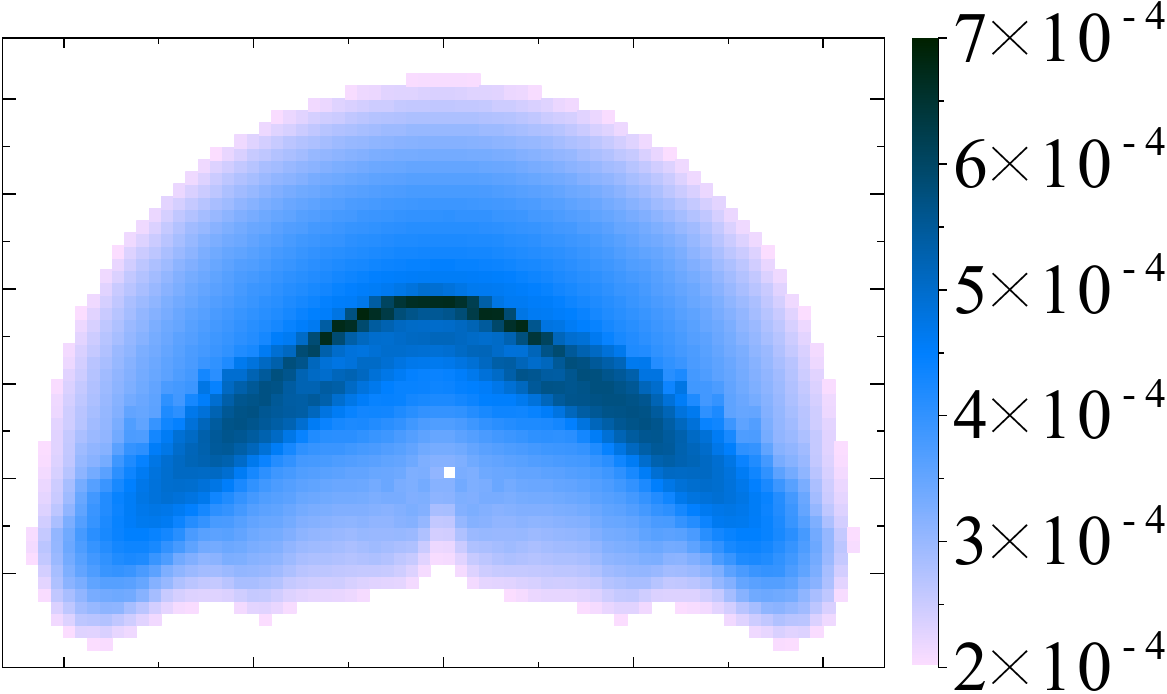}
    \includegraphics[scale=0.217]{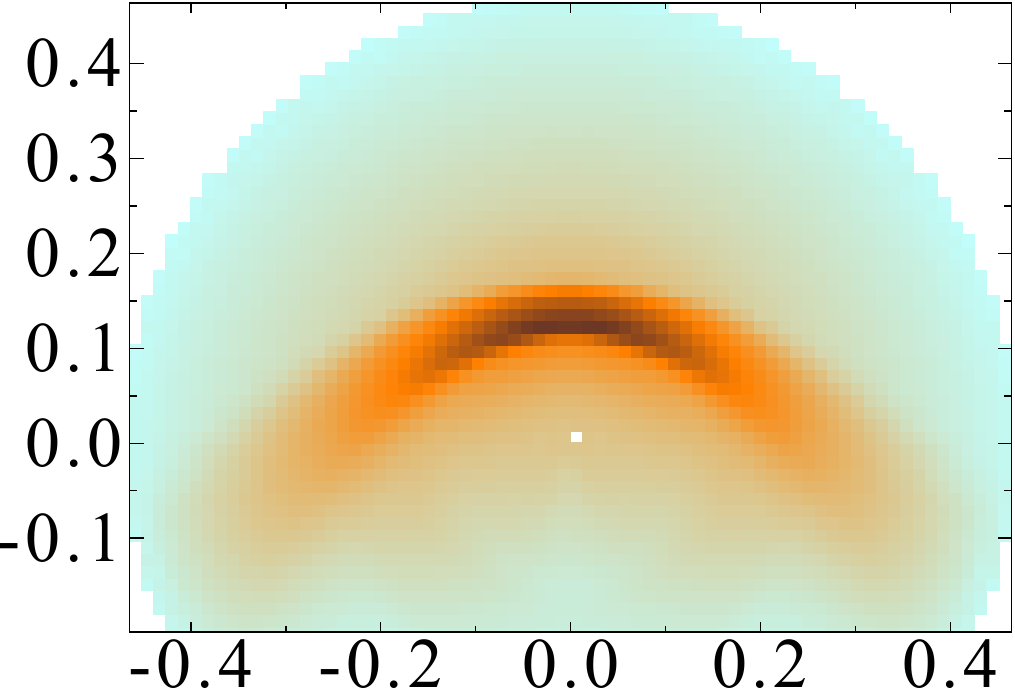}
    \includegraphics[scale=0.217]{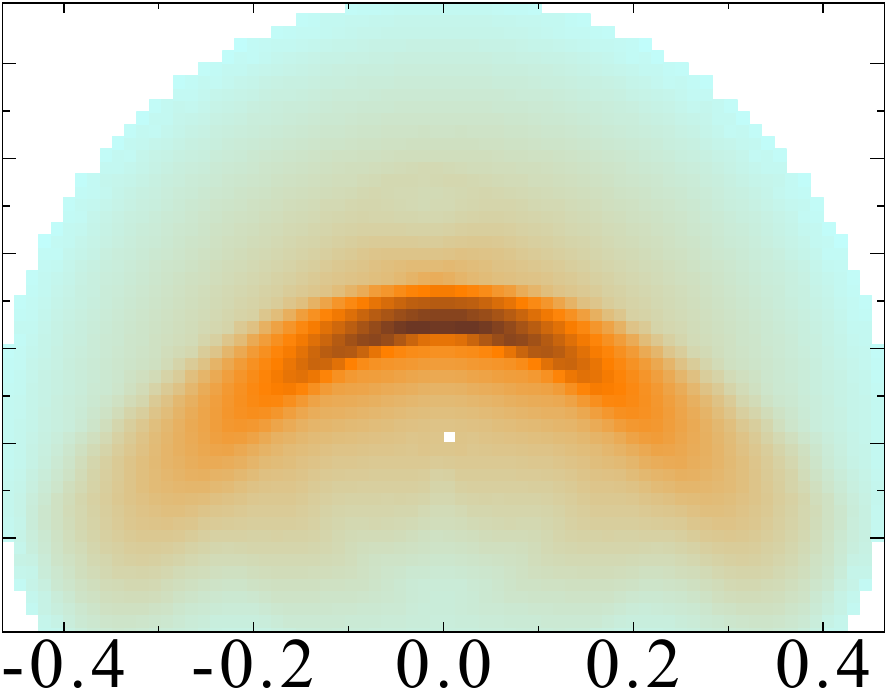}
    \includegraphics[scale=0.217]{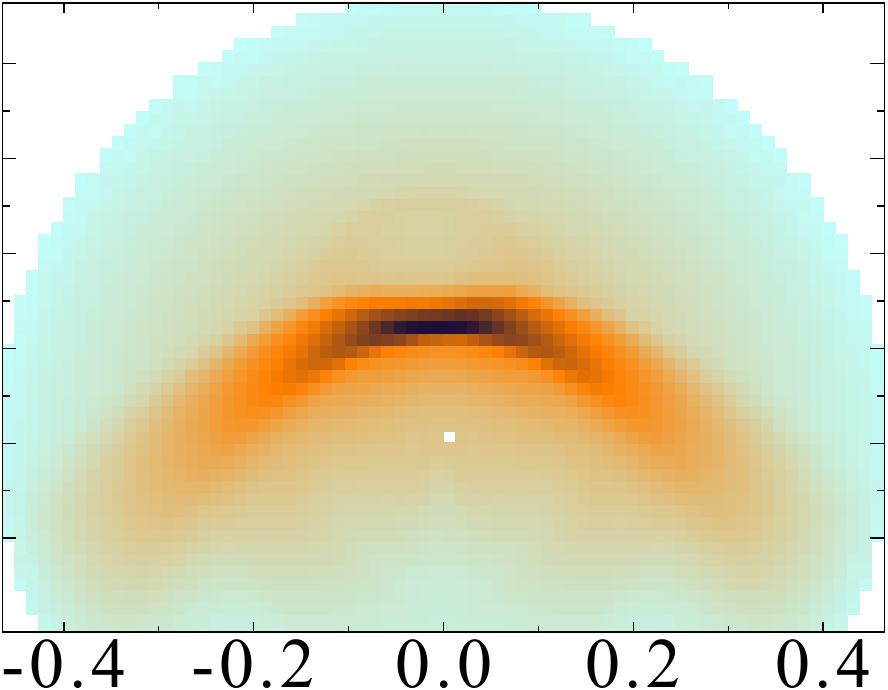}
    \includegraphics[scale=0.217]{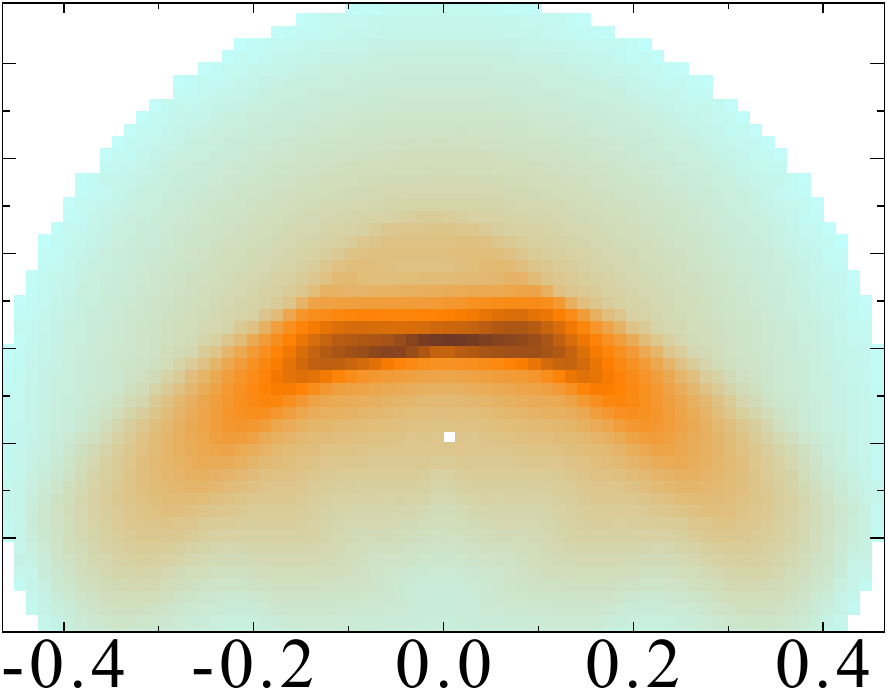}
    \includegraphics[scale=0.217]{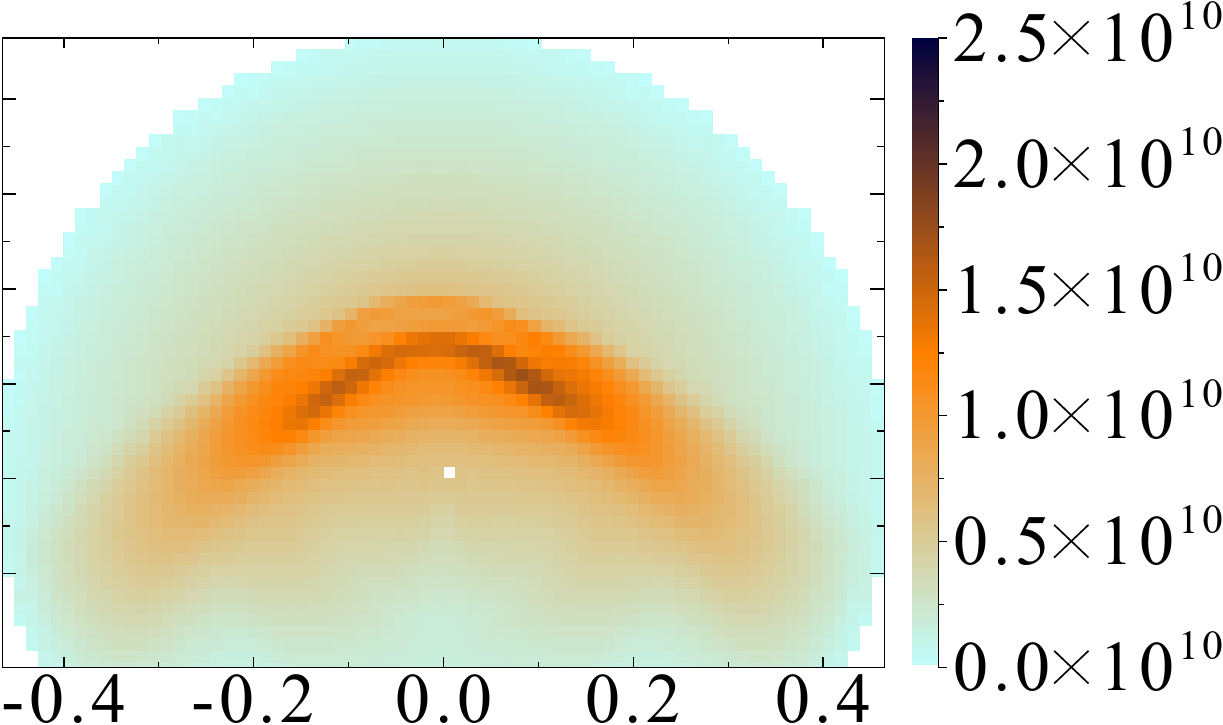}
    \caption{Evolution of the \texttt{v5blb} perturbation. Top row: Cross-sections of the \texttt{v5blb} model's ecliptic plane at $\Delta t\in\{0,1,2.5,4,6\}\times 4.4828\,\si{kyr}$ after injecting the perturbation in the absolute speed ($\log(u\,[\si{km/s}])$, left panel) and in the number density ($\log(n\,[\si{cm^{-3}}])$, four right panels;  distances in $\si{pc}$). The white arrow in the left panel indicates the movement direction of the perturbation with respect to the surrounding ISM. Centre row: Synthetic observations of the above models in H\textalpha\ $[\si{erg\,cm^{-2}\,s^{-1}\,sr^{-1}}]$ at a distance of $617\,\si{pc}$, face-on to the ecliptic (angular extent in degrees). The colour scale has been truncated at the bottom end; values below the minimum threshold are not displayed. Bottom row: Synthetic observations of the same configuration in $70\,\si{\micro m}$ dust emission $[\si{Jy\,sr^{-1}}].$ The solid angle per pixel is $5.07\times10^{-8}\,\si{sr}$.}
    \label{fig:v5blb}
\end{figure*}

The same does not hold true for a similar perturbation of the velocity. In the \texttt{v5blb} model (cf. Fig.~\ref{fig:v5blb}), the shape of the perturbation was set up in the same way as for the \texttt{t1blb} model but the speed was increased from $u_{\mathrm{ISM}}=80\,\si{km/s}$ to $u_1=800\,\si{km/s}$, keeping the velocity's orientation (leftmost column). As expected, the perturbation distends quickly, causing a cavity at the perturbation's co-moving position as well as an accumulation of material in front of and, to a lesser degree, behind it (centre-left column, $\Delta t=1\times4.4828\,\si{kyr}$). The cavity significantly distorts the shape of the outer astrosheath upon impacting the BS (centre column, $\Delta t=2.5\times4.4828\,\si{kyr}$), at first decreasing its density along the inflow axis (centre-right column, $\Delta t=4\times4.4828\,\si{kyr}$) before generating a shell-like structure of two high-density areas separated by a low-density one (rightmost column, $\Delta t=6\times4.4828\,\si{kyr}$). Afterwards, the shells fall together, and the configuration slowly returns to the stationary state (cf. Fig.~\ref{fig:basis}, $\Delta t\approx15\times4.4828\,\si{kyr}$). This behaviour can be observed in the synthetic H\textalpha\ flux densities as well (bottom row of Fig.~\ref{fig:v5blb}). The formed cavity results in lower flux densities, by roughly a factor of two, bordered by a region of higher flux densities, again by roughly a factor of two. In $70\,\si{\micro m}$ dust emission, the outer border of the cavity is only faintly visible due to its larger distance from the star. This is particularly prominent for the shell-like structure at $\Delta t=6\times4.4828\,\si{kyr}$ (right column), where the inner shell dominates over the outer one in dust emission but is much fainter in H\textalpha.

\begin{figure}
    \centering
    \includegraphics[scale=0.112]{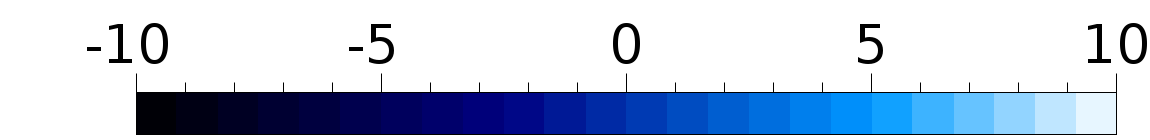}\hfill
    \includegraphics[scale=0.112,trim={2.5cm 0 0 0},clip]{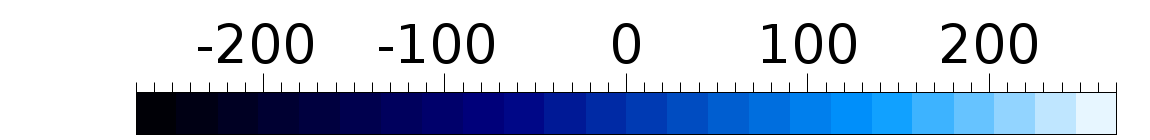}
    
    \includegraphics[scale=0.112]{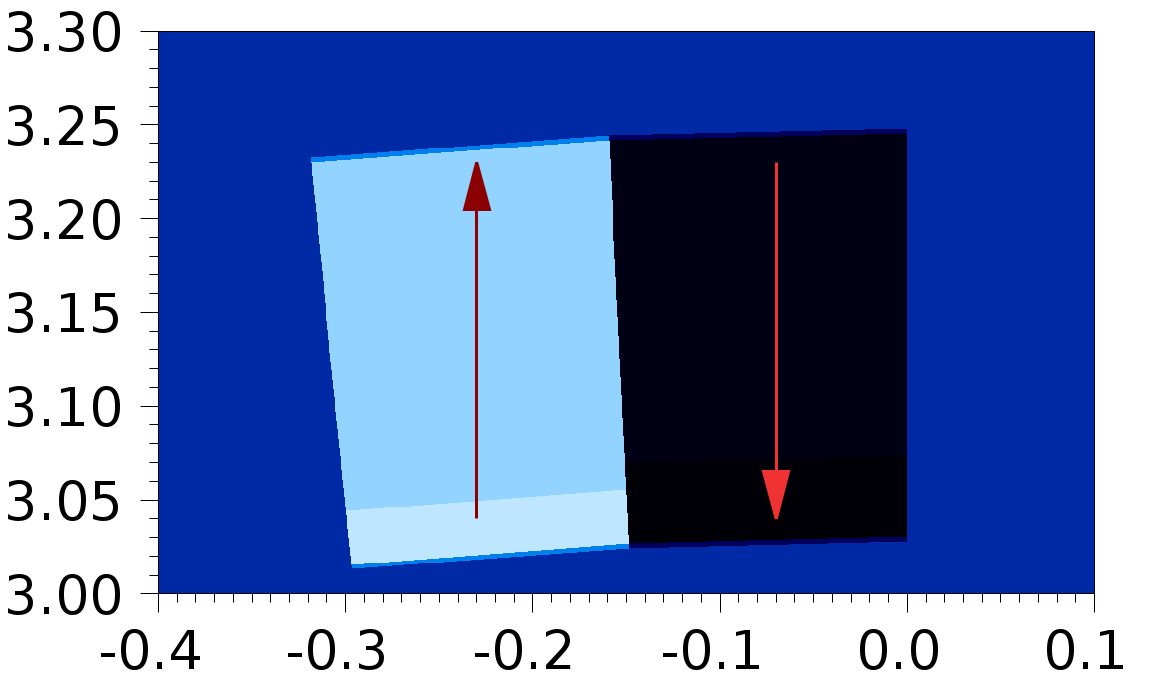}\hfill
    \includegraphics[scale=0.112,trim={2.5cm 0 0 0},clip]{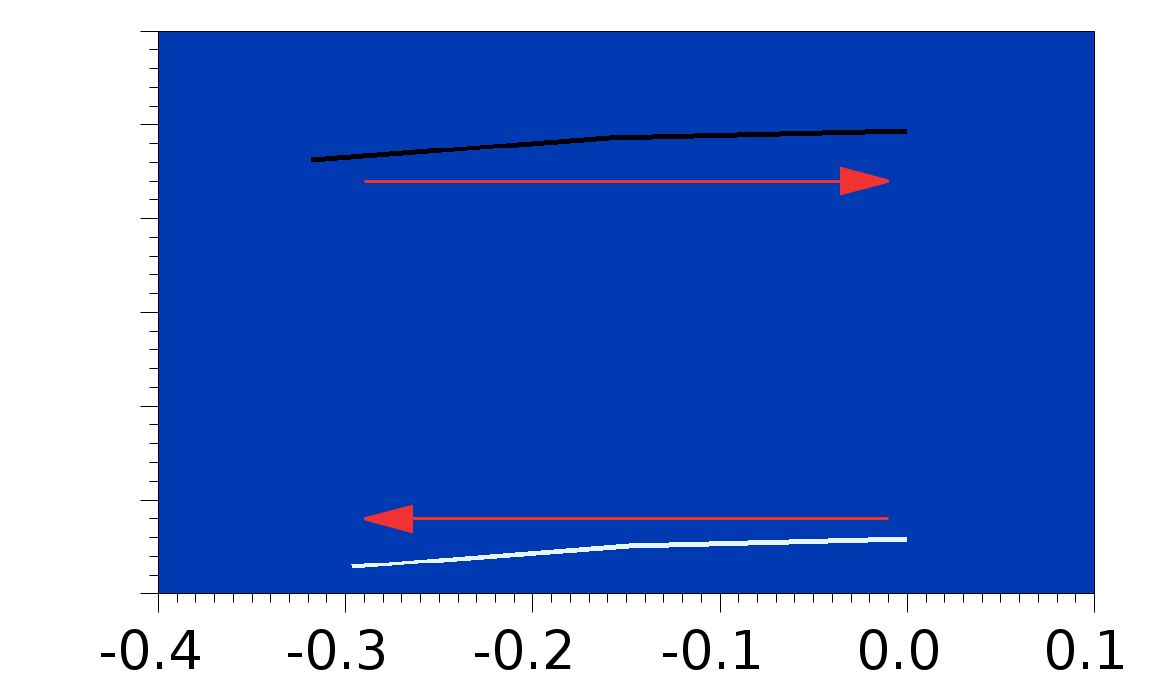}
    \caption{Magnetic structure of the \texttt{b4blb} perturbation: $r$-component (left) and $\varphi$-component (right) of the magnetic flux density $\vec{B}$ for $\vartheta=0\si{\degree}$ (linear scales of the magnetic flux density in $\si{nT}$, distances in $\si{pc}$). The red arrows indicate the orientation of the magnetic field.}
    \label{fig:b4blb_b}
\end{figure}

The \texttt{lores} grid cannot resolve the magnetic structure that evolves for a similar perturbation of the magnetic field, so the \texttt{hires} model was used instead. The \texttt{b4blb} model had a perturbation of the magnetic vector potential's $\vartheta$-component ($A_{\vartheta}=3.5\,\text{comp. units}$) injected at a position and size comparable to those of the perturbations of the \texttt{t1blb} and \texttt{v5blb} models, resulting in a perturbed magnetic field $\vec{B}_1=\vec{B}_0+\nabla\times\vec{A}$ inside a volume of $N_{r,1}\times N_{\vartheta,1}\times N_{\varphi,1}=90\times 2\times 2$ cells, injected $540\,\text{cells}$ in front of the BS ($r_1\in[1233,1322]\,\text{cells}$). In physical units, the perturbation's centre was located $1.41\,\si{pc}$ in front of the BS, with a radial thickness of $0.22\,\si{pc}$ and an average width of $0.31\,\si{pc}$. 
The resulting magnetic field $\vec{B}_1$ formed a boxy, clockwise vortex in the $(r,\varphi)$-plane for $\vartheta=0\si{\degree}$ (cf. Fig.~\ref{fig:b4blb_b}): The $\vartheta$-component of the magnetic flux density remained homogeneous, whereas the radial component increased to $B_{r,1}(\varphi=-4.22\si{\degree})=8.1\,\si{nT}$ on the `left' side and to $B_{r,1}(\varphi=-1.40\si{\degree})=-9.0\,\si{nT}$ on the `right' side (first panel); the $\varphi$-component increased to $B_{\varphi,1}(r=3.27\,\si{pc})=269\,\si{nT}$ on the `top' side and to $B_{\varphi,1}(r=3.24\,\si{pc})=-269\,\si{nT}$ on the `bottom' side, and also from $B_{\varphi,0}\approx0.3\,\si{nT}$ to $B_{\varphi,1}\in[0.46,0.49]\,\si{nT}$ `inside' the vortex at $r \in [3.24, 3.27]\,\si{pc}$ (second panel). On average, the perturbed magnetic flux density is ten times larger than the unperturbed one, $\bar{B}_1\approx 10B_0$, motivating the seemingly arbitrary choice of $A_{\vartheta}=3.5\,\text{computational units}$. Ideally, the magnetic vortex's outer edge would form a circle and not a trapezium; however, this requires either a finer resolution or a larger perturbation. 

\begin{figure*}
    \centering
    \includegraphics[scale=0.13]{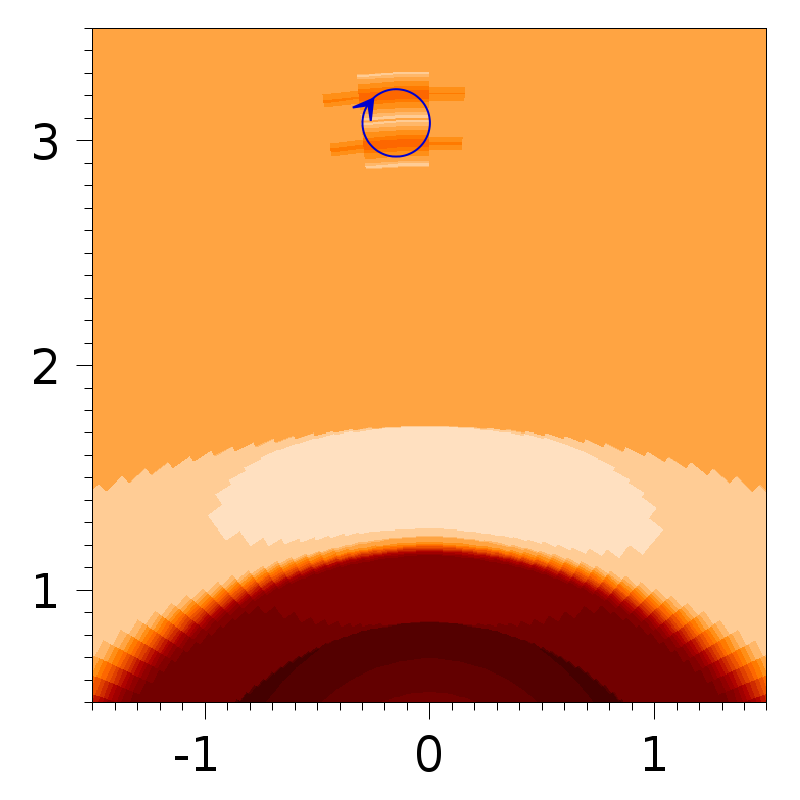}\hfill
    \includegraphics[scale=0.13,trim={18mm 0 0 0},clip]{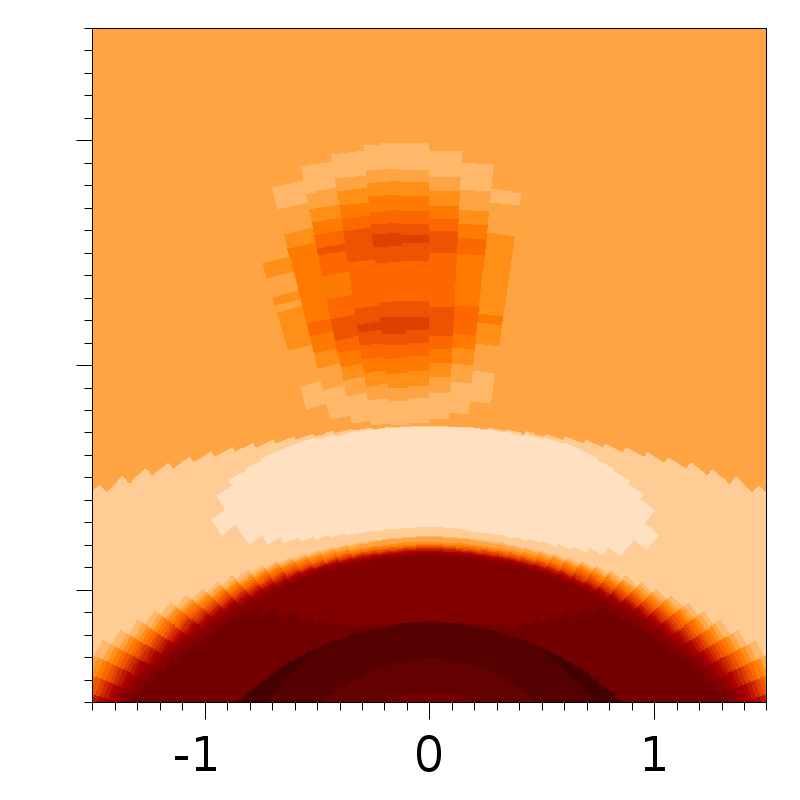}\hfill
    \includegraphics[scale=0.13,trim={18mm 0 0 0},clip]{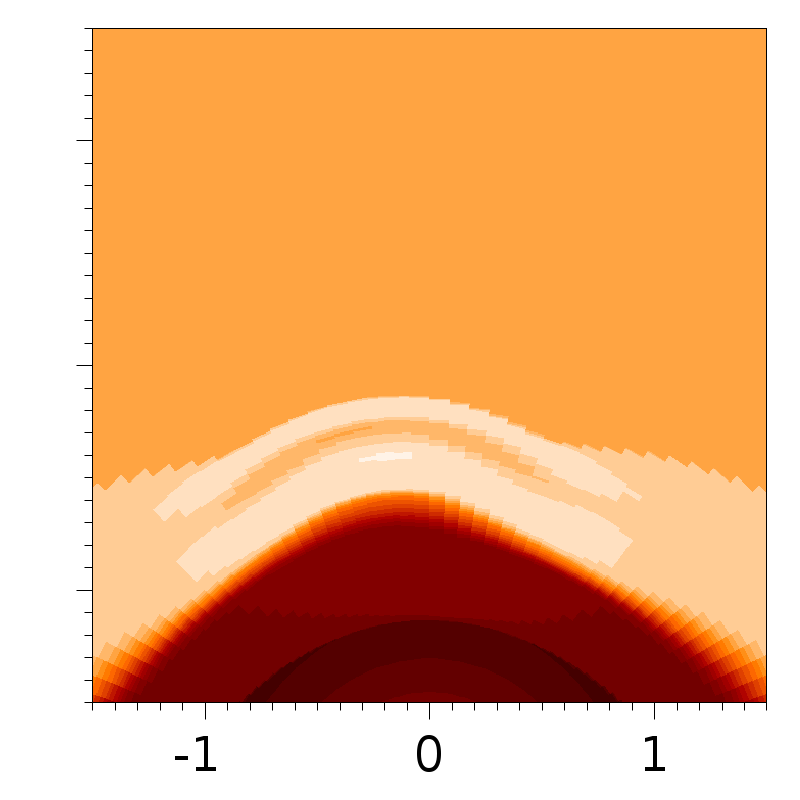}\hfill
    \includegraphics[scale=0.13,trim={18mm 0 0 0},clip]{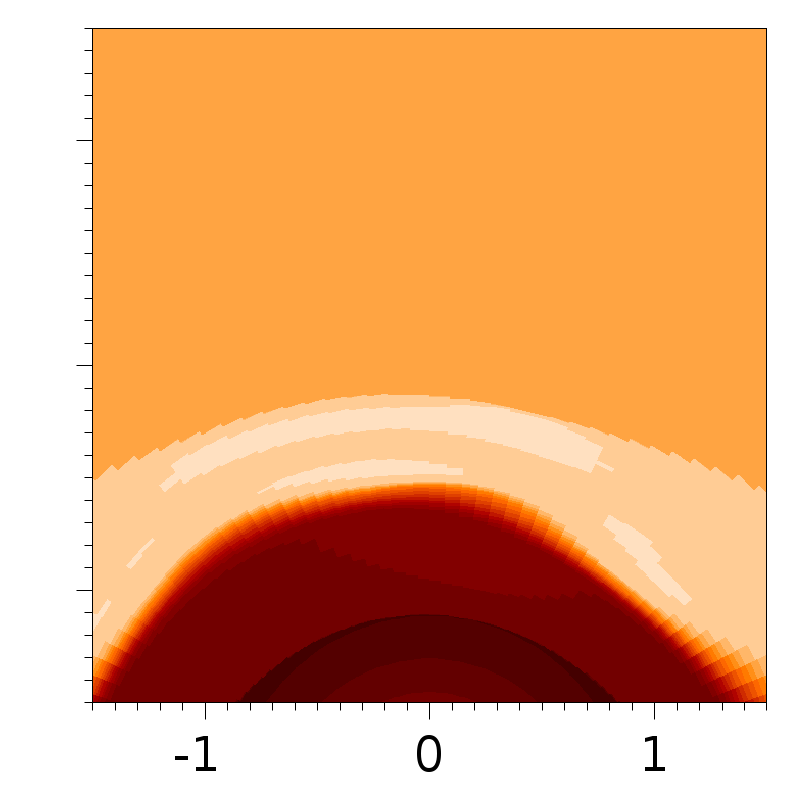}\hfill
    \includegraphics[scale=0.13,trim={18mm 0 0 0},clip]{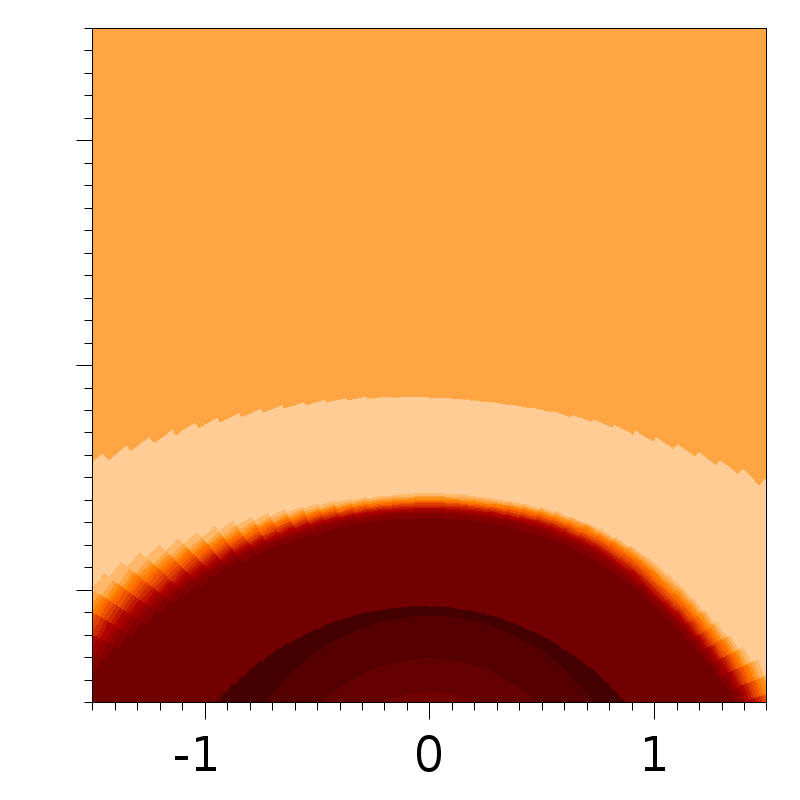}\hfill
    \includegraphics[scale=0.13]{img/e_redtemp-3+2.png}
    
    \raggedright
    \includegraphics[scale=0.233,align=c]{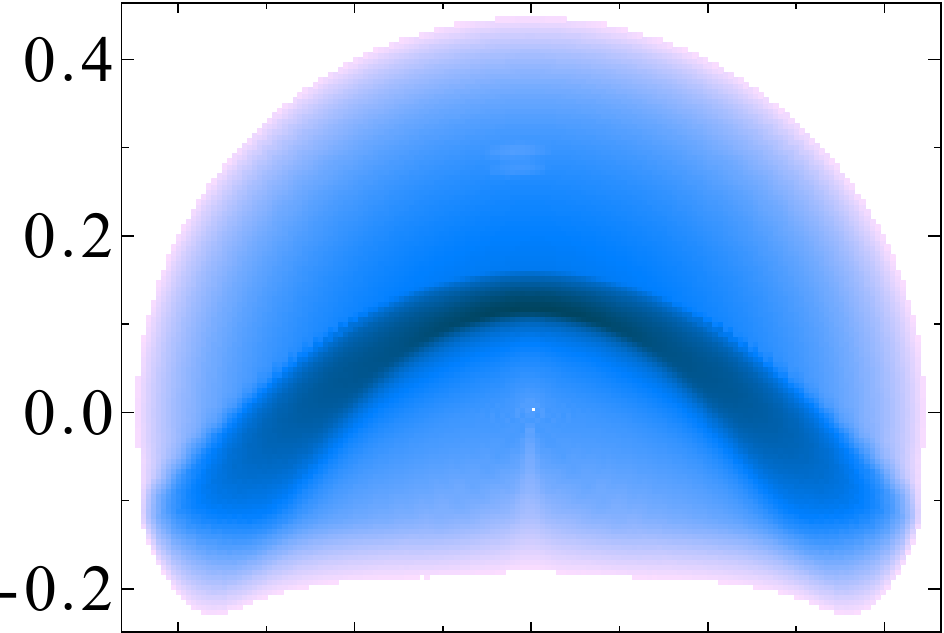}
    \includegraphics[scale=0.233,align=c]{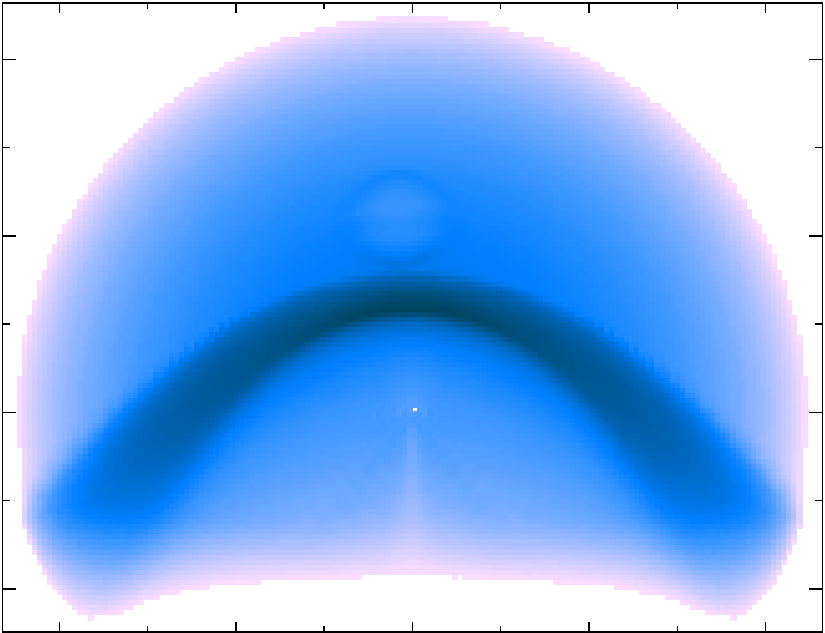}
    \includegraphics[scale=0.233,align=c]{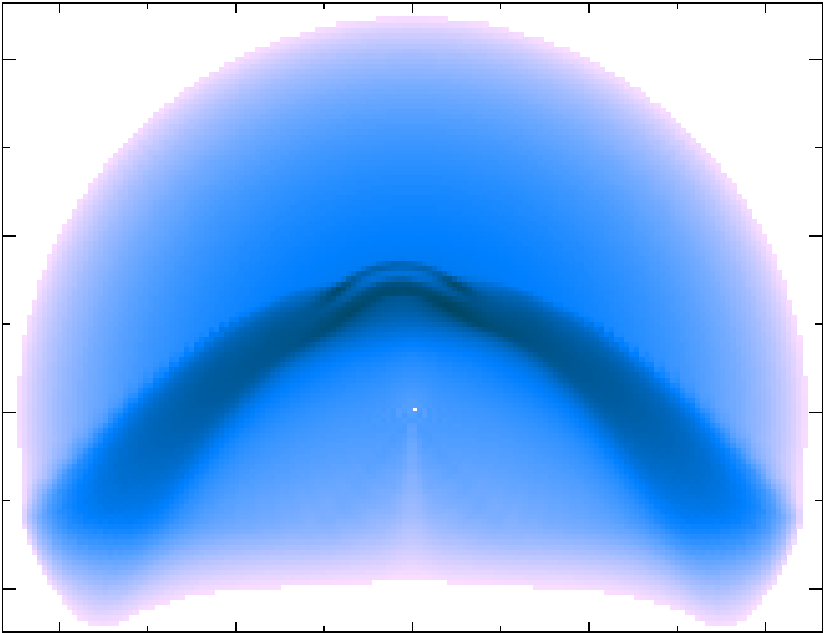}
    \includegraphics[scale=0.233,align=c]{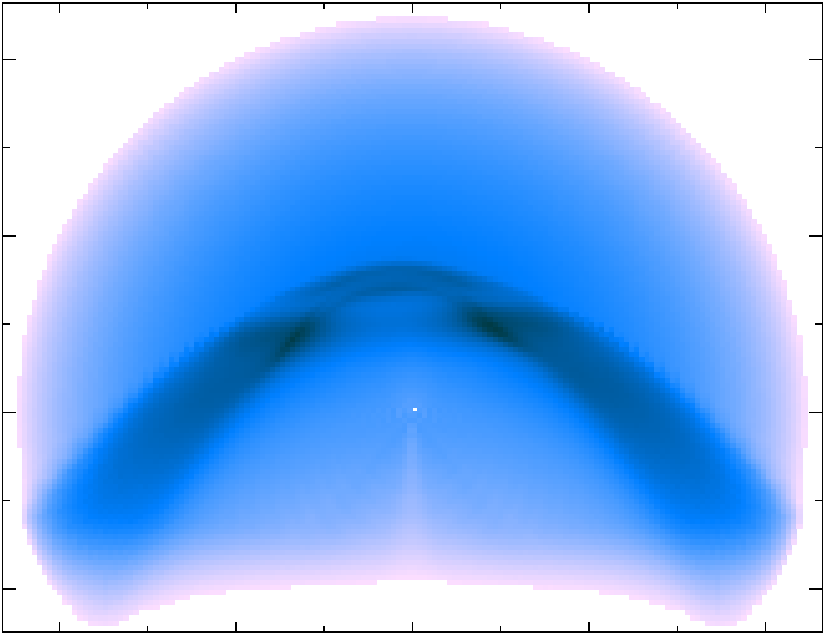}
    \includegraphics[scale=0.233,align=c]{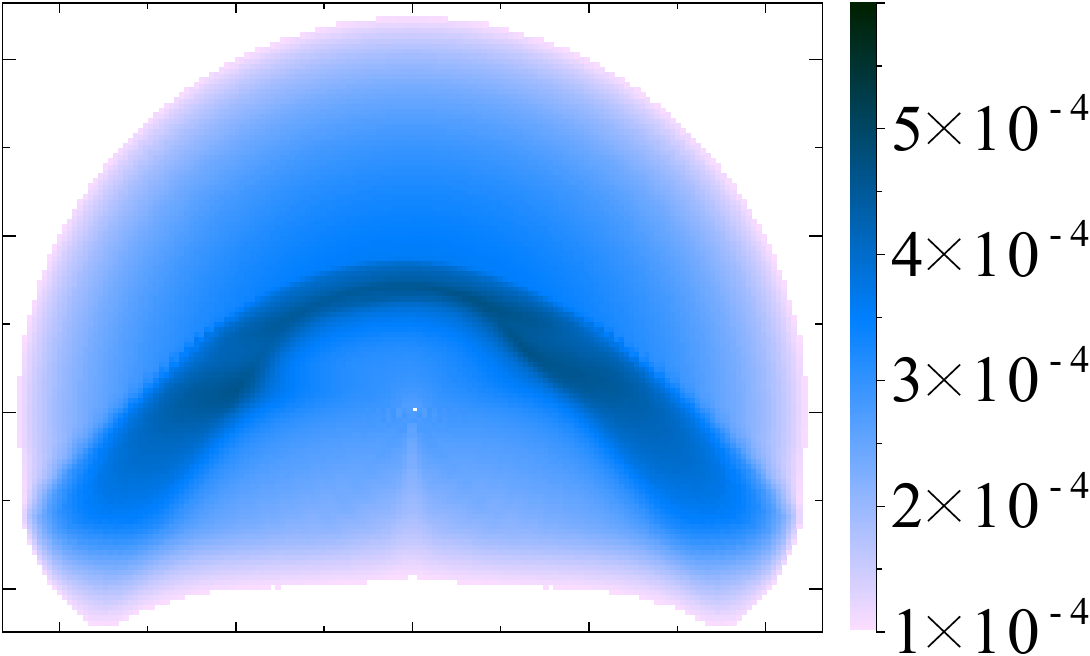}
    \includegraphics[scale=0.233]{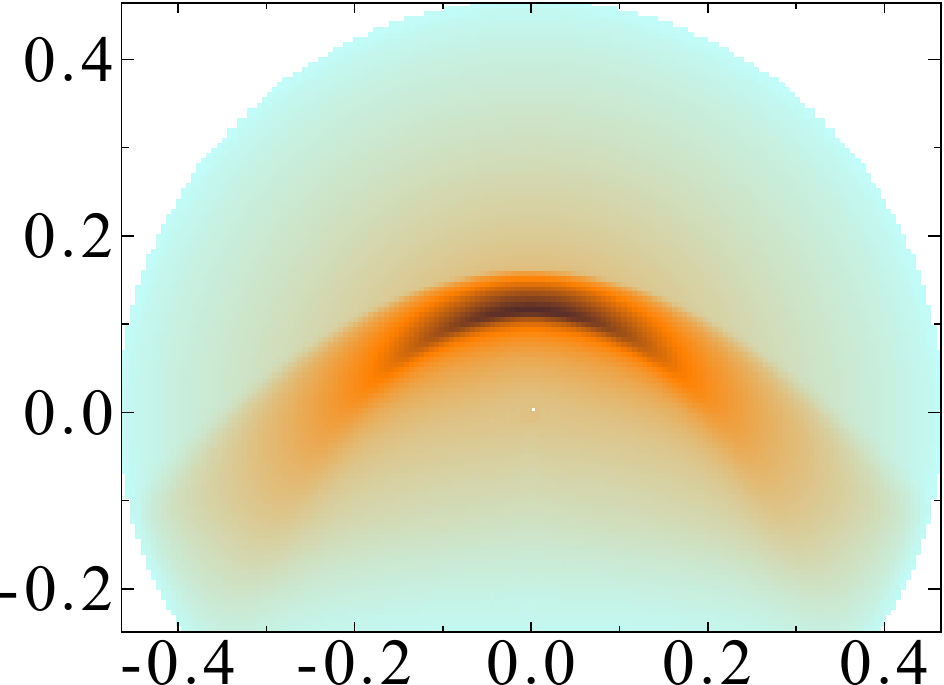}
    \includegraphics[scale=0.233]{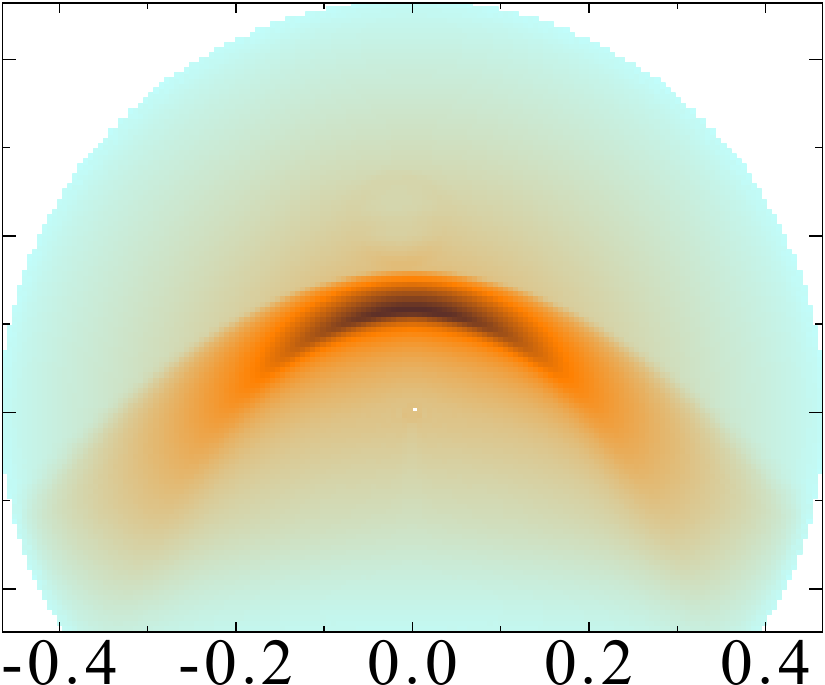}
    \includegraphics[scale=0.233]{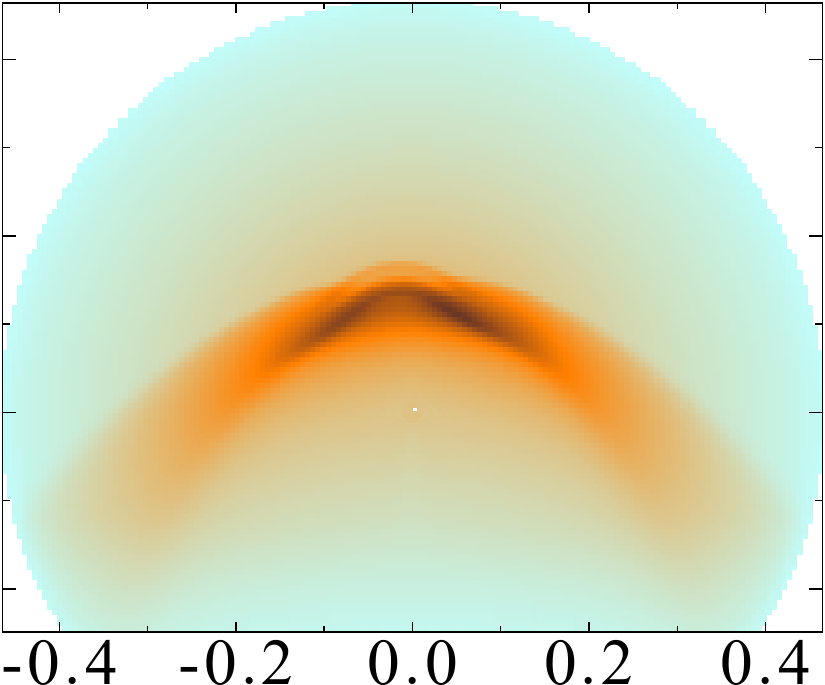}
    \includegraphics[scale=0.233]{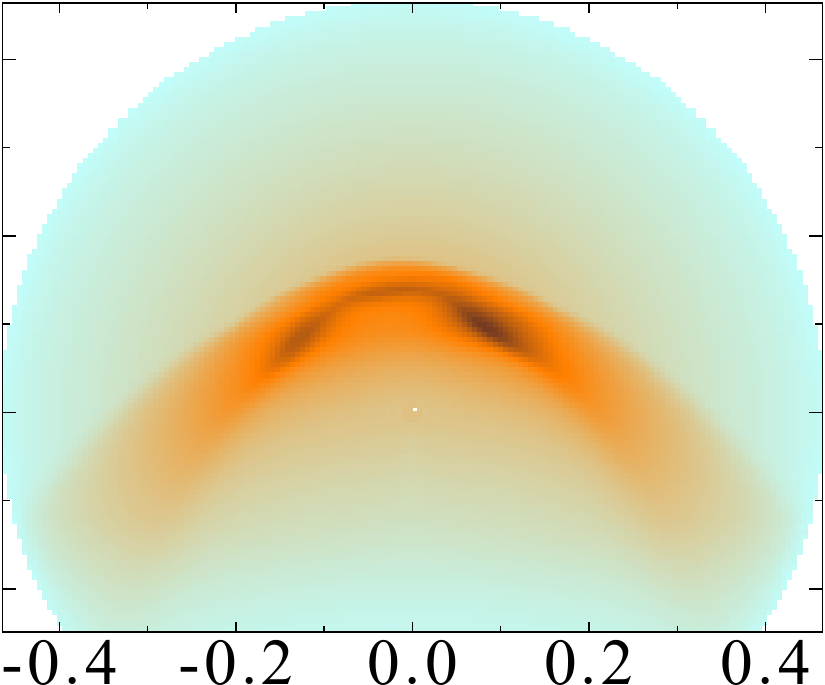}
    \includegraphics[scale=0.233]{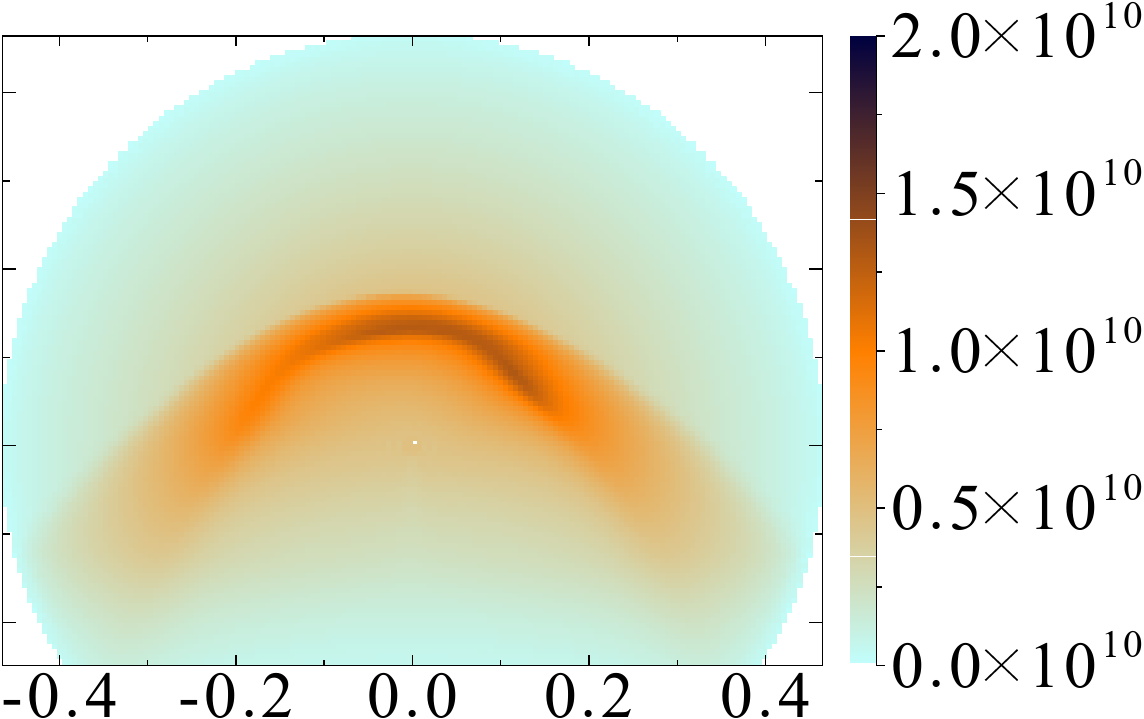}
    \caption{Evolution of the \texttt{b4blb} perturbation. Top row: Cross-sections of the log number density $\log(n \; [\si{cm^{-3}}])$ in the \texttt{b4blb} model's ecliptic plane at $\Delta t\in\{0.5, 2.5, 7, 11.5, 15\}\times 4.4828\,\si{kyr}$ after injecting the perturbation (distances in $\si{pc}$). The navy blue arrow in the left panel indicates the direction of the magnetic vortex. Centre row: Synthetic observations of the above models in H\textalpha\ $[\si{erg\,cm^{-2}\,s^{-1}\,sr^{-1}}]$ at a distance of $617\,\si{pc}$, face-on to the ecliptic (angular extent in degrees). The colour scale has been truncated at the bottom end; values below the minimum threshold are not displayed. Bottom row: Synthetic observations of the same configuration in $70\,\si{\micro m}$ dust emission $[\si{Jy\,sr^{-1}}].$ The solid angle per pixel is $1.00\times10^{-8}\,\si{sr}$.}
    \label{fig:b4blb}
\end{figure*}

After injection, the perturbation forms two local minima in the number density around the top and bottom side of the vortex (cf. Fig.~\ref{fig:b4blb}, leftmost column, $\Delta t=0.5\times4.4828\,\si{kyr}$). Due to the higher magnetic pressure inside the perturbation, it quickly expands to form a spherical cavity with an outer wall of higher number density (centre-left column, $\Delta t=2.5\times4.4828\,\si{kyr}$). After the impact of the perturbation with the BS, material from the outer astrosheath flows into the cavity, decreasing the number density downstream of the BS. When the perturbation's upwind upward slope of the number density impacts the BS, it causes a front of dense material to form. The impact of the upwind part of the perturbation's outer wall and the BS creates a second dense front, while a region of low density remains between these two fronts in a shell-like structure (centre column, $\Delta t=7\times4.4828\,\si{kyr}$). Afterwards, the BS and AP continue to move outwards to form a bump, reaching their outermost position at $\Delta t=1.15\times 4.4828\,\si{kyr}$ after injection (centre-right column). The bump slowly widens and decreases in density until it reaches its maximum extent (rightmost column, $\Delta t=15\times4.4828\,\si{kyr}$), from which point the structure returns to its stationary state (cf. Fig.~\ref{fig:basis}, $\Delta t\approx25\times4.4828\,\si{kyr}$). In total, the magnetic perturbation causes a behaviour similar to that of the velocity perturbation (v4blb), creating a cavity that modifies the shape of the outer astrosheath. 

The synthetic observations in H\textalpha\ are comparable to those of the v4blb model as well. The perturbation's cavity before impacting the BS is visible as a very slight decrease in H\textalpha\  brightness (centre-left column). After the perturbation impacts the BS, the small-scale shell-like structure is visible for a short time (centre column) but quickly smudges to a blurred distortion of the arc structure. In $70\,\si{\micro m}$ dust emission, the shell-like structure's outer shell is much fainter.

\subsection{Stronger perturbations}\label{sec:moreblobs}

\begin{figure*}
    \centering
    \includegraphics[scale=0.13]{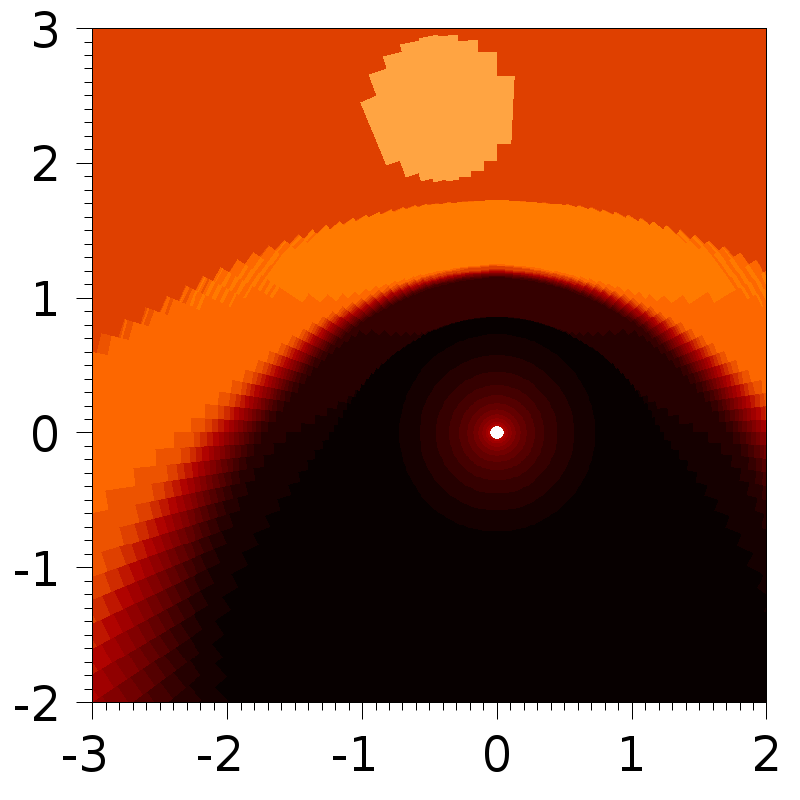}\hfill
    \includegraphics[scale=0.13,trim={18mm 0 0 0},clip]{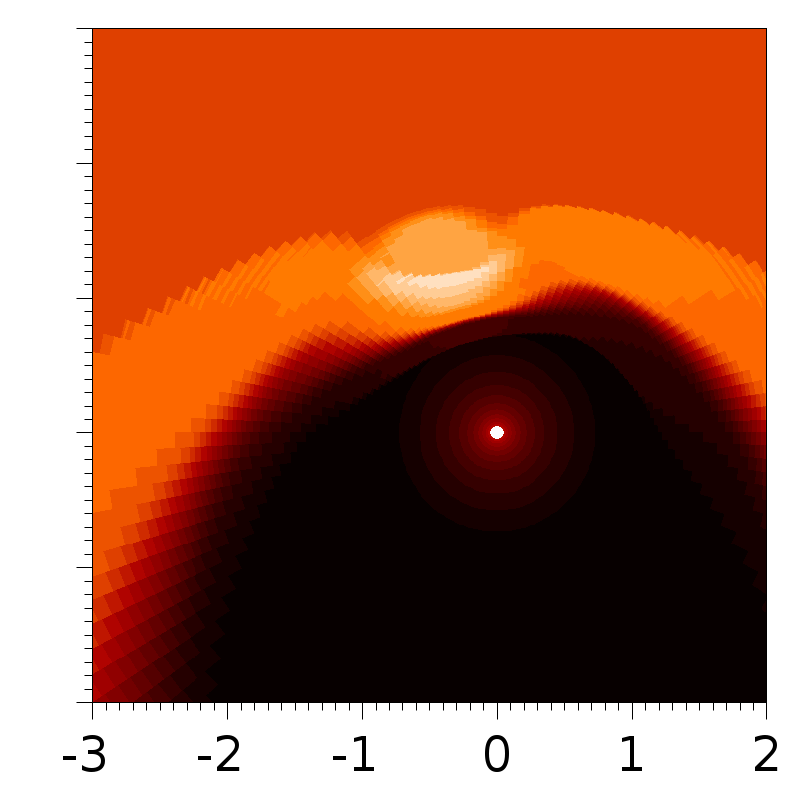}\hfill
    \includegraphics[scale=0.13,trim={18mm 0 0 0},clip]{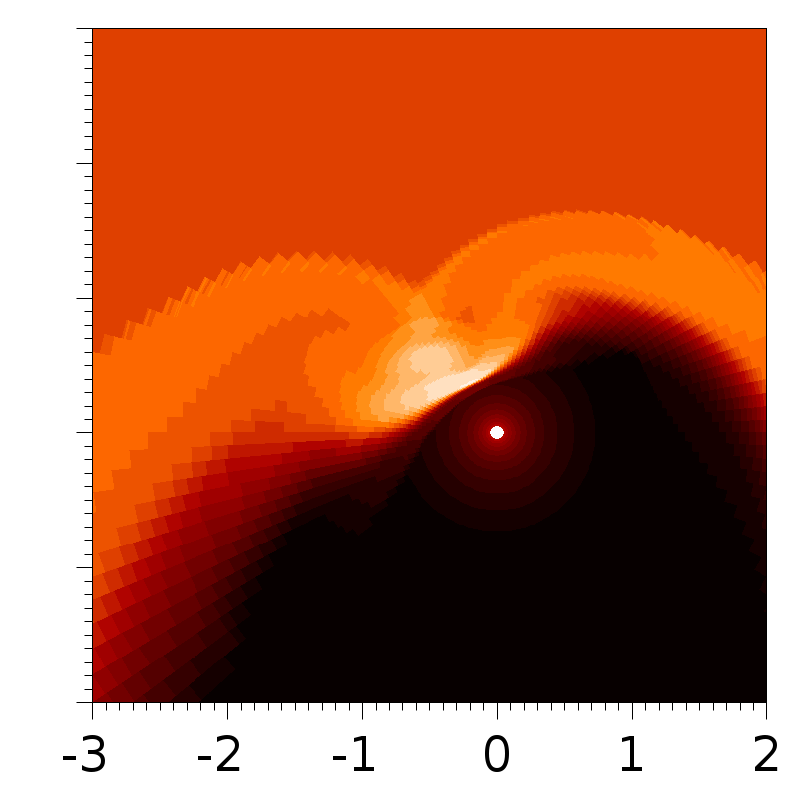}\hfill
    \includegraphics[scale=0.13,trim={18mm 0 0 0},clip]{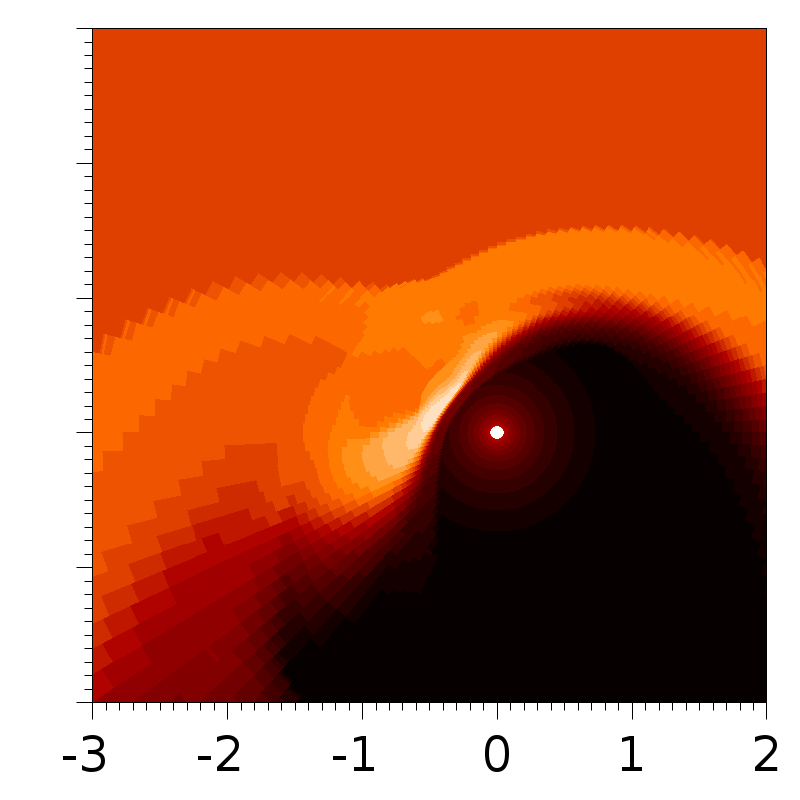}\hfill
    \includegraphics[scale=0.13,trim={18mm 0 0 0},clip]{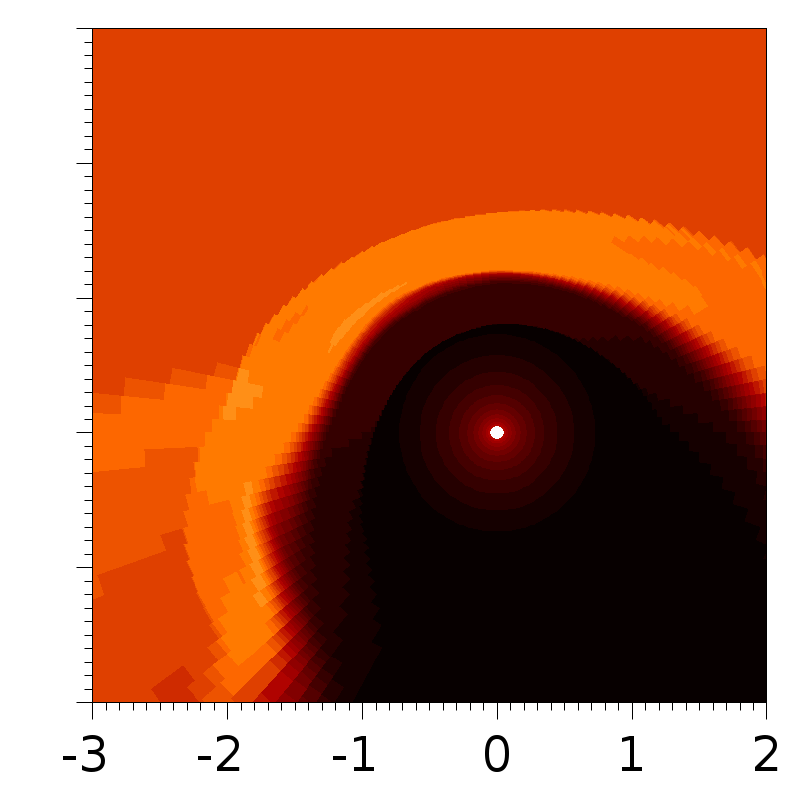}\hfill
    \includegraphics[scale=0.13]{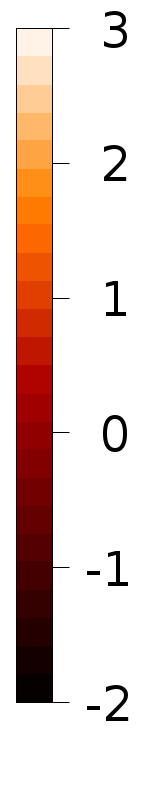}
    
    \raggedright
    \includegraphics[scale=0.263,align=t]{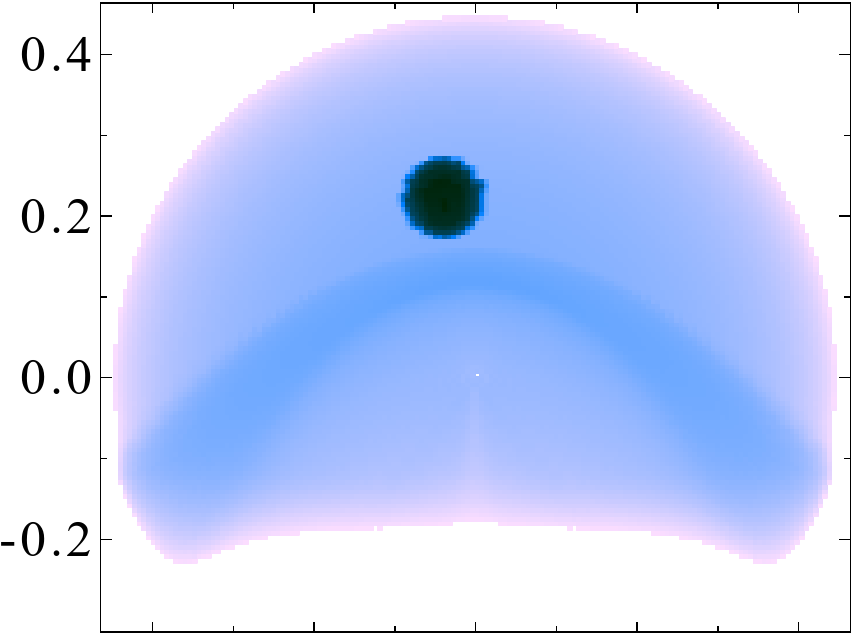}
    \includegraphics[scale=0.263,align=t]{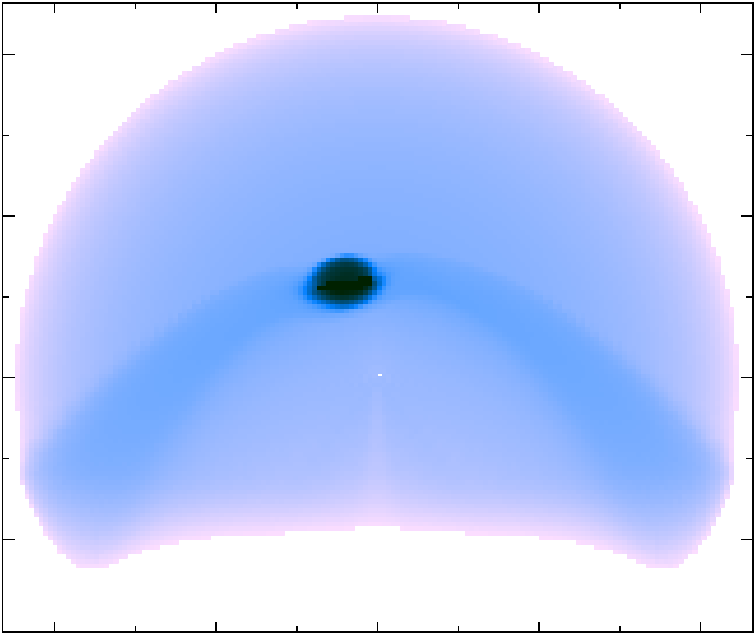}
    \includegraphics[scale=0.263,align=t]{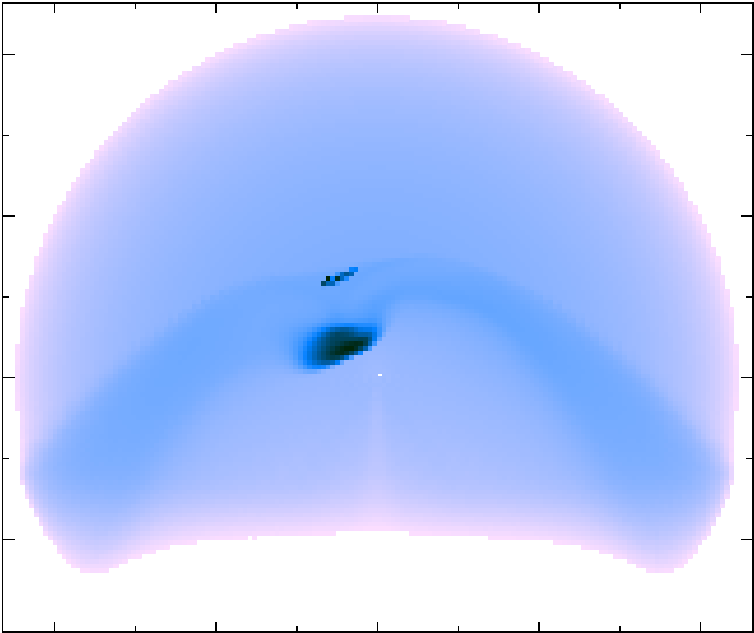}
    \includegraphics[scale=0.263,align=t]{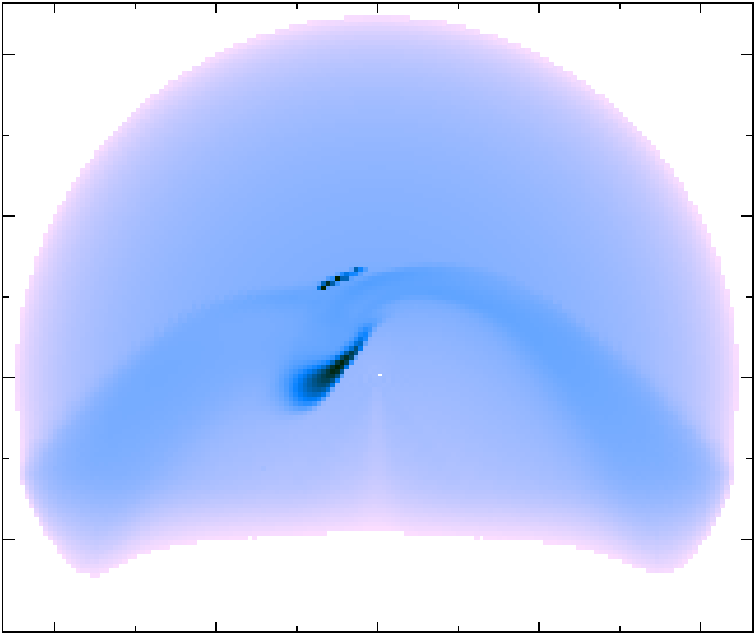}
    \includegraphics[scale=0.263,align=t]{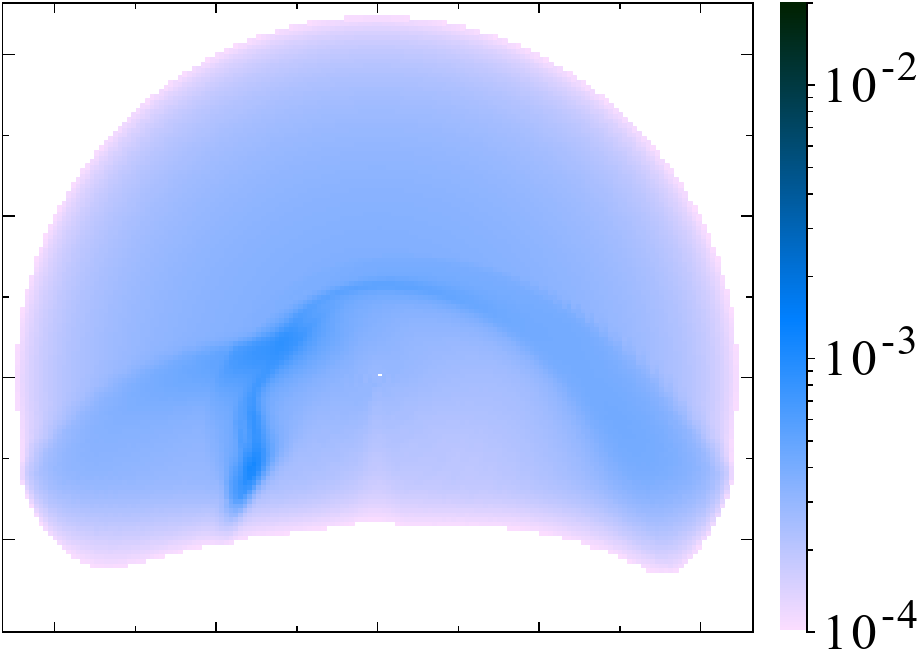}
    \includegraphics[scale=0.263,align=t]{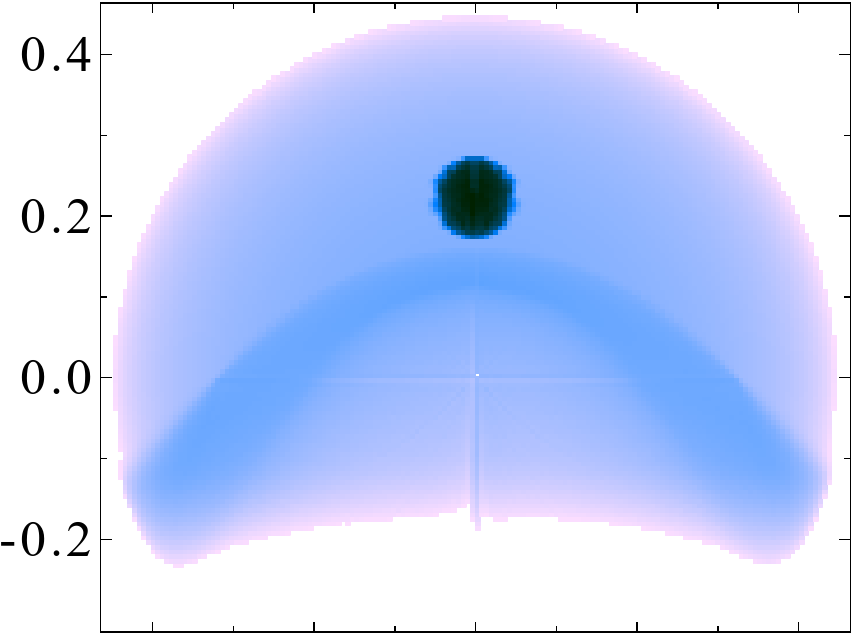}
    \includegraphics[scale=0.263,align=t]{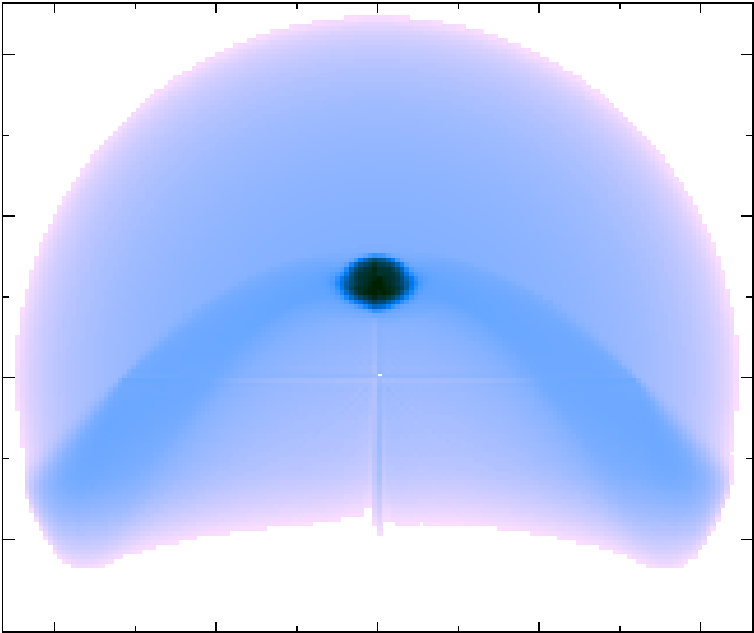}
    \includegraphics[scale=0.263,align=t]{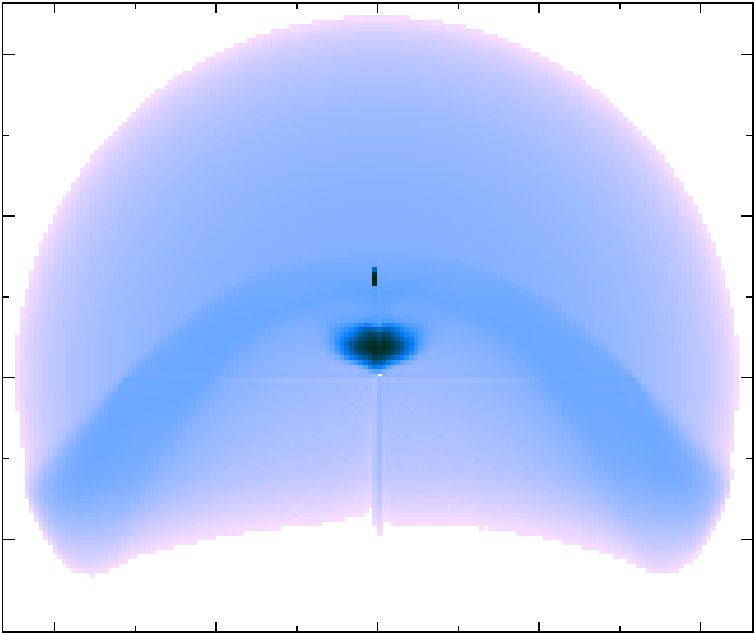}
    \includegraphics[scale=0.263,align=t]{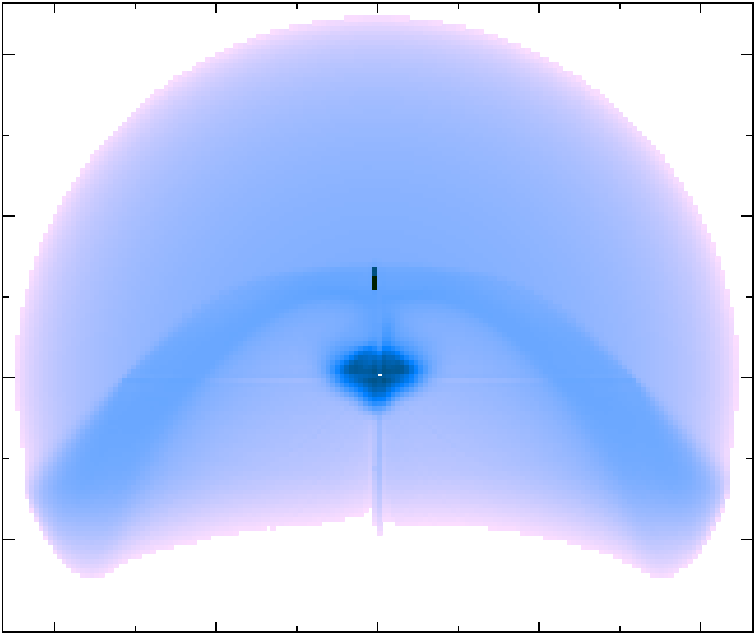}
    \includegraphics[scale=0.263,align=t]{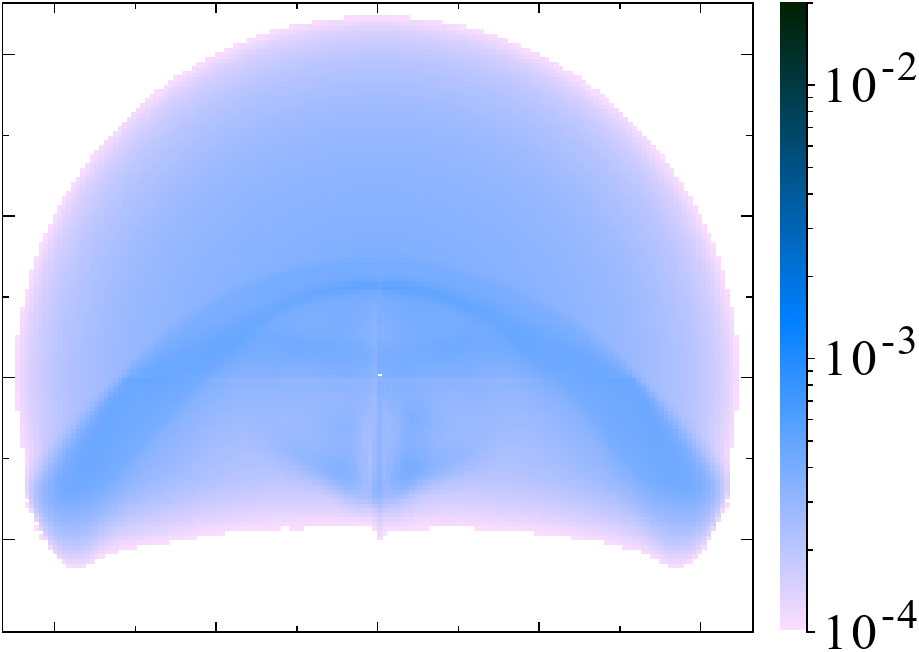}
    \includegraphics[scale=0.263,align=t]{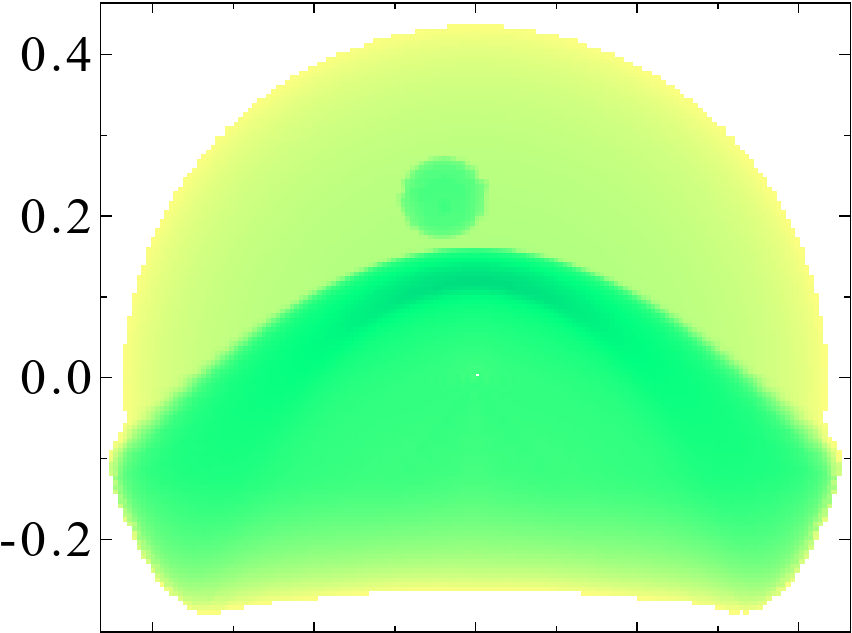}
    \includegraphics[scale=0.263,align=t]{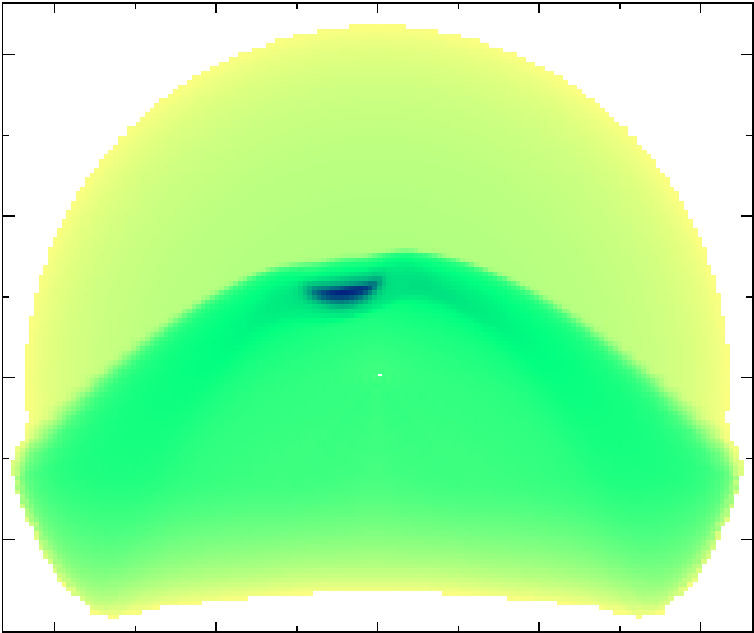}
    \includegraphics[scale=0.263,align=t]{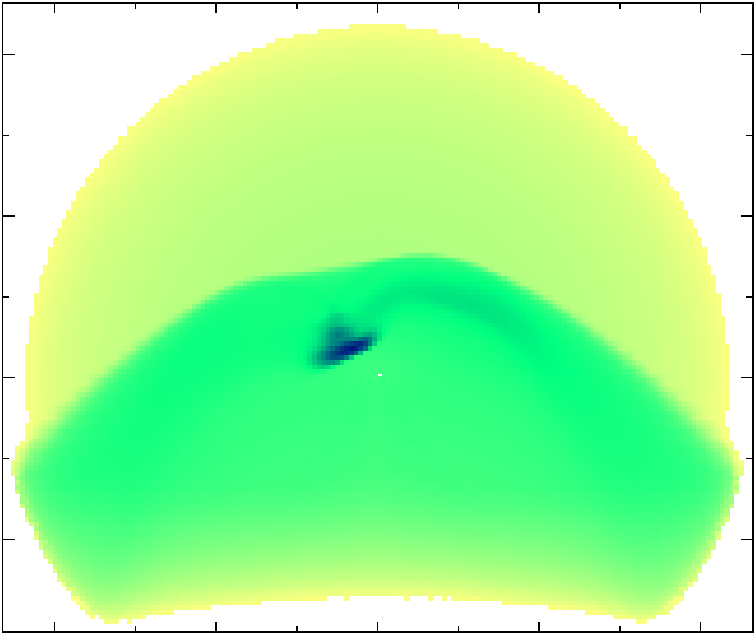}
    \includegraphics[scale=0.263,align=t]{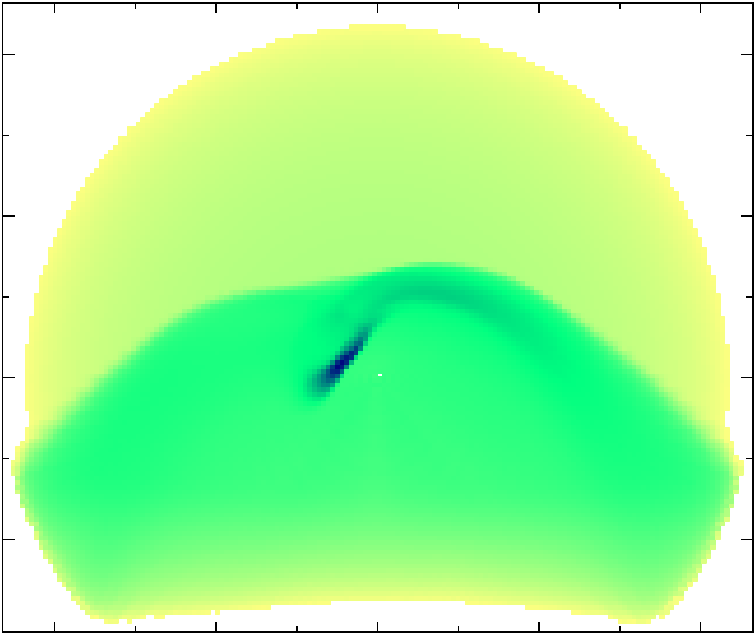}
    \includegraphics[scale=0.263,align=t]{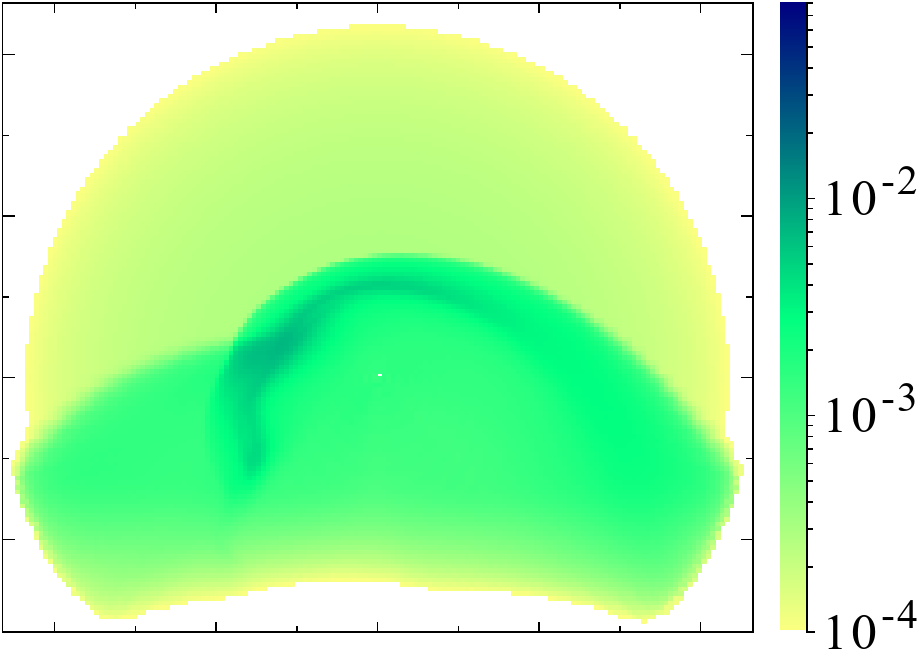}
    \includegraphics[scale=0.263,align=t]{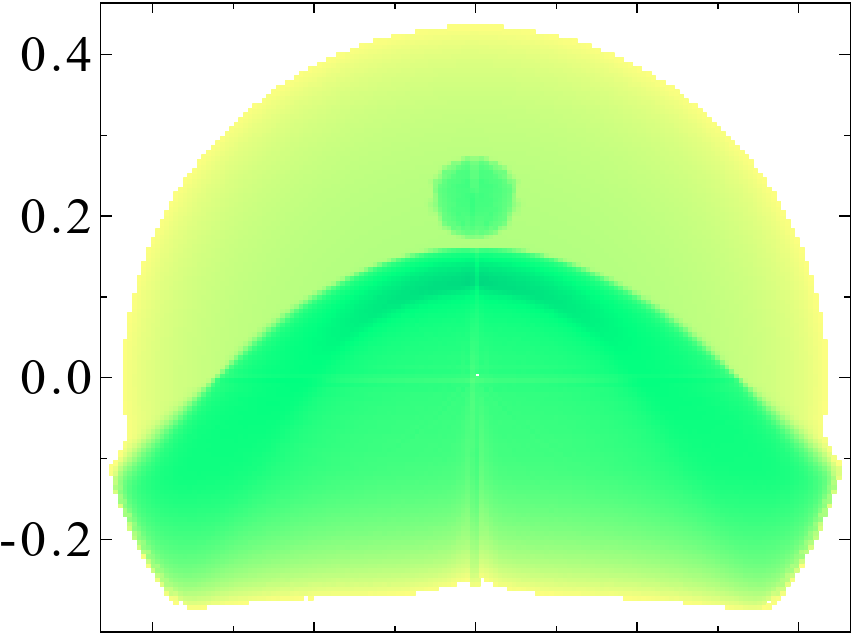}
    \includegraphics[scale=0.263,align=t]{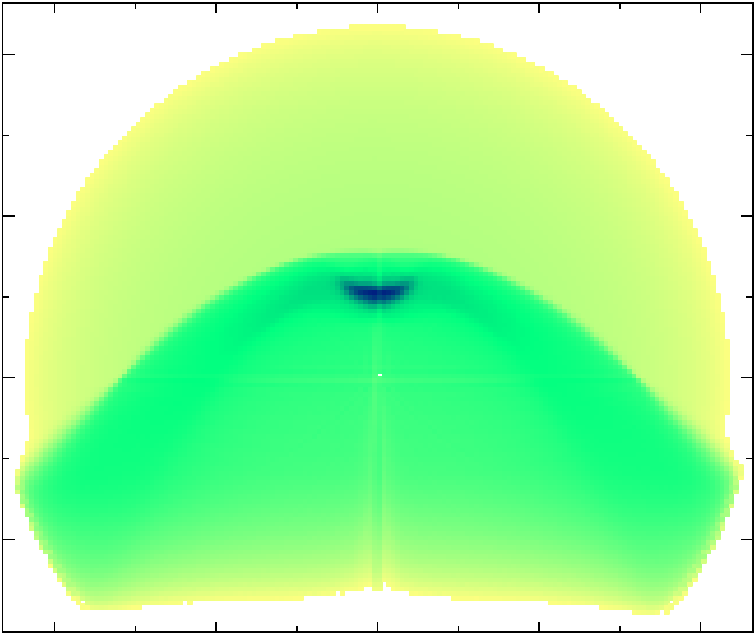}
    \includegraphics[scale=0.263,align=t]{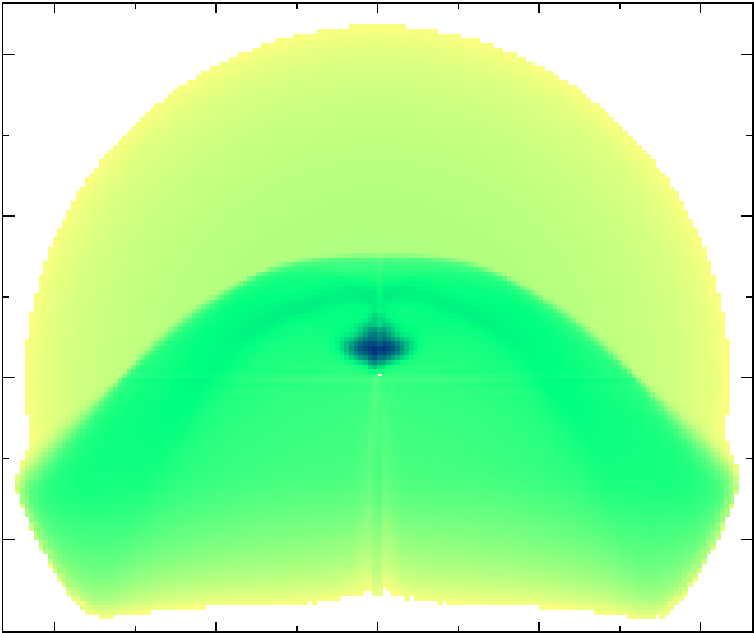}
    \includegraphics[scale=0.263,align=t]{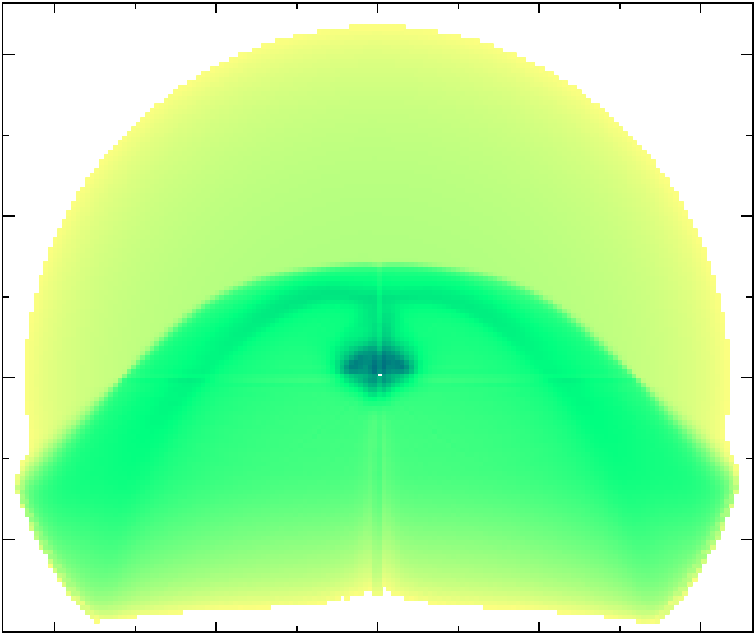}
    \includegraphics[scale=0.263,align=t]{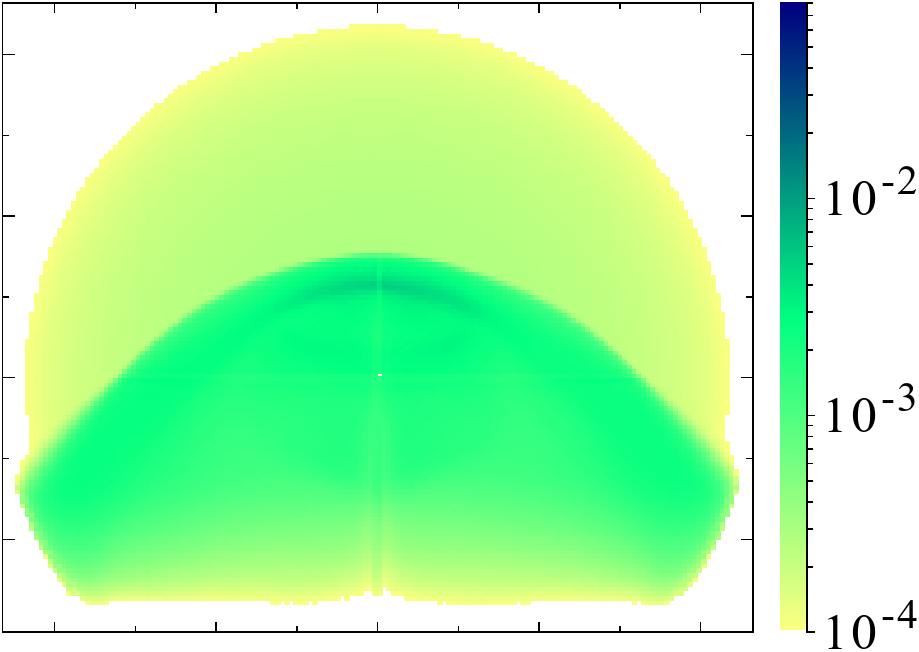}
    \includegraphics[scale=0.263,align=t]{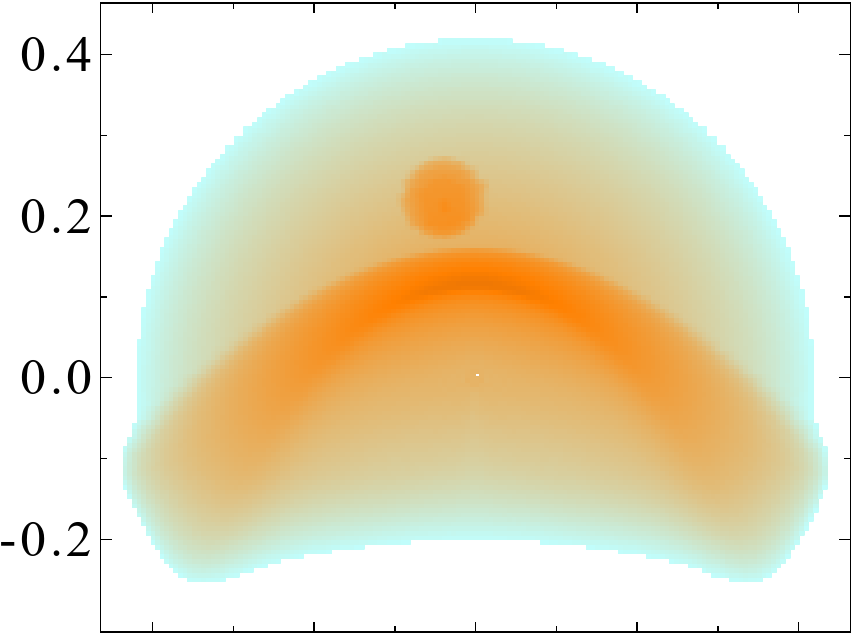}
    \includegraphics[scale=0.263,align=t]{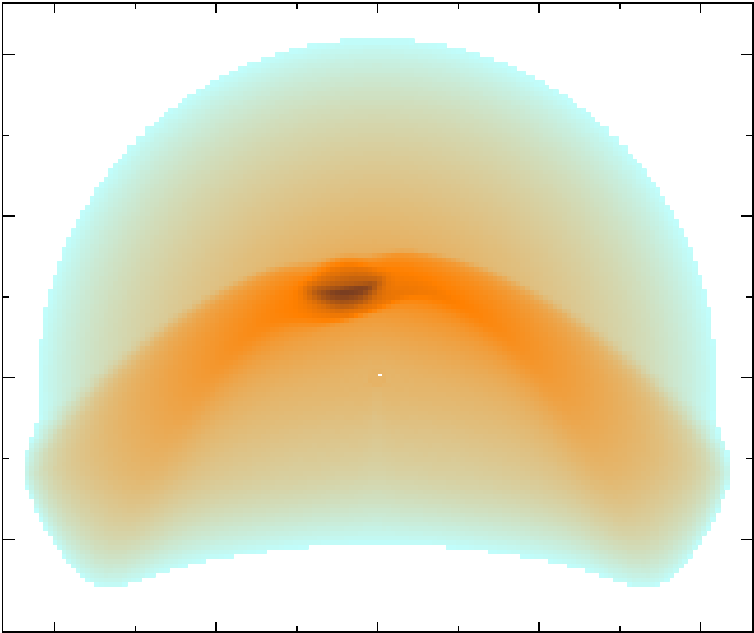}
    \includegraphics[scale=0.263,align=t]{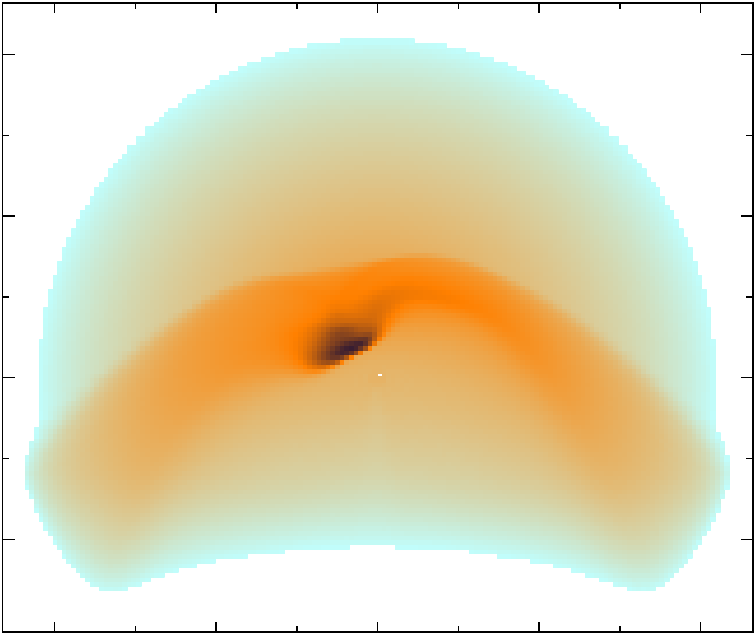}
    \includegraphics[scale=0.263,align=t]{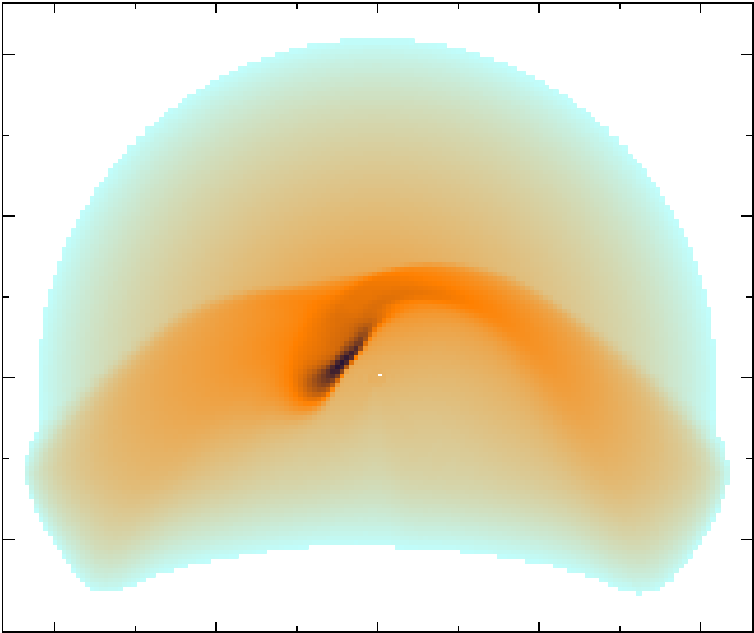}
    \includegraphics[scale=0.263,align=t]{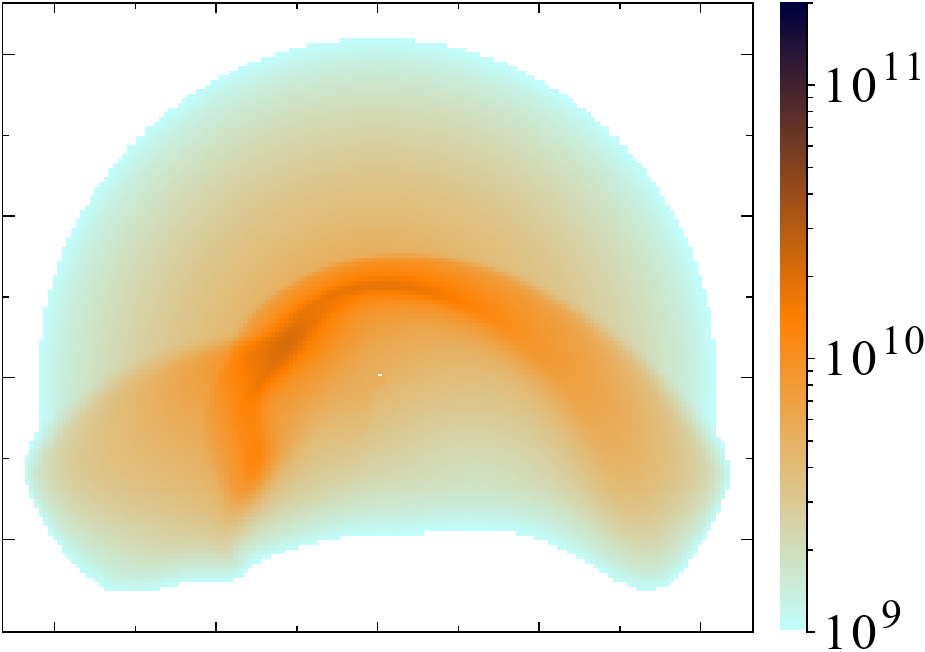}
    \includegraphics[scale=0.263]{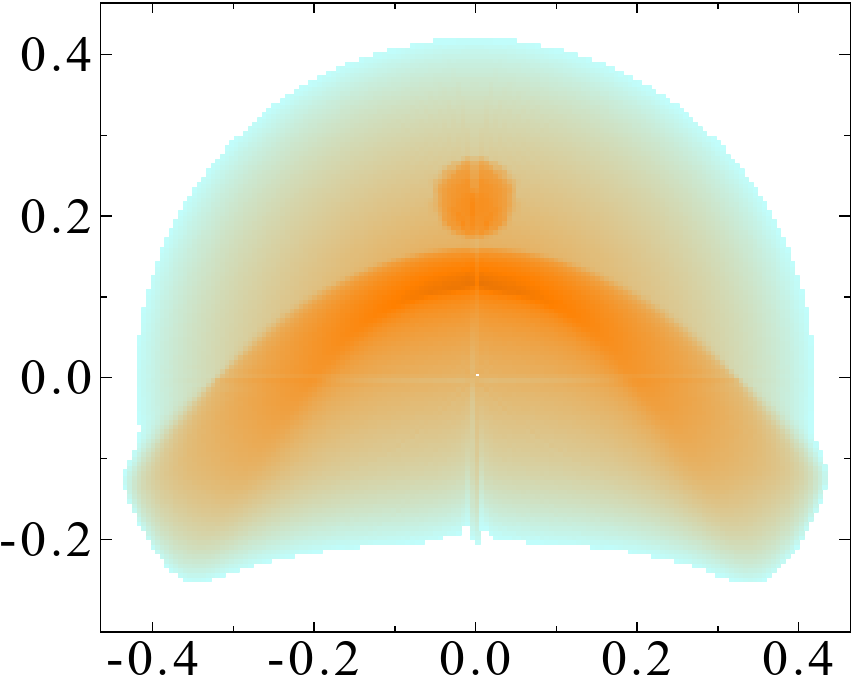}
    \includegraphics[scale=0.263]{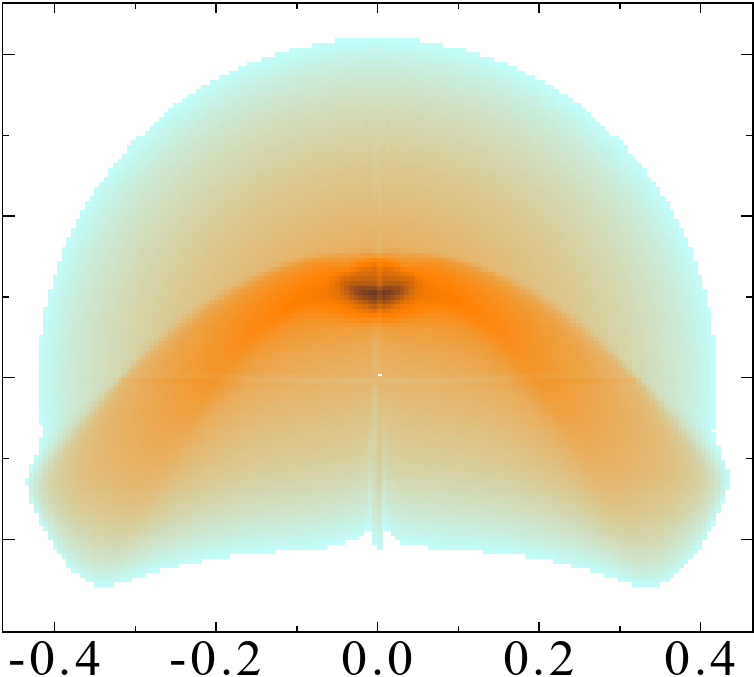}
    \includegraphics[scale=0.263]{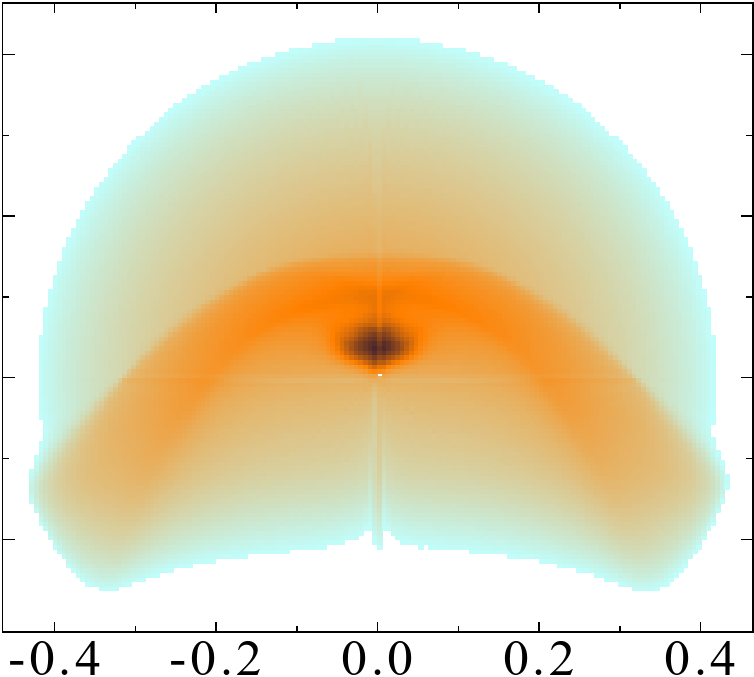}
    \includegraphics[scale=0.263]{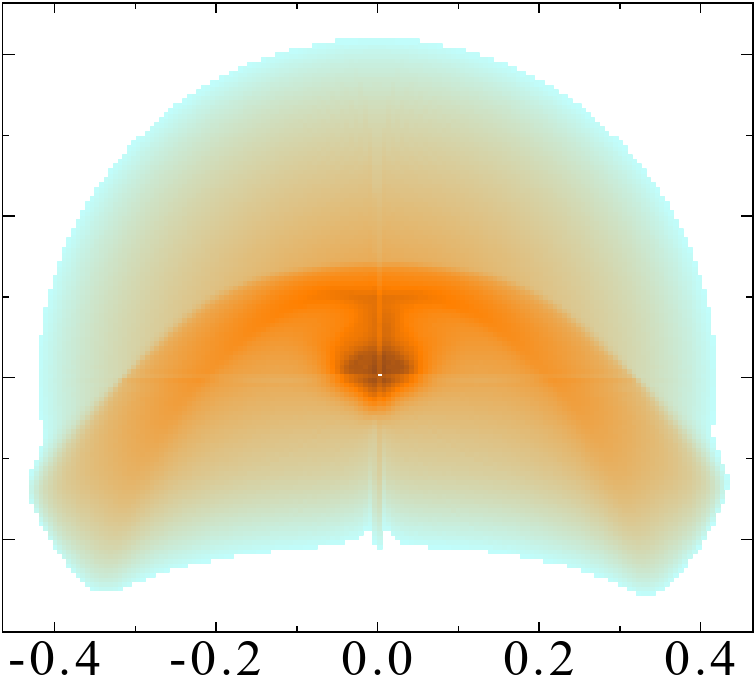}
    \includegraphics[scale=0.263]{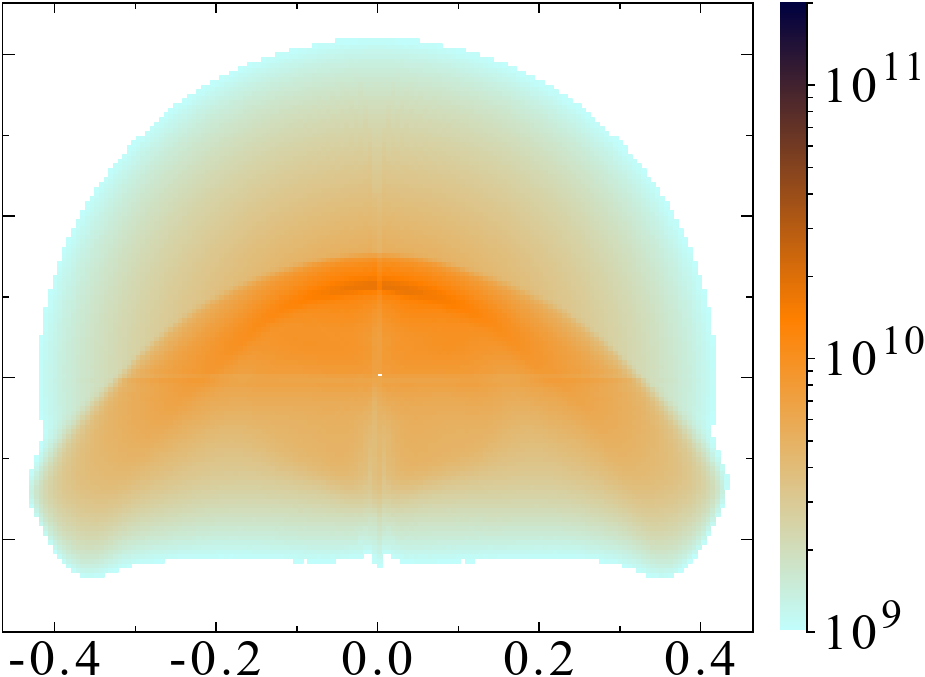}
    \caption{Evolution of the \texttt{n2blb} perturbation. Top row: Cross-sections of the log number density $\log(n \; [\si{cm^{-3}}])$ in the \texttt{n2blb} model's ecliptic plane at $\Delta t\in\{0, 4, 7, 9, 14\}\times 4.4828\,\si{kyr}$ after injecting the perturbation (distances in $\si{pc}$). Six bottom rows: Synthetic observations of the above models in H\textalpha\ (rows 2 and 3) and bremsstrahlung (rows 4 and 5) $[\si{erg\,cm^{-2}\,s^{-1}\,sr^{-1}}]$ and $70\,\si{\micro m}$ dust emission $[\si{Jy\,sr^{-1}}]$ (rows 6 and 7) at the same times and at a distance of $617\,\si{pc}$, face-on (rows 2, 4, and 6) and edge-on (rows 3, 5, and 7)  to the ecliptic (angular extent in degrees). The colour scales have been truncated at the bottom end; values below the minimum threshold are not displayed. The solid angle per pixel is $1.00\times10^{-8}\,\si{sr}$.}
    \label{fig:n2blb}
\end{figure*}

A larger perturbation was injected in the \texttt{n2blb} model, computed in high resolution. As in the \texttt{nblob} model, the perturbation's number density was increased by an order of magnitude to $n_1=110\,\si{cm^{-3}}$, while all other simulated values remained unchanged ($p=\mathit{const.}\Rightarrow T_1=T_0/10$). The perturbation's shape was a sphere with a radius $R_1=225\Delta r\approx 0.544\,\si{pc}$, injected $300\,\text{cells}\ \hat{\approx}\ 0.725\,\si{pc}$ in front of the BS at $\varphi_1=3\,\text{cells}\ \hat{\approx} -9.84\si{\degree}$, meaning that the perturbation's centre was $y_1=(r_{\mathrm{BS}}+0.725\,\si{pc})\tan\varphi_1\approx 0.425\,\si{pc}$ off the inflow axis. Ecliptic cross-sections at selected times are presented in the top row of Fig.~\ref{fig:n2blb}; the leftmost column shows the model at injection.

Moving parallel to the inflow axis, the perturbation impacts the BS off-centre and severely indents it, causing a buildup of material in the outer astrosheath by almost an order of magnitude in the number density (centre-left column, $\Delta t=4\times4.4828\,\si{kyr}$). The perturbation continues to move parallel to the inflow axis, distorting both the AP and the TS and causing a second buildup of material (centre column, $\Delta t=7\times4.4828\,\si{kyr}$). Both buildups move farther in the tail direction and combine (centre-right column, $\Delta t=9\times4.4828\,\si{kyr}$). At this point in time, the perturbation is at its closest position to the star. Its momentum towards the star is overcome by the SW, which causes the inner astrosheath to extend outwards again (right column, $\Delta t=14\times4.4828\,\si{kyr}$). As the perturbation continues to move in the tail direction, the astrosphere slowly returns to its stationary state, which is regained after about $\Delta t=29\times 4.4828\,\si{kyr}\approx 130\,\si{kyr}$.

In rows two and three of Fig.~\ref{fig:n2blb}, synthetic observations of the H\textalpha\ flux density are presented for face-on (row 2) and edge-on (row 3) orientations of the model's ecliptic plane; the two bottom rows show the same for bremsstrahlung. In H\textalpha, the perturbation is brighter than the outer astrosheath by roughly an order of magnitude, darkening only after the perturbation has mostly dissolved (rightmost column). The two discrete buildups of dense material corresponding to the centre column of Fig.~\ref{fig:n2blb} blur together into one H\textalpha-bright clump with a distinct edge at the side towards the star, which marks the TS. A second clump of H\textalpha-bright material, roughly $0.1\,\si{pc}$ upwind of the first clump, is located just upwind of the BS, where the perturbation's movement has caused the BS to bifurcate. The local maximum of H\textalpha\ emission is not caused by a local maximum of the number density, which remains at $n=n_\textrm{ISM}=11\,\si{cm^{-3}}$, but by a drastic decrease in the temperature to values as low as $T=10^{-3}\,\si{K}$. This clump appears to be very narrow in edge-on observations of the ecliptic (lower row), covering only a single pixel column. Both the unphysically low temperatures and the singular extent of this region indicate that it likely is a numerical artefact and should be dismissed for observational purposes. The movement and dissolution of the clump made of material from the perturbation are easily visible in H\textalpha, leaving a bright trail (rightmost column) that is particularly distinctive in face-on observations (upper row).

In bremsstrahlung (rows 4 and 5 in Fig.~\ref{fig:n2blb}), the movement of the perturbation can be clearly followed as well. While the perturbation is not as bright compared to the ISM emission in bremsstrahlung as it is in H\textalpha\ at injection (leftmost column), the buildup of dense material after the perturbation impacts the BS causes said clump to emit in bremsstrahlung just as brightly as it does in H\textalpha. The second clump, caused by low temperatures, cannot be seen in bremsstrahlung, as can be expected from Eq.~(\ref{eq:brs}). While the bremsstrahlung flux density is of the same order of magnitude compared to the H\textalpha\ flux density, one must keep in mind that H\textalpha\ is line emission, whereas bremsstrahlung is continuous, as seen in Fig.~\ref{fig:brsspec}. Thus, the spectral bremsstrahlung flux density is far lower compared to that of H\textalpha.

The apparent structure in $70\,\si{\micro m}$ dust emission is similar to that of bremsstrahlung. The most notable difference is that the dust emission arc is hollow, as it is for H\textalpha: Because little emission comes from the inner astrosheath, the arc structure has a distinct inner edge. This is not true for bremsstrahlung, where the arc structure is filled out.

\section{Observational classification}\label{sec:obs}

\begin{figure}
    \centering
    \includegraphics[scale=0.16]{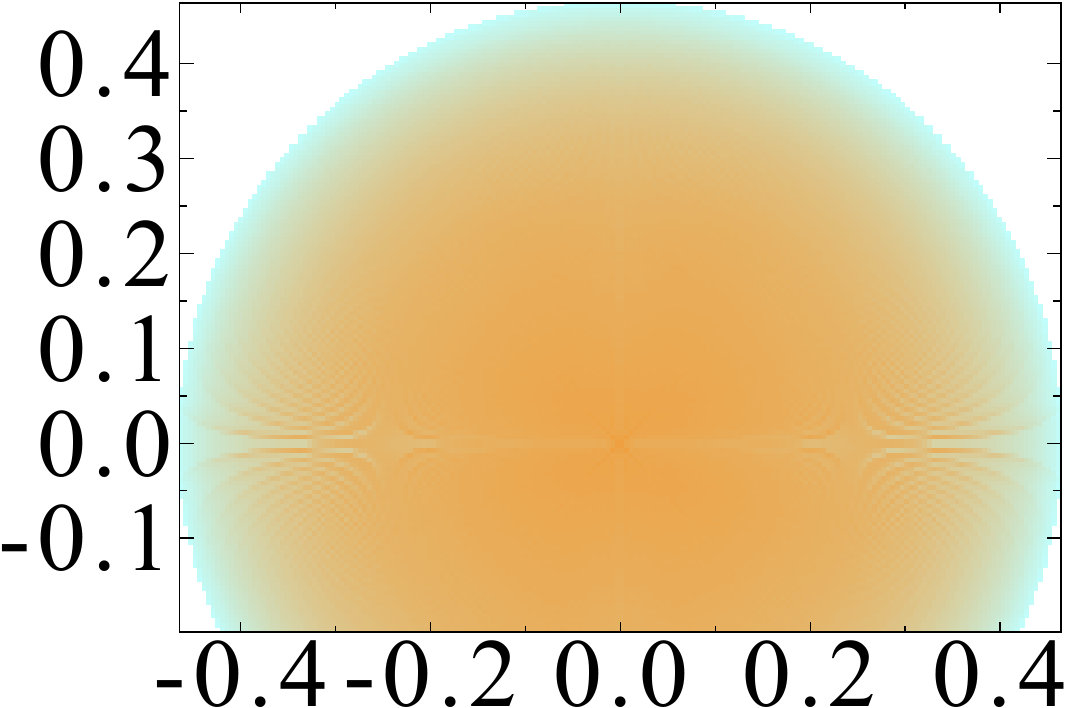}\hfill
    \includegraphics[scale=0.16]{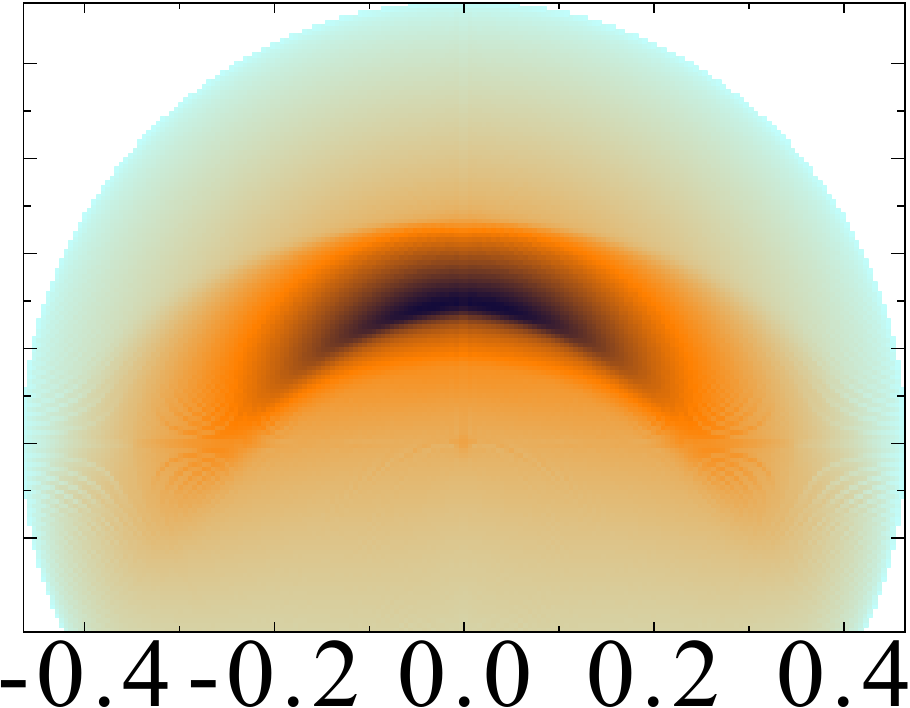}\hfill
    \includegraphics[scale=0.16]{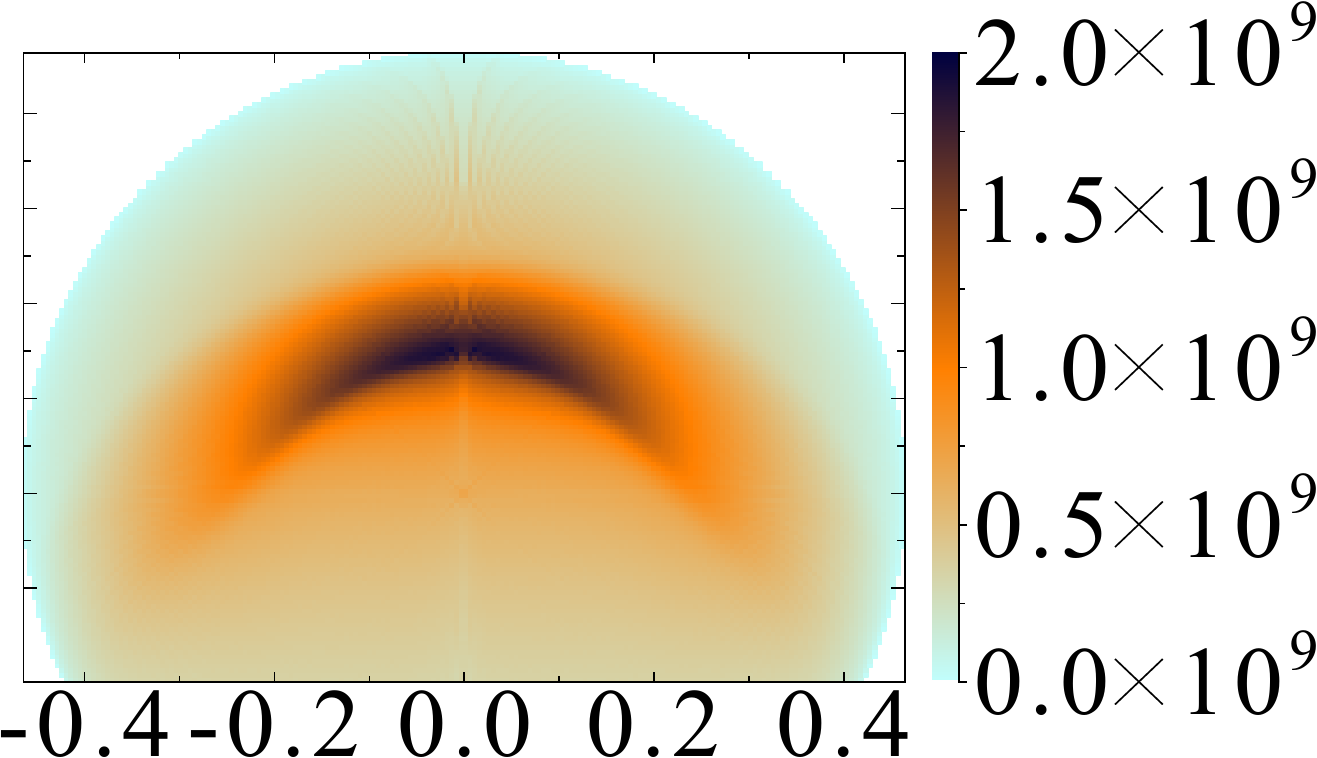}
    \caption{Synthetic observations of the stationary astrosphere on the \texttt{hires} grid in $70\,\si{\micro m}$ dust emission $[\si{Jy\,sr^{-1}}]$ at a distance of $617\,\si{pc}$ at different angles: with the LOS in the direction of the star's relative motion (left), at $45\si{\degree}$ to the ecliptic's face-on view (centre), and at $45\si{\degree}$ to the ecliptic's edge-on view (right). The solid angle per pixel is $4.00\times 10^{-8}\,\si{sr}.$}
    \label{fig:basrot}
\end{figure}

To facilitate comparing the perturbed models to observational data, synthetic observations for additional viewing angles were generated. Figure~\ref{fig:basrot} shows the unperturbed astrosphere at stationarity in the \texttt{hires} grid in $70\,\si{\micro m}$ dust emission from three different perspectives: In the left panel, the LOS is oriented parallel to the star's relative motion; because the astrosphere is reasonably symmetric about this axis, the resulting image shows only a slight radial gradient. The dust emission is not significantly higher compared to the homogeneous ISM. The two right panels both show the astrosphere at an inclination of $45\si{\degree}$ between the axis of stellar motion and the LOS. The centre panel displays a tilted view of the face-on ecliptic; this rotation is halfway between the zero-inclination perspective of the left panel and the face-on perspective used above (see e.g. Fig.~\ref{fig:basis}). The right panel displays a tilted view of the edge-on ecliptic, halfway between the zero-inclination perspective and the edge-on perspective only used for analysing the \texttt{n2blb} model (rows 3, 5, and 7 of Fig.~\ref{fig:n2blb}). Because of the astrosphere's symmetry, the two images only differ in their geometry, most notably the Moir\'e patterns caused by the projection.

\begin{figure*}
    \raggedright
    \includegraphics[scale=0.3,align=c]{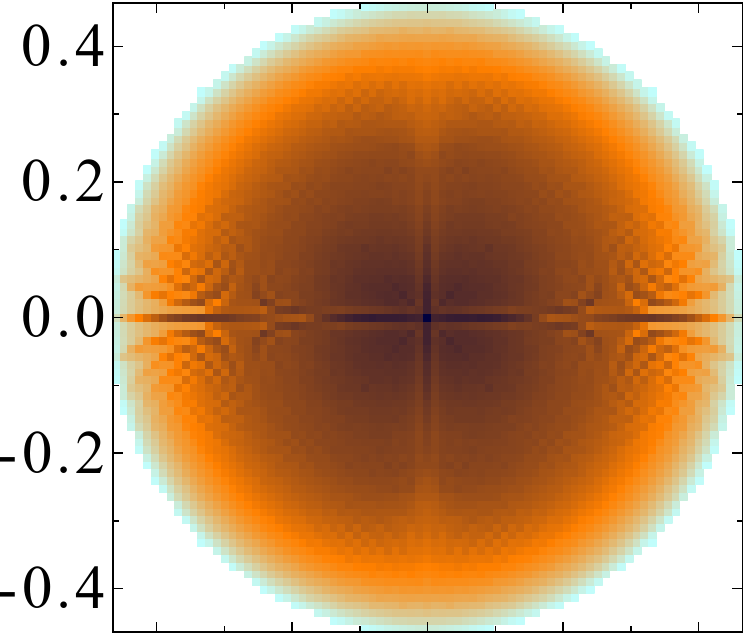}
    \includegraphics[scale=0.3,align=c]{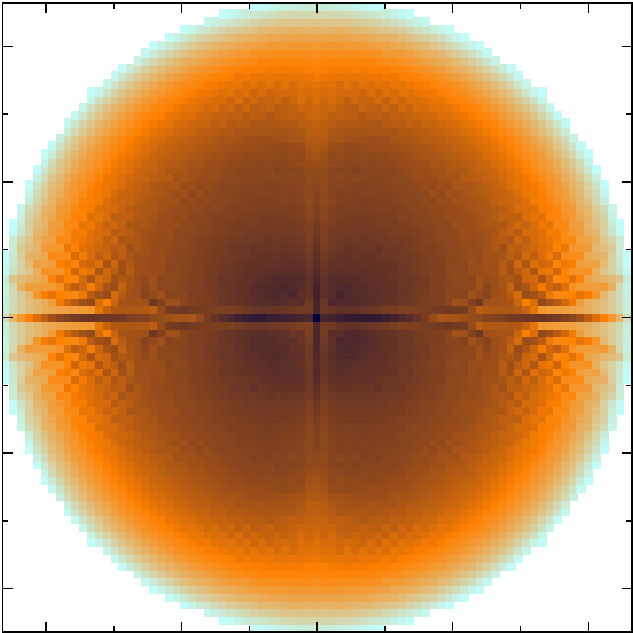}
    \includegraphics[scale=0.3,align=c]{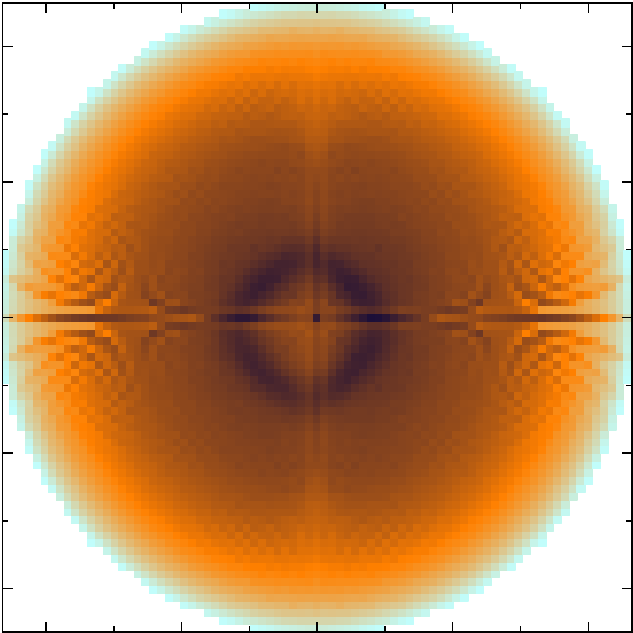}
    \includegraphics[scale=0.3,align=c]{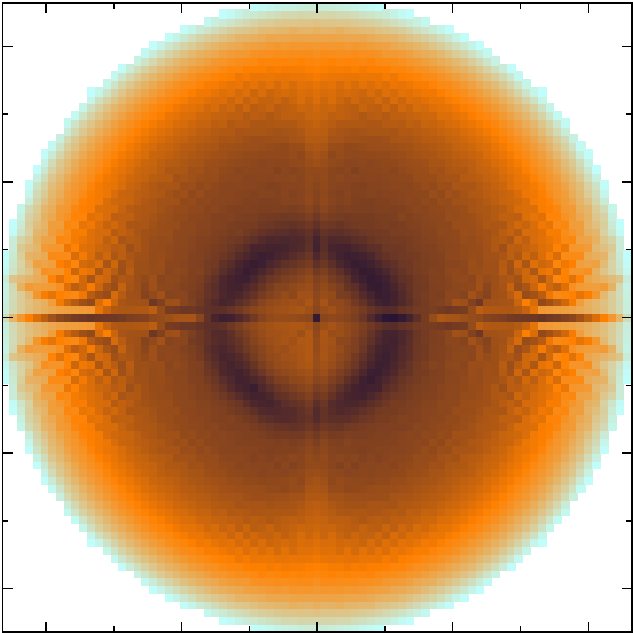}
    \includegraphics[scale=0.3,align=c]{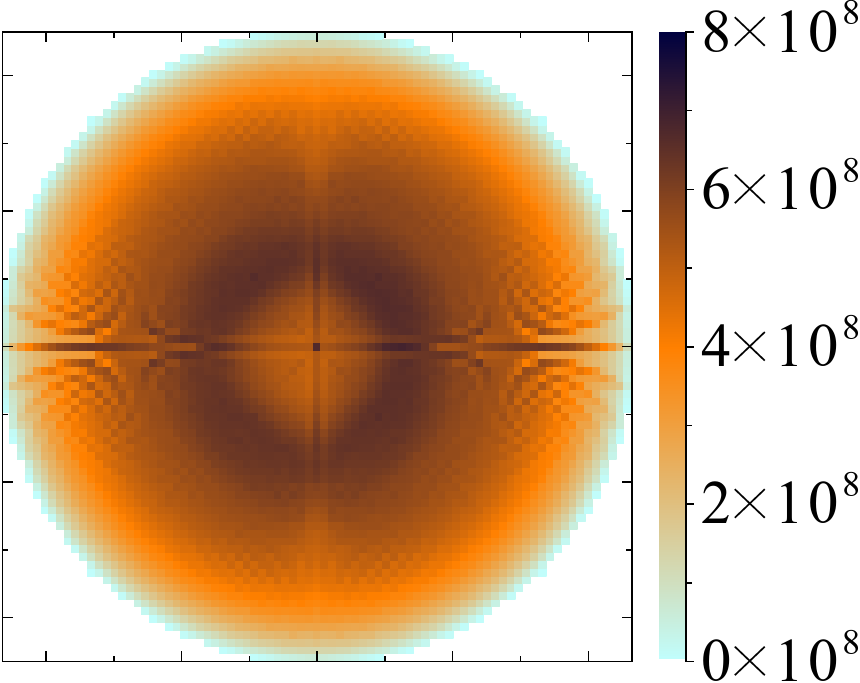}
    \includegraphics[scale=0.21]{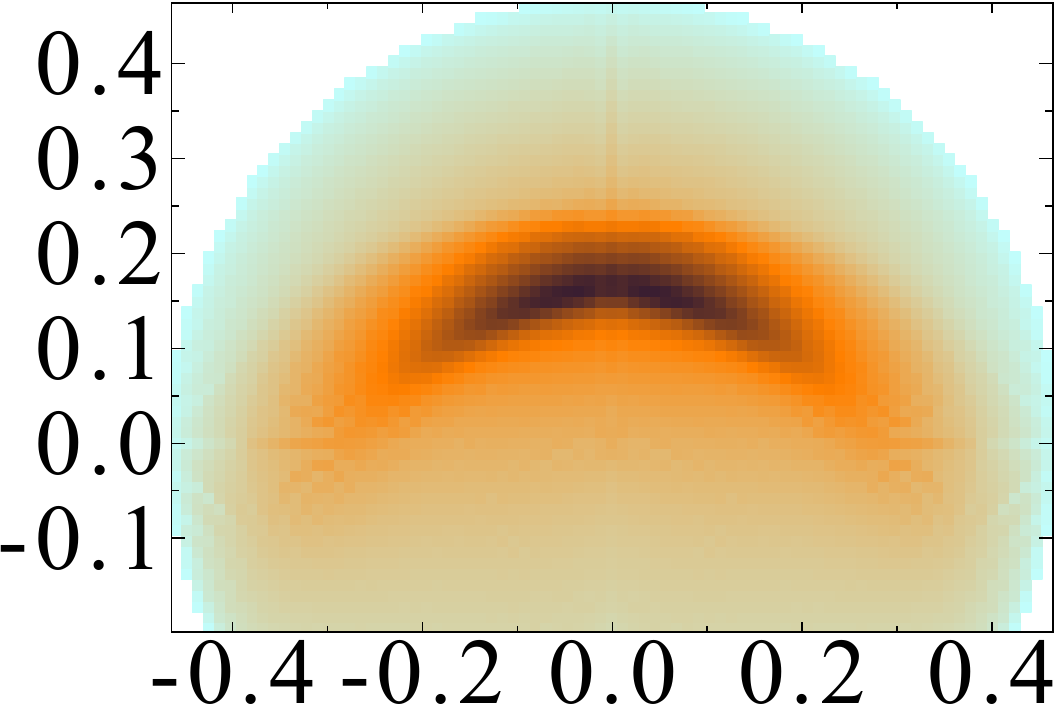}
    \includegraphics[scale=0.21]{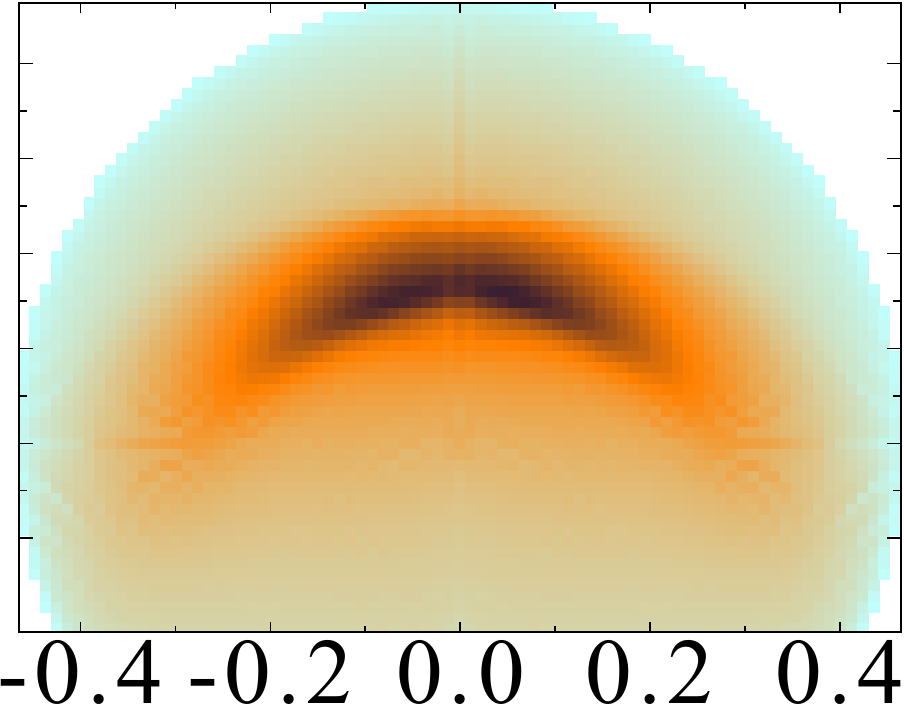}
    \includegraphics[scale=0.21]{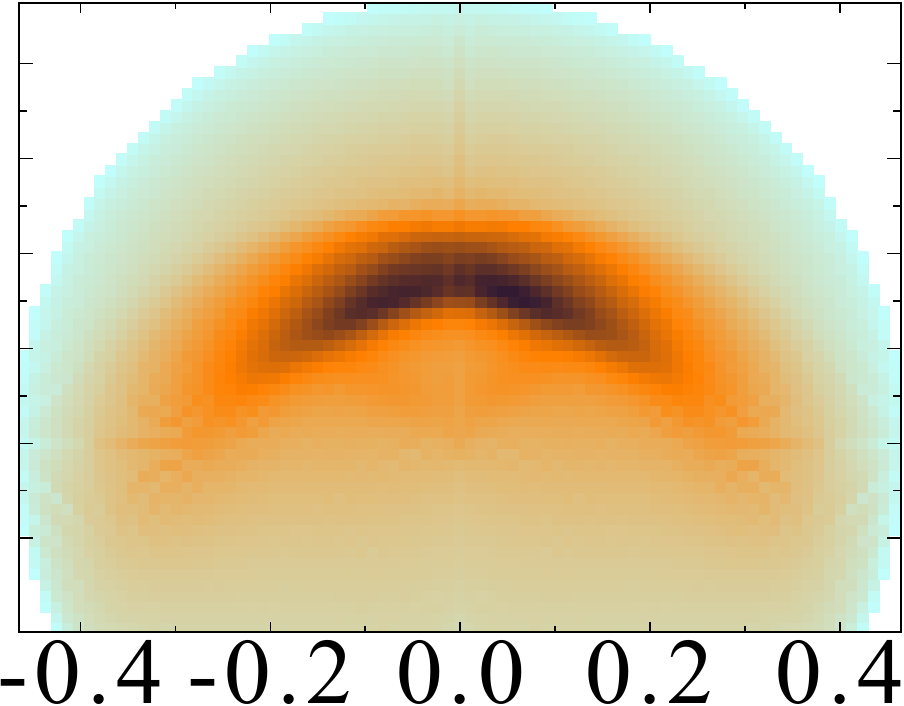}
    \includegraphics[scale=0.21]{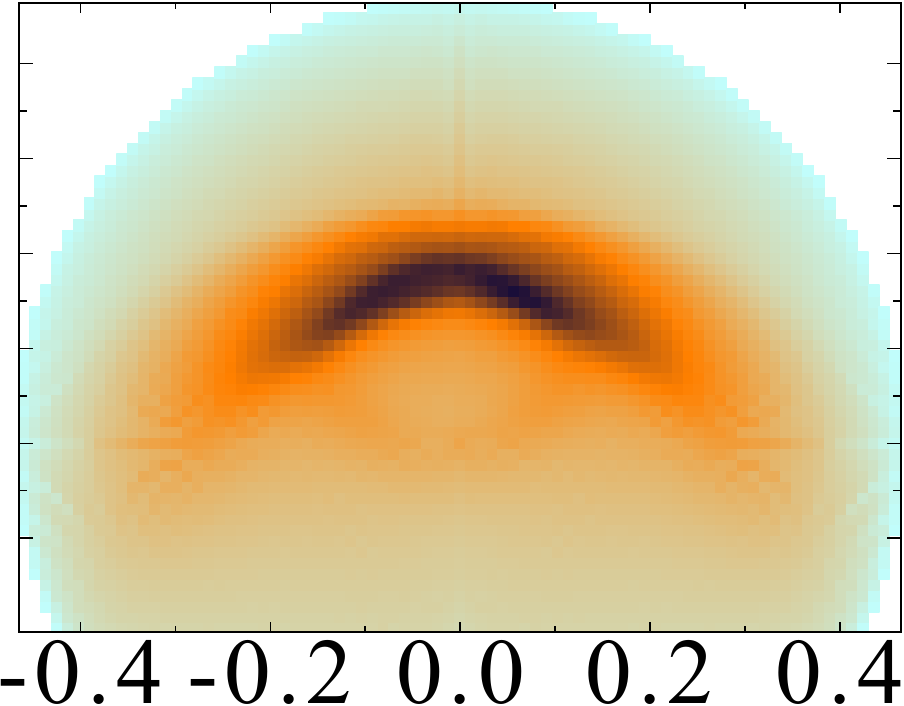}
    \includegraphics[scale=0.21]{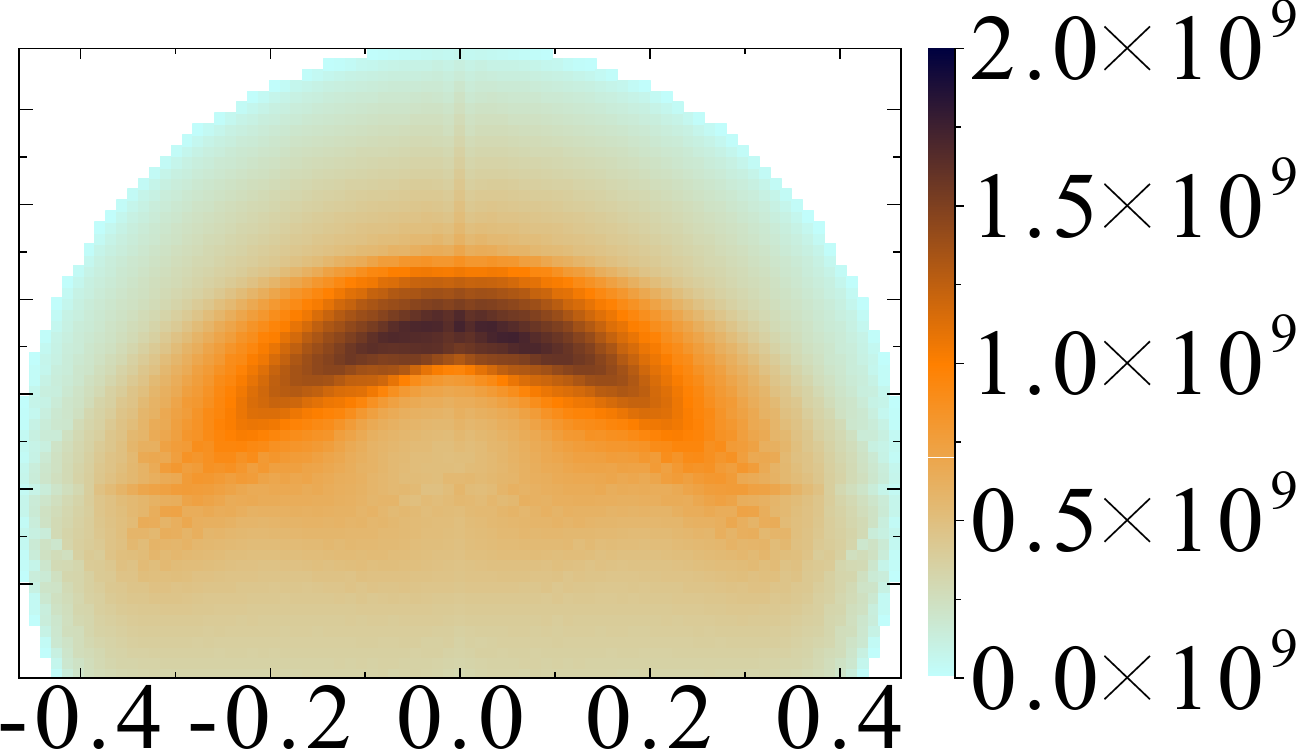}
    \caption{Synthetic observations of the \texttt{v5blb} model at $\Delta t\in\{0,1,2.5,4,6\}\times 4.4828\,\si{kyr}$ after injecting the perturbation in $70\,\si{\micro m}$ dust emission $[\si{Jy\,sr^{-1}}]$ at a distance of $617\,\si{pc}$ at different angles: with the LOS in the direction of the star's relative motion (top) and at $45\si{\degree}$ to the ecliptic's face-on view (bottom). The solid angle per pixel is $4.00\times 10^{-8}\,\si{sr}.$}
    \label{fig:v5rot}
\end{figure*}

In Fig.~\ref{fig:v5rot}, further rotations of the \texttt{v5blb} model in $70\,\si{\micro m}$ dust emission are displayed. The configuration is that of Fig.~\ref{fig:v5blb} but rotated to the zero-inclination perspective in the top row and the $45\si{\degree}$-inclination face-on perspective in the bottom row. Because of the rotational symmetry about the inflow axis, the $45\si{\degree}$-inclination edge-on perspective does not significantly differ from the face-on view and is therefore not plotted. For both rows of images, the colour scales have been readjusted to utilise the full contrast. The Moir\'e patterns, fairly distinct at the images' outer edges, are projection effects.

At zero inclination, the total emission is reduced by roughly three-quarters compared to the $90\si{\degree}$-inclination perspective of Fig.~\ref{fig:v5blb}. The perturbation is distinctly visible as a ring of high emission, similar to Class III ring-type interactions as described by \citet{cox12}. At $45\si{\degree}$ inclination, the perturbation's outer border is not visible as a shell-like structure as it is at $90\si{\degree}$ (Fig.~\ref{fig:v5blb}). Instead, the `bow shock' arc shows a double structure, with an inverted and much fainter arc below. For inclination angles $\in\left]0\si{\degree}, 45\si{\degree}\right[$, the two arcs would appear more similar in intensity and more circular in shape. A denser or hotter outer border of the perturbation would further increase the intensity, resulting in an image similar to Class II eye-type interactions as described by \citet{cox12}.

\begin{figure*}
    \raggedright
    \includegraphics[scale=0.308,align=c]{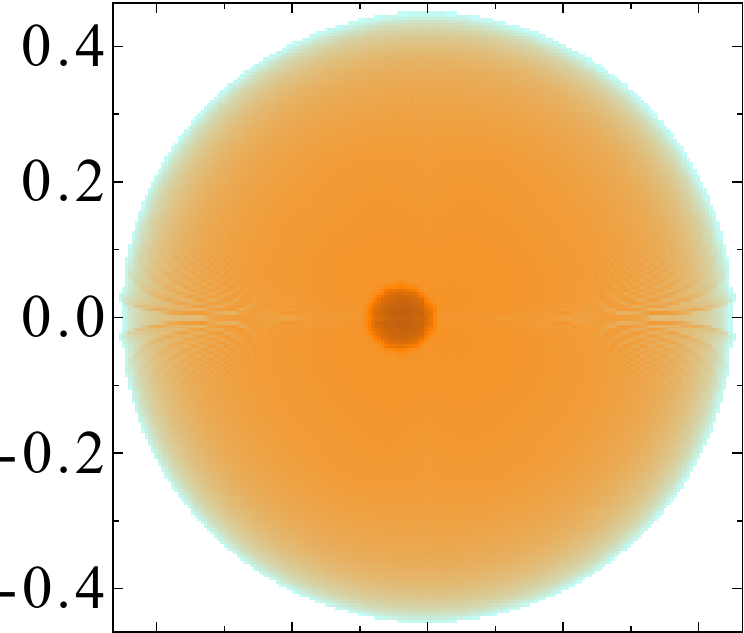}
    \includegraphics[scale=0.308,align=c]{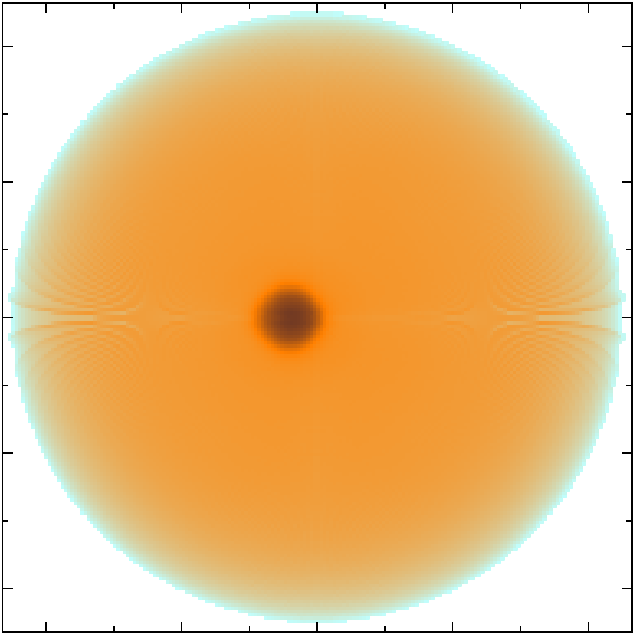}
    \includegraphics[scale=0.308,align=c]{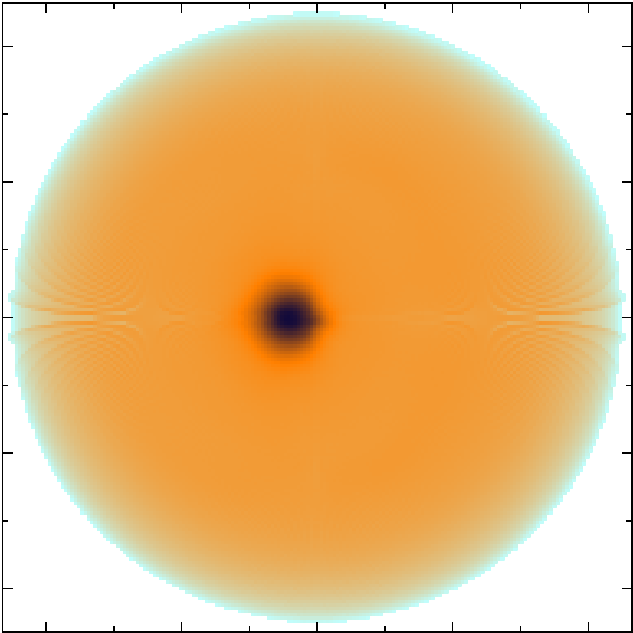}
    \includegraphics[scale=0.308,align=c]{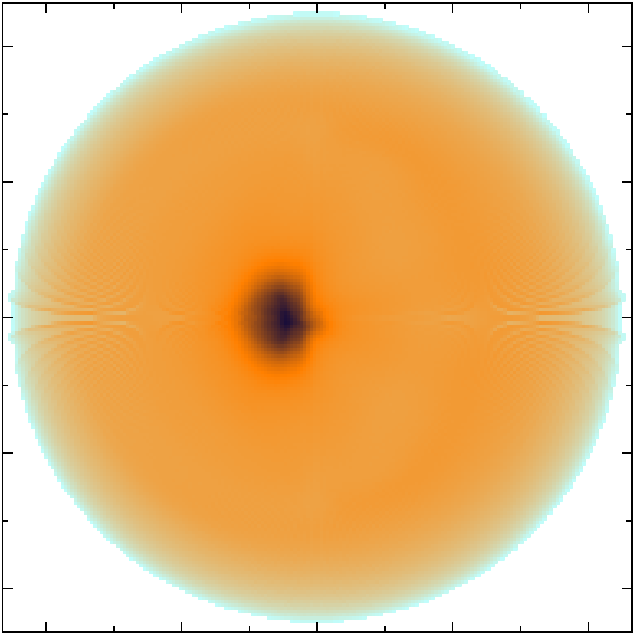}
    \includegraphics[scale=0.308,align=c]{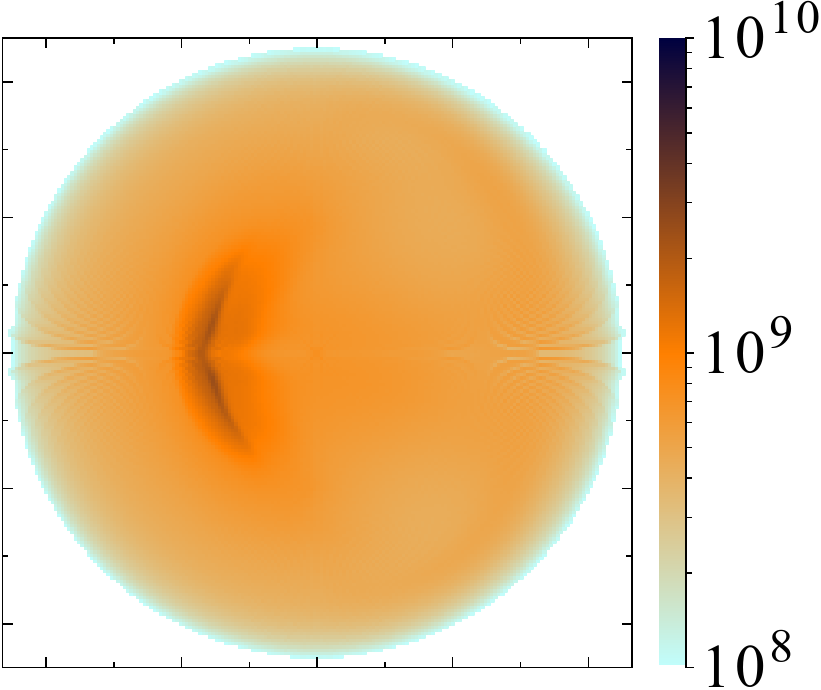}
    \includegraphics[scale=0.26,align=c]{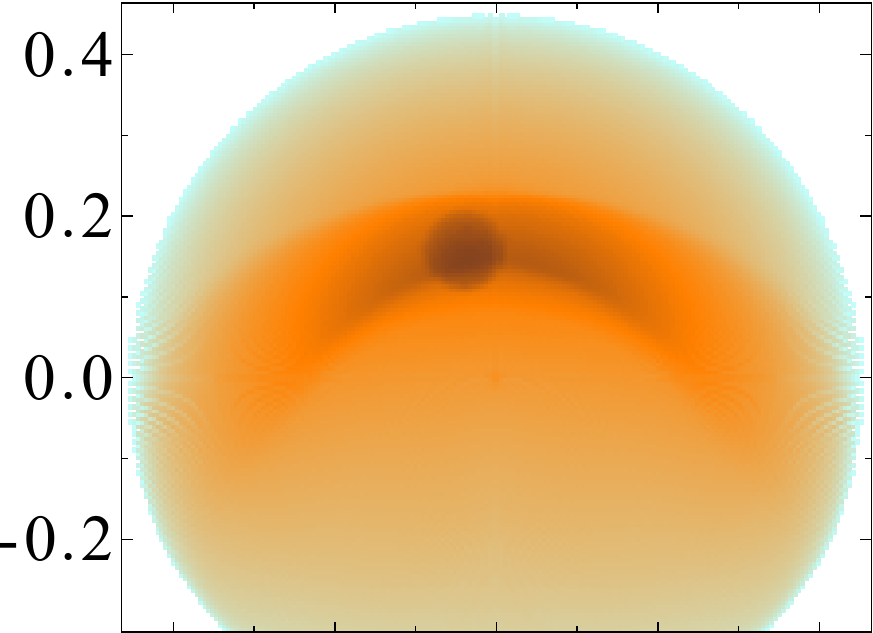}
    \includegraphics[scale=0.26,align=c]{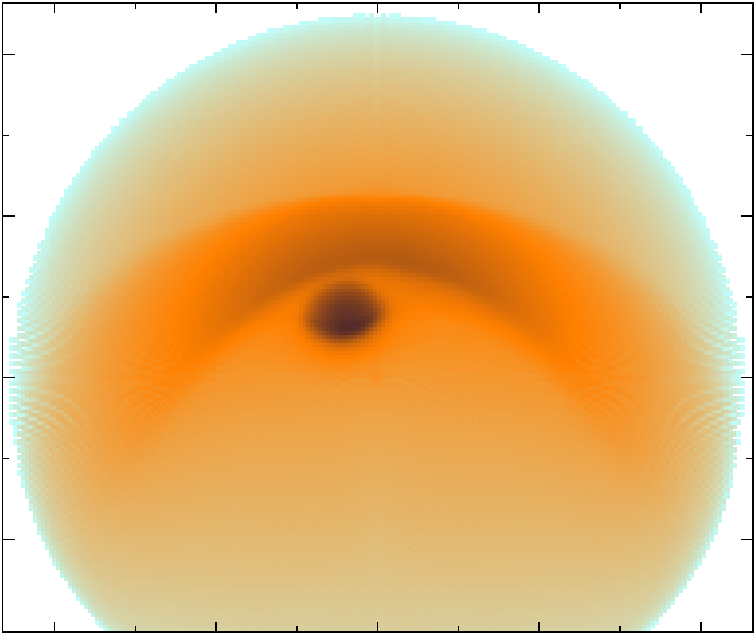}
    \includegraphics[scale=0.26,align=c]{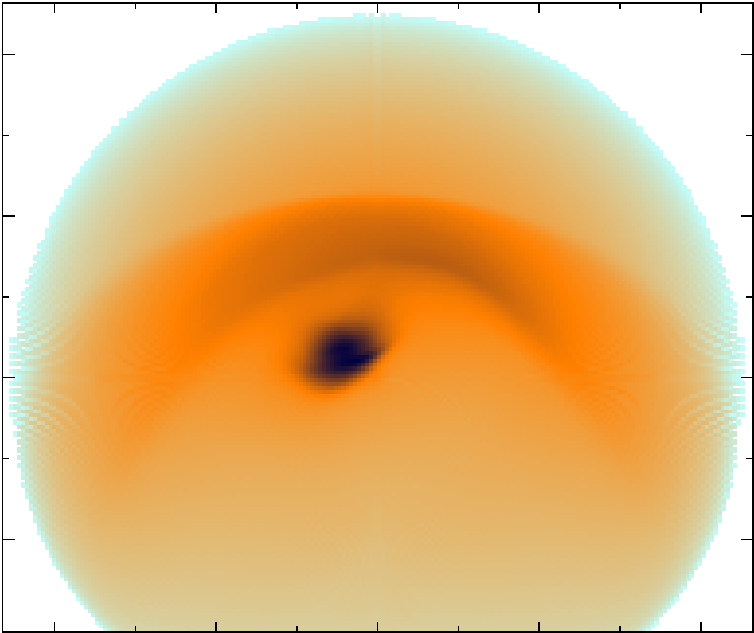}
    \includegraphics[scale=0.26,align=c]{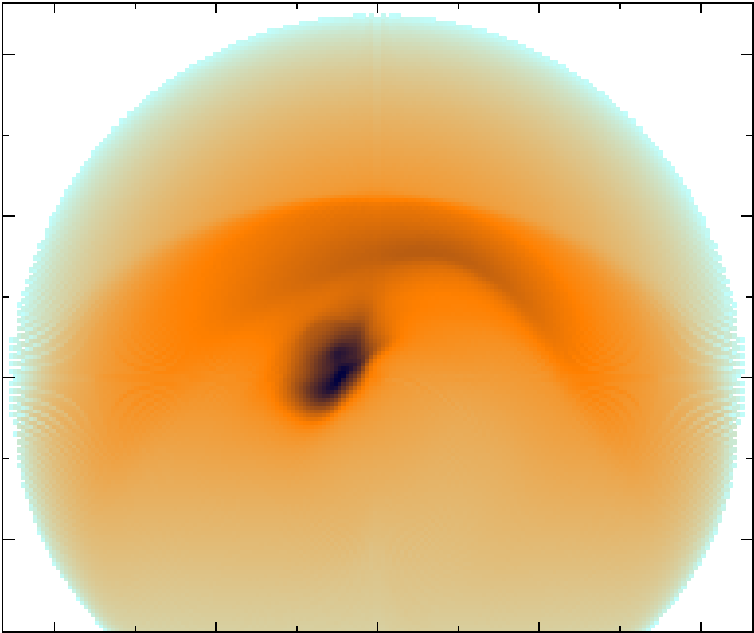}
    \includegraphics[scale=0.26,align=c]{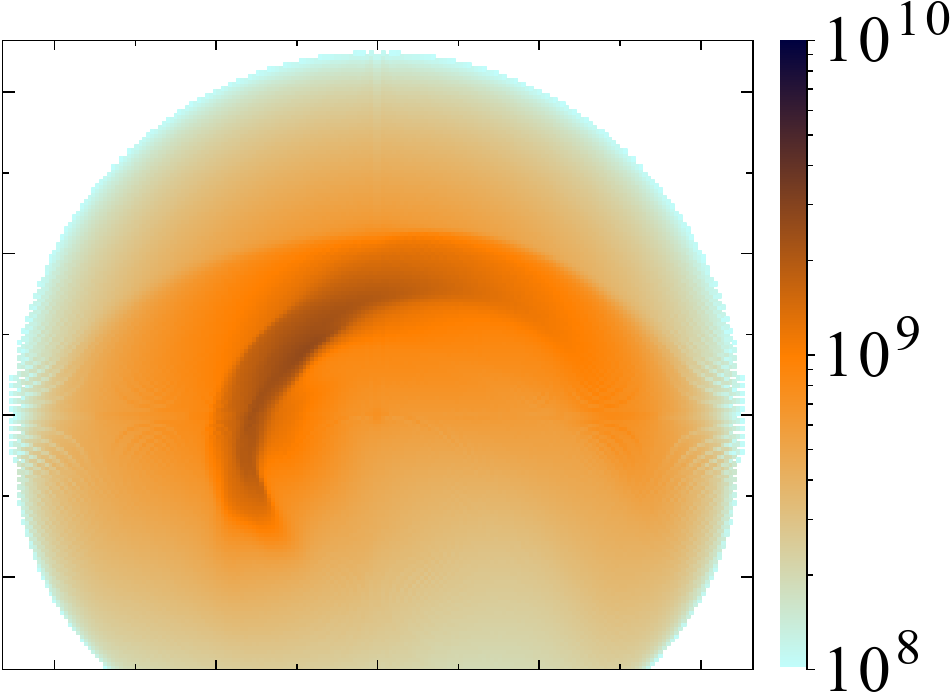}
    \includegraphics[scale=0.26]{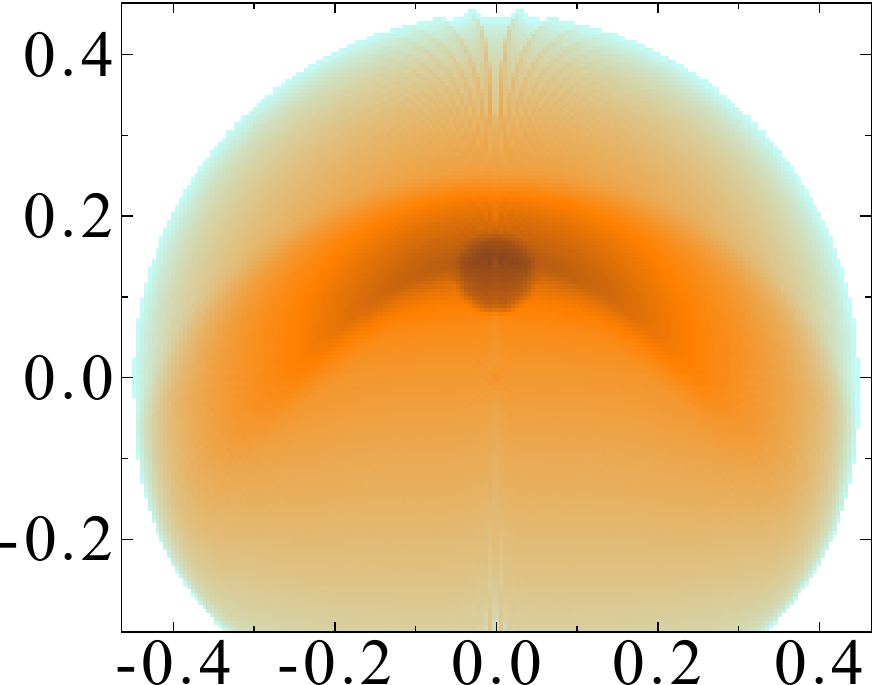}
    \includegraphics[scale=0.26]{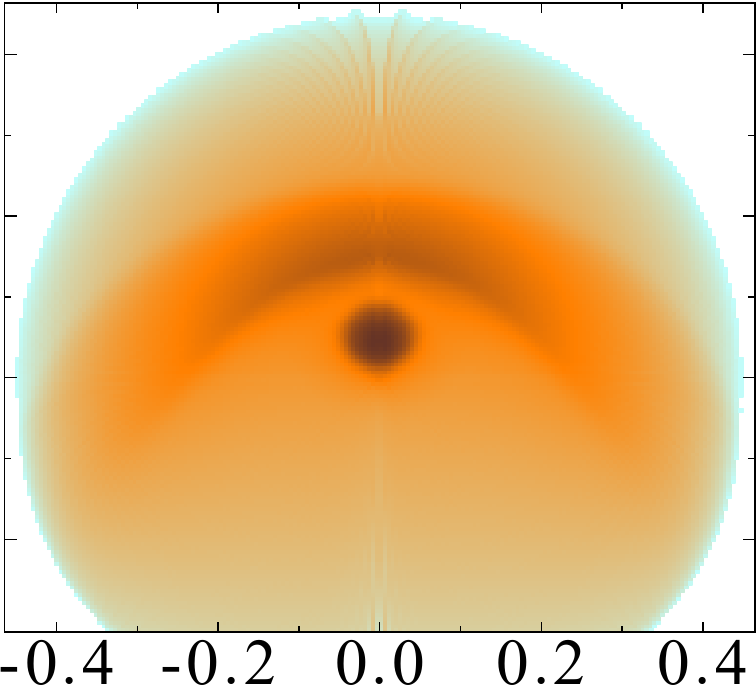}
    \includegraphics[scale=0.26]{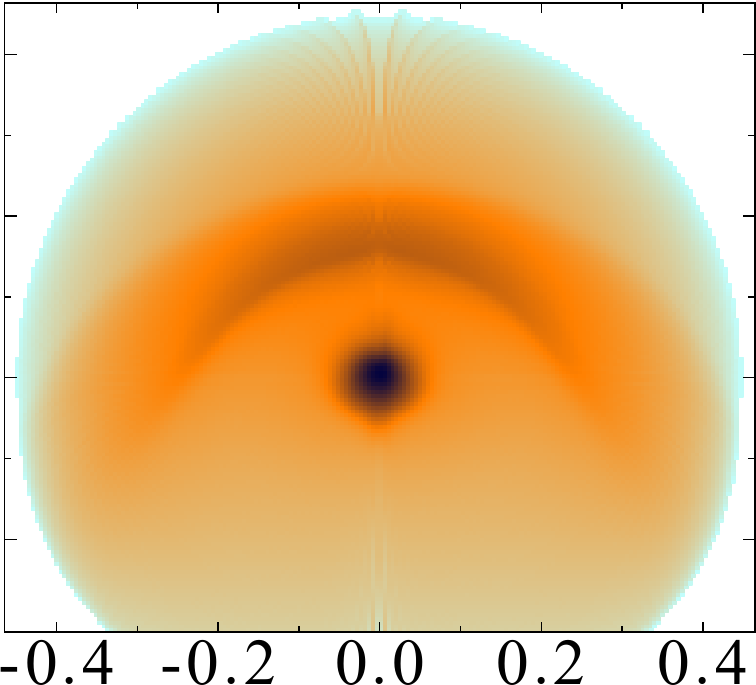}
    \includegraphics[scale=0.26]{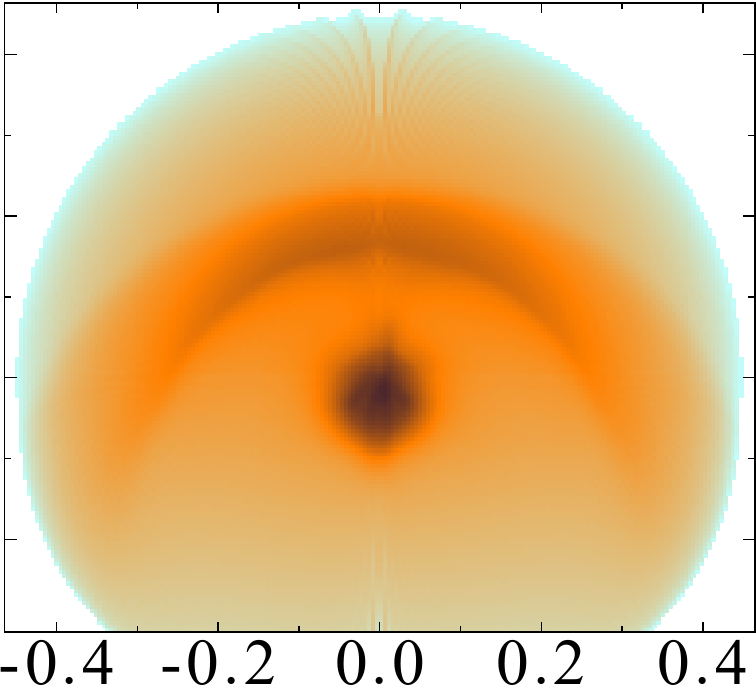}
    \includegraphics[scale=0.26]{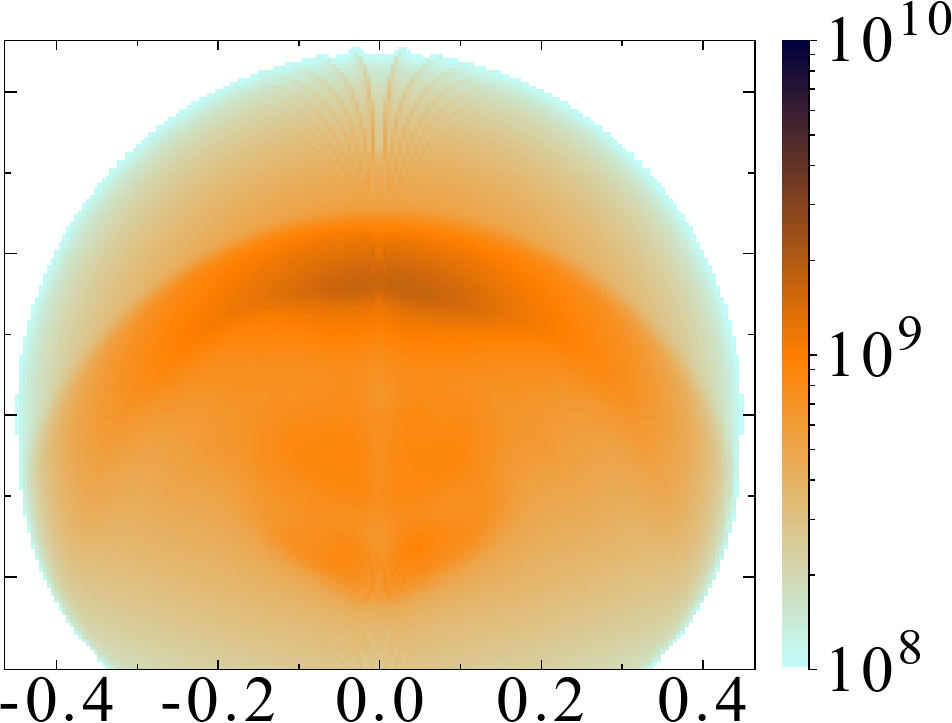}
    \caption{Synthetic observations of the \texttt{n2blb} model at $\Delta t\in\{0, 4, 7, 9, 14\}\times 4.4828\,\si{kyr}$ after injecting the perturbation in $70\,\si{\micro m}$ dust emission $[\si{Jy\,sr^{-1}}]$ at a distance of $617\,\si{pc}$ at different angles: with the LOS in the direction of the star's relative motion (top), at $45\si{\degree}$ to the ecliptic's face-on view (centre), and at $45\si{\degree}$ to the ecliptic's edge-on view (bottom). The solid angle per pixel is $7.13\times 10^{-9}\,\si{sr}.$}
    \label{fig:n2rot}
\end{figure*}

Further projections of the \texttt{n2blb} model in $70\,\si{\micro m}$ dust emission are displayed in Fig.~\ref{fig:n2rot}; the configuration is that of Fig.~\ref{fig:n2blb} for different rotational angles. As in Fig.~\ref{fig:v5rot}, the first two rows are the zero-inclination and the $45\si{\degree}$-inclination face-on ecliptic perspective. Because of the model's asymmetry, the $45\si{\degree}$-inclination edge-on ecliptic perspective is additionally displayed in the bottom row.

At zero inclination, only the perturbation is visible as a disk of high emission that becomes more and more distorted as the perturbation loses its spherical shape. At a time of $\Delta t=9\times 4.4828\,\si{kyr}$ after injection, the former disk has lost all symmetry but remains fairly compact, akin to a Class IV irregular-type interaction as described by \citet{cox12}. The perturbation's shape distorts further, resulting in an arc-like shape at $\Delta t=14\times 4.4828\,\si{kyr}$ after injection that looks similar to the arc-like shapes typically identified as BSs. However, following the classification of \citet{cox12}, unlike  true BSs, this structure would not be classified as a Class I fermata-type interaction because its opening angle is far below $120\si{\degree}$. At an inclination of $45\si{\degree}$, the perturbation's distorted disk detaches from the outer astrosheath's arc shortly after impact ($\Delta t=4\times 4.4828\,\si{kyr}$) and remains discrete until the two structures seemingly coalesce in the face-on perspective ($\Delta t=14\times 4.4828\,\si{kyr}$), producing a spiral-like shape. In the edge-on perspective, this does not occur; instead, the perturbation diffuses to a larger but fainter structure compared to the outer astrosheath's distinct arc. By themselves, the perturbation's distorted disk and the outer astrosheath's arc would be classified as Class IV irregular-type and Class I fermata-type interactions, respectively, whereas together the arc shape prevails for the purpose of classification. However, it should be noted that the arc shapes of the $45\si{\degree}$-inclination face-on perspective can lead to a misidentification of the star's relative motion: By the shape of the arc alone, the star's relative motion appears to point to the top right and the top left at $\Delta t=9\times 4.4828\,\si{kyr}$ and $\Delta t=14\times 4.4828\,\si{kyr}$, respectively.

\section{Conclusions}\label{sec:conc}

Simple perturbations of the number density, the temperature (thermal pressure), the bulk velocity, and the magnetic flux density were injected in front of the model BS (cf. Sect.~\ref{sec:simpleblobs}). While the thermal perturbation dissolved almost immediately and the density perturbation caused no significant changes to the astrosphere's structure, both the velocity and the magnetic perturbations drastically altered the BS shape by adding a bubble-like structure connected to the BS. These modified structures were visible for about $10\,\si{kyr}$, roughly the travel time through the perturbed domain with the ISM's bulk speed (i.e. the speed of the star's relative motion). A larger perturbation of the number density likewise modified the observable structure by creating a bifurcation of the BS. In all examined cases, the astrosphere returned to its previous stationary state after few tens of thousands of years, comparable to the travel time given by the astrosphere's extent and the perturbation's speed of motion. Depending on the orientation of the star's relative motion with respect to the observer's LOS, the resulting images of the perturbed astrospheres differ strongly. Using the classification by \citet{cox12}, the above-mentioned bubble-like structures resulted in Class I fermata-type, Class II ring-type, or Class III eye-type interactions, whereas the bifurcation perturbation was visible as a Class I or a Class IV irregular-type interaction.

From a numerical standpoint, it has become clear that higher angular resolutions are preferable to reduce the uncertainty introduced by numerical diffusion. However, doubling the number of angular cells, for example, would necessitate an increase in computational resources by a factor of eight. Interpolating the model before generating synthetic observations can drastically improve the image resolution, though at the risk of introducing Moir\'e patterns, but it has no impact on the problem of numerical diffusion.

From the presented images, it is apparent that even simple perturbations can cause a variety of different observational shapes. More complex perturbations, as examined for example by \citet{gvaramadze18}, would likely allow the replication of many distorted BS structures. The exact reproduction of a single object, as performed in the aforementioned study, requires high resolutions and precise insights into the respective object and its immediate surroundings, whereas this study's aim is to provide part of that insight from multiple low-resolution simulations.

\begin{acknowledgements}
We appreciate the valuable suggestions and comments of the referee, William Henney. KS is grateful to the \textit{Deutsche Forschungsgemeinschaft} (DFG), funding the project SCHE334/9-2. JK acknowledges financial support through the \textit{Ruhr Astroparticle and Plasma Physics (RAPP) Center}, funded as MERCUR project St-2014-040.
\end{acknowledgements}

\bibliographystyle{aa} 
\bibliography{2020paper}

\end{document}